\documentclass[11pt,a4paper]{article}

\usepackage{epsfig}
\usepackage{graphicx}

\usepackage{latexsym,times}
\usepackage{amsmath}

\usepackage{amsxtra}
\usepackage{amscd}
\usepackage{amsthm}
\usepackage{amsfonts}
\usepackage{amssymb}
\usepackage{dsfont}
\usepackage{xcolor}
\usepackage{cancel}
\usepackage{framed}

\usepackage{esint}
\usepackage{comment}

\usepackage{enumitem}
\usepackage{float}

\usepackage{hyperref}
\usepackage[square,sort&compress,comma,numbers]{natbib}

\makeatletter

\@addtoreset{equation}{section} \makeatother

\input amssym.def
\input amssym.tex

\newcommand\To{T_{\scriptscriptstyle{(0)}}}

\newcommand\brTo{\bar{T}_{\scriptscriptstyle{(0)}}}

\newcommand{\be}{\begin{equation}}
	\newcommand{\ee}{\end{equation} }
\newcommand{\beqa}{\begin{eqnarray} }
	\newcommand{\eeqa}{\end{eqnarray} }
\newcommand{\ba}{\begin{array}}
	\newcommand{\ea}{\end{array}}
\newcommand{\bpm}{\begin{pmatrix}}
	\newcommand{\epm}{\end{pmatrix}}

\newcommand{\rmd}{{\rm d}}

\newcommand{\ODD}{\mathbf{O}(D,D)}

\newcommand\rd{{\rm d}}

\newcommand\cA{{\cal A}}

\newcommand\cC{{\cal C}}

\newcommand\cE{{\cal E}}

\newcommand\cH{{\cal H}}
\newcommand\cI{{\cal I}}
\newcommand\cJ{{\cal J}}

\newcommand\cM{{\cal M}}

\newcommand\cR{{\cal R}}

\newcommand{\Lm}{L_{\rm{m}}}
\newcommand{\cLm}{{\cal L}_{\rm{m}}}

\newcommand\hcB{{\hat{\cal B}}}

\newcommand\hcG{{\hat{\cal G}}}
\newcommand\hcI{{\hat{\cal I}}}
\newcommand\hcJ{{\hat{\cal J}}}

\def\tA{\tilde{A}}
\def\tcA{\tilde{\cal A}}
\def\tB{\tilde{B}}
\def\tC{\tilde{C}}
\def\tcC{\tilde{\cal C}}

\def\tcH{\tilde{\cal H}}

\def\tN{\tilde{N}}

\def\tU{{\tilde{U}}}
\def\tf{\tilde{f}}
\def\tx{\tilde{x}}

\def\tT{\tilde{T}}
\def\tTh{\tilde{\Theta}}
\def\tsi{\tilde{\sigma}}
\def\tSigma{\tilde{\Sigma}}

\def\tZ{\tilde{Z}}

\def\ta{\tilde{a}}
\def\tc{\tilde{c}}
\def\tg{\tilde{g}}
\def\th{\tilde{h}}
\def\tn{\tilde{n}}
\def\tp{\tilde{p}}
\def\ts{\tilde{s}}
\def\tv{\tilde{v}}
\def\tw{\tilde{w}}
\def\tDelta{\tilde{\Delta}}
\def\tdelta{\tilde{\delta}}
\def\trho{\tilde{\rho}}
\def\tPi{\tilde{\Pi}}
\def\tpi{\tilde{\pi}}
\def\tlambda{\tilde{\lambda}}
\def\tzeta{\tilde{\zeta}}

\def\bre{\bar{e}}

\def\breta{\bar{\eta}}

\def\brrho{\bar{\rho}}

\def\brsigma{\bar{\sigma}}

\def\brp{{\bar{p}}}
\def\brq{{\bar{q}}}

\def\brpi{{\bar{\pi}}}

\def\brPhi{{{\bar{\Phi}}}}

\def\brB{\bar{B}}

\def\brH{\bar{H}}
\def\brK{\bar{K}}

\def\brV{{\bar{V}}}

\def\brP{{\bar{P}}}
\def\brT{{\bar{T}}}

\def\brTheta{\bar{\Theta}}

\def\brtp{\bar{\tilde{p}}}
\def\brtrho{\bar{\tilde{\rho}}}
\def\brtsigma{\bar{\tilde{\sigma}}}

\def\ckrho{\check{\rho}}

\def\tcI{\tilde{\cI}}
\def\tcJ{\tilde{\cJ}}
\def\tGamma{\tilde{\Gamma}}

\newcommand\p\partial

\newcommand{\nn}{\nonumber}

\newcommand{\brg}{\bar{g}}

\newcommand{\brphi}{\bar{\phi}}
\newcommand{\brd}{\bar{d}}

\begin{document}
	
\begin{titlepage}
\title{Perturbations in $\ODD$ string cosmology \\ from double field theory}

\author{Stephen Angus${}^{1}$ and Shinji Mukohyama${}^{2,3}$}
		
\date{}
\maketitle 
\vspace{-1.0cm}
\begin{center}
	\mbox{\!\!\!\!\!\!\!\!${}^{1}$Center for Quantum Spacetime, Sogang University, 35 Baekbeom-ro, Mapo-gu, Seoul 04107, Republic of Korea}\\
	\mbox{\!\!\!\!\!\!${}^{2}$Center for Gravitational Physics and Quantum Information, 
		Yukawa Institute for Theoretical Physics,}\\ 
	\mbox{\!\!\!Kyoto University,
		606-8502, Kyoto, Japan}\\
	\mbox{\!\!\!\!\!\!${}^{3}$Kavli  Institute  for  the  Physics  and  Mathematics  of  the  Universe  (WPI),}\\
	\mbox{\!\!\!\!\!\!\!\!\!\!\!The  University  of  Tokyo  Institutes  for  Advanced  Study, The  University  of  Tokyo,  Kashiwa,  Chiba  277-8583, Japan}\\
	~\\
	\texttt{sangus@sogang.ac.kr\qquad 
		shinji.mukohyama@yukawa.kyoto-u.ac.jp}\\
	~\\
\end{center}
\begin{abstract}
	\noindent
	The low-energy limit of string theory contains additional gravitational degrees of freedom, a skew-symmetric tensor $B$-field and a scalar dilaton, that are not present in general relativity.  Together with the metric, these three fields are naturally embedded in the $\ODD$-symmetric framework of double field theory.  The $\ODD$ symmetry uniquely prescribes the interactions between the extended gravitational sector and other matter, leading to novel features beyond conventional string cosmology.  In this work we  
	present the equations of motion for linear perturbations around $\ODD$ string cosmological backgrounds in $D=4$ 
	under a scalar-vector-tensor decomposition.  We obtain analytic solutions in the superhorizon limit for scalar perturbations around various homogeneous and isotropic background solutions, including some candidate models for bouncing cosmologies.  The generalized energy-momentum tensor includes source terms for the $B$-field and dilaton, and we show how the resulting generalized conservation laws modify the conditions for conservation of curvature perturbations.
\end{abstract}
\thispagestyle{empty}
{\small
	\begin{flushright}
		\textit{Preprint}: CQUeST-2024-0743, YITP-24-102, IPMU24-0035
\end{flushright}}
\end{titlepage}
\newpage

\tableofcontents 


\section{Introduction}
The standard $\Lambda$CDM model of cosmology, based on general relativity (GR) with dark matter and a cosmological constant $\Lambda$, provides a simple and powerful model of the universe which is in excellent agreement with data~\cite{Planck:2018vyg,DESI:2024mwx}.  However, several key features of the model remain mysterious, including the nature of dark matter, the origin and isotropy of cosmic microwave background (CMB) fluctuations, and the smallness of $\Lambda$ relative to naive expectations from quantum field theory.  Furthermore, in recent years new tensions have emerged, for example, between estimations of the Hubble constant $H_{0}$ inferred from early-time (CMB) versus late-time (astrophysical) observations~\cite{Riess:2021jrx}.  As the quality and volume of data about the early universe continues to increase~\cite{Labbe:2022ahb}, it is pertinent to study whether modified gravity beyond GR can alleviate such tensions and reveal how the early universe evolved to its present state.

The question is, how should we modify GR?  Is there any guiding principle we can use to select a proposed modified gravity framework from the plethora of possible candidates?  
In the very early universe, whose imprint we observe in the CMB, the quantum nature of gravity may become relevant.  To this end, string theory remains the leading candidate for a consistent framework of quantum gravity; 
however, direct evidence for the stringy nature of gravity remains elusive.  It is therefore worth considering a modified gravity scenario that captures universal features of string theory that arise at low energies.  Such a model may be used to address mysteries in cosmology through the perspective of a well-established theoretical framework.  Meanwhile, comparison with data provides an avenue to potentially assess the applicability of string theory to our universe.

The original impetus for considering string theory as quantum gravity is that the massless modes of the closed string include a spin-$2$ mode, which can be identified as a graviton, i.e. fluctuations of the spacetime metric $g_{\mu\nu}$.  However, string theory does not automatically reduce to GR alone: the closed-string massless sector (or the NS-NS sector of the closed superstring) also includes a skew-symmetric tensor `$B$-field' $B_{\mu\nu}$ and a dilaton $\phi$.  It is these three fields together, $\{g_{\mu\nu},B_{\mu\nu},\phi\}$, that constitute the generic `gravitational' multiplet arising in the low-energy macroscopic limit of string theory.  In the case where $B_{\mu\nu}$ vanishes and $\phi$ is constant, which may be achieved in concrete string compactifications via the presence of appropriate fluxes in the compact geometry~\cite{Giddings:2001yu},
at low energies and late times the dynamics of the theory can be reduced to those of GR.  However, in the early universe it is conceivable that all three modes could be active, and together they would constitute a theory of `stringy gravity'.

Moreover, the relationship between $g_{\mu\nu}$, $B_{\mu\nu}$ and $\phi$ extends beyond mere co-appearance.  The closed-string effective action that encapsulates their dynamics is invariant under a hidden $\ODD$ symmetry, where $D$ is the spacetime dimension,\footnote{This should not be confused with $\mathbf{O}(d,d)$ symmetry, where $d = D - 1$ is the spatial dimension, which is a property of the particular spacetime backgrounds considered in string cosmology~\cite{Meissner:1991zj,Meissner:1991ge,Gasperini:1991ak}.} and the three fields transform into each other non-trivially under the Buscher rules~\cite{Buscher:1987sk,Buscher:1987qj}.  The $\ODD$ symmetry can be made manifest via the framework of \textit{double field theory} (DFT)~\cite{Siegel:1993xq,Siegel:1993th,Hull:2009mi,Hull:2009zb,Hohm:2010jy,Hohm:2010pp}.  In double field theory the spacetime dimension is formally doubled, such that fields act on doubled coordinates $X^{A} = (x^{\mu},\tx_{\nu})$.  While this may appear to dramatically increase the number of degrees of freedom, consistency demands a constraint, 
the so-called `section condition', which
can be satisfied by simply requiring all fields to be independent of the additional $\tx_{\nu}$ coordinates, and thus ultimately $D$-dimensional physics is recovered.
 
In DFT gravitational physics is characterized by two fields: a generalized metric $\cH_{AB}$ and a scalar $d$ known as the DFT dilaton, which play a role analogous to the spacetime metric $g_{\mu\nu}$ in GR. On ordinary spacetime backgrounds, $\cH_{AB}$ and $d$ encapsulate all the degrees of freedom of the closed-string fields $\{g_{\mu\nu},B_{\mu\nu},\phi\}$, and furthermore
a stringy geometry can be constructed from them~\cite{Hitchin:2003cxu,Gualtieri:2003dx,Hitchin:2010qz,Coimbra:2011nw,Coimbra:2012yy,Vaisman:2012ke,Berman:2013uda,Garcia-Fernandez:2013gja,Cederwall:2014kxa,Cederwall:2016ukd,Deser:2016qkw,Sakamoto:2017cpu,Cederwall:2017fjm,Freidel:2018tkj,Chatzistavrakidis:2019huz,Jeon:2010rw,Jeon:2011cn,Hohm:2011si}, whose Ricci scalar reduces to the NS-NS action of the closed string.
In this description, the Buscher transformations turn out to be simply $\ODD$ rotations of $\cH_{AB}$.
It is important to emphasize that general $\ODD$ transformations do not preserve the spacetime metric, but rather the symmetry is broken spontaneously by a particular choice of vacuum.  Nevertheless, its presence suggests that one may consider it as a symmetry of the underlying fundamental theory.  Moreover, various other types of matter --- scalars, spinors, point particles, strings, etc. --- can be incorporated into the DFT formalism, and explicit $\ODD$-covariant Lagrangians have been constructed~\cite{Choi:2015bga,Jeon:2011vx,Jeon:2011kp,Hohm:2011ex,Hohm:2014sxa,Cho:2018alk,Jeon:2012kd,Ko:2016dxa,Blair:2017gwn,Hull:2006va,Lee:2013hma,Park:2016sbw,Sakamoto:2018krs}.  In conjunction with the gravitational action, this leads to a DFT generalization of Einstein's equations, dubbed the \textit{Einstein double field equations}~\cite{Angus:2018mep},
\begin{equation}
	G_{AB} = 8\pi G T_{AB} \, , \label{eq:EDFEDFT}
\end{equation}
where $G_{AB}$ is a DFT analogue of the Einstein tensor depending on $\cH_{AB}$ and $d$, and the \textit{DFT energy-momentum tensor} $T_{AB}$ takes a specific form prescribed by the $\ODD$ covariance of the underlying action.

Thus we see that beyond being merely an exercise in academic formalism, DFT restricts the allowed couplings between the generalized gravity sector and additional matter via the requirement of consistency with the underlying $\ODD$ symmetry.  In addition, the DFT energy-momentum tensor includes components corresponding to sources for $B_{\mu\nu}$ and $\phi$, beyond the usual $T_{\mu\nu}$ sourcing the metric.\footnote{Source terms for the dilaton have also been considered in~\cite{Gasperini:2007ar,Gasperini:2004ss}.}  Applying these ideas in a cosmological setting, DFT thus provides a rigorous and expanded framework that encompasses, and at times extends beyond, the well-studied string cosmology~\cite{Gasperini:2002bn,Tseytlin:1991xk,Brustein:1997ny,Gasperini:2007ar,Gasperini:2004ss,Gasperini:2007zz,Meissner:1991zj,Meissner:1991ge,Gasperini:1991ak,Copeland:1994vi,Lidsey:1999mc,Cicoli:2023opf}.  In~\cite{Angus:2019bqs} a study of $\ODD$-covariant cosmology was initiated by applying equation~\eqref{eq:EDFEDFT} to homogeneous and isotropic backgrounds of DFT, leading to an $\ODD$ version of the Friedmann equations, and various cosmological solutions were considered.  For related recent work on DFT cosmology, see~\cite{Wu:2013sha,Wu:2013ixa,Ma:2014ala,Brandenberger:2017umf,Brandenberger:2018xwl,Brandenberger:2018bdc,Bernardo:2019pnq,Hohm:2019jgu,Hohm:2019ccp,Bernardo:2019bkz,Codina:2021cxh,Bernardo:2020zlc,Quintin:2021eup,Bernardo:2022nex,Nunez:2020hxx,Bernardo:2021xtr,Gasperini:2023tus,Lescano:2021nju,Lescano:2023gge,Arapoglu:2024umz} as well as the review~\cite{Brandenberger:2023ver}.

Our goal in this paper is to analyse perturbations in the $\ODD$-covariant DFT cosmological framework,\footnote{Cosmological perturbations in DFT have been studied in~\cite{Hohm:2022pfi} for the case of a flat FLRW background with vanishing $B$-field, constant dilaton, and in the absence of external matter.}  
and in doing so to take the next major step towards confronting data.  Observables such as CMB fluctuations and structure formation depend upon the evolution of perturbations in the early history of the universe, for which DFT cosmology hints at significant departures from the standard assumptions of GR cosmology.  For example, the modified energy-momentum conservation equation in DFT implies that the conditions for conservation of adiabatic superhorizon perturbations should be revisited in this framework.  Moreover, 
in~\cite{Angus:2019bqs} backgrounds were found with non-vanishing $H$-flux (the field strength of $B_{\mu\nu}$), including bouncing solutions with a frozen dilaton at late times.  It turns out that the presence of background $H$-flux leads to non-trivial mixing between different types of perturbations, which may play a dominant role in the early-universe evolution.  
That being said, we will endeavour to remain agnostic on on the specific interpretation and implementation: our main focus in this work is to develop the formalism for DFT cosmological perturbations, which can then be readily applied to particular scenarios as required. 

The structure of this paper is as follows.  In section~\ref{sec:rev} we provide an accessible overview of DFT coupled to matter, establish our conventions for the DFT energy-momentum tensor, and review the application to cosmological backgrounds.  Following this, in section~\ref{sec:pert} we introduce perturbations in the gravity and matter sectors, obtain their equations of motion, and assess the implications of the generalized energy-momentum conservation for early-universe physics.  We consider various examples of cosmological backgrounds in section~\ref{sec:ex} and obtain explicit solutions to the perturbation equations.  Finally, we discuss implications for future work in section~\ref{sec:discuss}.

\section{$\ODD$ string cosmology from double field theory} \label{sec:DFT}
We begin with a brief overview of double field theory and its coupling to matter in section~\ref{sec:rev}.  Here we aim for a balanced approach, giving the reader a flavor of DFT and introducing necessary ingredients while skipping all technical derivations --- for a full exposition see, for example, section 2 of~\cite{Angus:2018mep}.  Following this, we discuss the DFT energy-momentum tensor and its relation to fluid variables in section~\ref{sec:DFTemtensor}, in order to establish and motivate our conventions.  Finally, in section~\ref{sec:cosmobkgds} we review the application of the DFT framework to cosmological backgrounds, as presented in~\cite{Angus:2019bqs}.

\subsection{Double field theory coupled to matter} \label{sec:rev}
In double field theory we describe $D$-dimensional physics using $D+D$ coordinates, $X^{A} = (x^{\mu},\tx_{\nu})$, where doubled indices, $A = 1,\ldots, 2D$, are raised and lowered by contracting with the $\ODD$ invariant
\begin{equation}
	\cJ_{AB}={\bf\left(
		\begin{array}{cc} 
			0 & \mathbf{1}_{D} \\ \mathbf{1}_{D} & 0
		\end{array}\right)}
	\normalsize = \cJ^{AB} \; , \label{eq:ODDinv}
\end{equation}
where $\mathbf{1}_{D}$ is the $D\times D$ unit matrix.  For consistency,\footnote{Symmetries of NS-NS gravity include diffeomorphisms and $B$-field gauge transformations, which in DFT are unified into \textit{doubled diffeomorphisms}.  Closure of the symmetry algebra of doubled diffeomorphisms requires the section condition.} fields in DFT are constrained to satisfy the section condition,
\begin{equation}
	\p_{A}\p^{A}=2\,\p_{\mu}\tilde{\p}^{\mu} = 0 \, , \label{eq:sectioncondition}
\end{equation}
where each derivative is understood to act on a field or product of fields.  This is conveniently solved by choosing $\tilde{\p}^{\mu} = 0$, such that all fields are independent of the $\tx_{\mu}$ coordinates.

The gravitational fields in double field theory are
the generalized metric $\cH_{AB}$ and the DFT dilaton $d$.  The DFT metric is symmetric, $\cH_{AB} = \cH_{BA}$, and satisfies the $\ODD$ property
\begin{equation}
	\cH_{A}{}^{C}\cH_{B}{}^{D}\cJ_{CD}=\cJ_{AB}\,. \label{eq:ODDcond}
\end{equation}
Meanwhile, the DFT dilaton $d$ provides a scalar density of unit weight, $e^{-2d}$, that serves as the DFT volume element.
On ordinary Riemannian/Lorentzian spacetime backgrounds,\footnote{DFT also admits \textit{non-Riemannian backgrounds}~\cite{Morand:2017fnv}, on which the usual spacetime metric becomes singular while the DFT metric remains regular.  Such backgrounds encompass various `non-relativistic' gravities such as Newton--Cartan and Carrollian gravity, and their properties hint at surprising connections to exotic phases of matter such as fractons~\cite{Angus:2021jvm}.} the three NS-NS sector fields $\{g_{\mu\nu}, B_{\mu\nu}, \phi\}$ are packaged together into $\cH_{AB}$ and $d$ as
\begin{equation}
	\cH_{AB} = \left(\begin{array}{cc}g^{\mu\nu} & -g^{\mu\sigma}B_{\sigma\nu} \\ B_{\mu\rho}g^{\rho\nu} & g_{\mu\nu} - B_{\mu\rho}g^{\rho\sigma}B_{\sigma\nu}
	\end{array}\right) \, , \qquad
	e^{-2d} = e^{-2\phi}\sqrt{-g} \, ,
\end{equation}
where $g \equiv \det(g_{\mu\nu})$.
From $\cH_{AB}$ and $d$ a doubled geometry can be formulated,\footnote{It is often convenient to decompose the DFT metric into left and right projectors,
\begin{equation}
	P_{AB} \equiv \frac{1}{2}(\cJ_{AB} + \cH_{AB}) \, , \qquad \brP_{AB} \equiv \frac{1}{2}(\cJ_{AB} - \cH_{AB}) \, , \label{eq:PbrPdef}
\end{equation}
where the $2D\times2D$ matrix $\cJ_{AB}$ is the $\ODD$ invariant element~\eqref{eq:ODDinv}.
Moreover, it is sometimes necessary to introduce a pair of DFT vielbeins, $V_{Ap}$ and $\brV_{A\brp}$, which are defined implicitly as
\begin{equation}
	P_{AB} \equiv V_{A}{}^{p}V_{Bp} \, , \qquad \brP_{AB} = \brV_{A}{}^{\brp}\brV_{B\brp} \, , \label{eq:VbrVdef}
\end{equation}
where $p$ and $\brp$ are independent $D$-dimensional local Lorentz indices.
This structure should be used, for example, when dealing with spinors in DFT.  Although we will not consider fermionic matter explicitly at a Lagrangian level in this paper, the fact that there are two independent local Lorentz groups in DFT could have interesting phenomenological implications, for example due to a possible coupling between fermions and $H$-flux.}
whose Ricci scalar $\cR$ provides the gravitational action
\begin{equation}
	S_{\rm grav} \equiv \frac{1}{16\pi G}\int_{\Sigma}\,e^{-2d}\cR \, , \label{eq:DFTgravaction}
\end{equation}
where $\Sigma$ is a $D$-dimensional section of the doubled spacetime.  On Riemannian/Lorentzian backgrounds, the DFT Ricci scalar reduces to the form
\begin{equation}
	\cR = R + 4\Box\phi - 4\nabla_{\mu}\phi\nabla^{\mu}\phi - \frac{1}{12}H_{\lambda\mu\nu}H^{\lambda\mu\nu} \, , \label{eq:NSNSLagrangian}
\end{equation}
where $R$ is the usual Ricci scalar constructed from $g_{\mu\nu}$ and the $H$-flux, $H_{\lambda\mu\nu} = 3\p_{[\lambda}B_{\mu\nu]}$, 
is the field strength of $B_{\mu\nu}$.

We can extend the $\ODD$-covariant DFT formalism to include additional matter fields, denoted $\{\Upsilon_{a}\}$, via an action of the form
\begin{equation}
	S = \int_{\Sigma}e^{-2d}\left[\,\frac{1}{16\pi G}\cR+L_{\rm{m}}(\Upsilon_{a})\,\right]\;, \label{eq:DFTtotalaction}
\end{equation}
where the DFT matter Lagrangian $L_{\rm{m}}$ is known for various types of matter embedded into DFT, including scalars~\cite{Choi:2015bga}, fermions~\cite{Jeon:2011vx}, vectors~\cite{Jeon:2011kp,Hohm:2011ex,Hohm:2014sxa,Cho:2018alk}, point particles~\cite{Ko:2016dxa,Blair:2017gwn}, strings~\cite{Hull:2006va,Lee:2013hma,Park:2016sbw}, etc.  The fact that integral volumes in DFT must be written as $e^{-2d}$, and not simply $\sqrt{-g}$ as in GR, has significant implications for the allowed couplings between the stringy gravity of DFT and additional matter.  Specifically, since all matter in DFT should be uniformly coupled in string frame with a factor $e^{-2d}$, this means that depending on the type of matter, `minimal coupling' in Einstein frame is not guaranteed.

The variation of~\eqref{eq:DFTtotalaction} gives the equations of motion~\eqref{eq:EDFEDFT} for DFT coupled to matter.
On spacetime backgrounds,
these \textit{Einstein double field equations} can be decomposed into the component equations~\cite{Angus:2018mep}
\beqa
R_{\mu\nu} + 2\nabla_{\mu}\nabla_{\nu}\phi - \frac{1}{4}H_{\mu\rho\sigma}H_{\nu}{}^{\rho\sigma} &=& 8\pi G K_{(\mu\nu)} \, , \label{eq:EDFER1} \\
\nabla^{\rho}\left(e^{-2\phi}H_{\rho\mu\nu}\right) &=& 16\pi Ge^{-2\phi}K_{[\mu\nu]} \, , \label{eq:EDFER2} \\
R + 4\Box\phi - 4\p_{\mu}\phi\p^{\mu}\phi - \frac{1}{12}H_{\lambda\mu\nu}H^{\lambda\mu\nu} &=& 8\pi G\To \label{eq:EDFER3} \, ,
\eeqa
where $K_{\mu\nu}$ and $\To$ arise from the DFT energy-momentum tensor $T_{AB}$, which will be discussed further in the next section.  For now, we remark that the DFT Einstein tensor in~\eqref{eq:EDFEDFT} satisfies a Bianchi identity of the form $\nabla^{A}G_{AB} = 0$~\cite{Park:2015bza}, which implies that the DFT energy-momentum tensor is conserved on-shell,
\begin{equation}
	\nabla^{A}T_{AB} = 0 \, . \label{eq:DFTcon}
\end{equation}
This conservation law includes additional components beyond GR, which will have implications for the evolution of perturbations on cosmological backgrounds.

\subsection{The DFT energy-momentum tensor} \label{sec:DFTemtensor}
A central feature of the $\ODD$ theory is that the energy-momentum tensor is generalized.  Whereas in GR the energy-momentum tensor is characterized as the response of matter to metric perturbations, in DFT the response due to the extended gravitational sector also includes terms sourcing the $B$-field and dilaton.  Here we examine in more detail the energy-momentum tensor in DFT and its relation with the standard energy-momentum tensor in GR.

\subsubsection{Preliminaries}
First recall that in GR, the energy-momentum tensor is defined as
\begin{equation}
	T_{\mu\nu} \equiv \frac{-2}{\sqrt{-g}}\frac{\delta S_{\rm m}}{\delta g^{\mu\nu}} \, ,
\end{equation}
where the matter action is expressed in terms of a Lagrangian density as
\begin{equation}
	S_{\rm{m}} = \int\rd^{D}x\,\sqrt{-g}\cLm \, .
\end{equation}
The variation of the matter action with respect to the (inverse) metric $g^{\mu\nu}$ is
\begin{equation}
	\delta S_{\rm{m}} 
	= \int\rd^{D}x\,\frac{\delta S_{\rm m}}{\delta g^{\mu\nu}}\delta g^{\mu\nu}
	= \int\rd^{D}x\,\sqrt{-g}\left[-\frac{1}{2}T_{\mu\nu}\delta g^{\mu\nu}\right] \, . \label{eq:deltaSGR}
\end{equation}

In DFT the gravitational multiplet is given by the $\ODD$ metric $\cH_{AB}$ and the DFT dilaton $d$.  When we couple DFT to additional matter as in~\eqref{eq:DFTtotalaction}, the corresponding matter action 
takes the form
\begin{equation}
	S_{\rm{m}} = \int_{\Sigma}e^{-2d}\Lm \, ,
\end{equation}
where in general the matter Lagrangian $\Lm$ is a functional of $\cH_{AB}$, $d$ and the additional matter fields 
$\{\Upsilon_{a}\}$.  Note that the covariant integration measure is given by $e^{-2d}$.  The total variation of the matter action includes terms proportional to $\delta\Upsilon_{a}$, which yield the equations of motion for the matter fields.  Meanwhile, the variation with respect to the gravitational fields is given by
\begin{equation}
	\delta S_{\rm{m}}
	= \int_{\Sigma}\left[\frac{\delta S_{\rm m}}{\delta\cH_{AB}}\delta\cH_{AB} + e^{-2d}\To\delta d\right] \,  \label{eq:deltaSDFT},
\end{equation}
where we have defined
\begin{equation}
	\To \equiv e^{2d}\frac{\delta S_{\rm m}}{\delta d} \, .
\end{equation}

\subsubsection{Riemannian backgrounds}
We would now like to evaluate~\eqref{eq:deltaSDFT} on Riemannian spacetime backgrounds and thereby understand the various components of the DFT energy-momentum tensor.
For now, and in the remainder of this paper, let us focus on situations where the matter Lagrangian does not depend explicitly on the local frame structure 
(in particular, we will not consider DFT spinors in this work).  
In such cases, 
we may define the spacetime `kinetic' DFT energy-momentum tensor as \cite{Angus:2018mep}\footnote{See Appendix~\ref{sec:KgB} for explicit definitions of the DFT vielbeins $V_{Ap}$ and $\brV_{B\brq}$ as well as the spacetime vielbeins $e_{\mu}{}^{p}$ and $\bre_{\nu}{}^{\brq}$.
}
\begin{equation}
	K_{\mu\nu} 
	\equiv -4e_{\mu}{}^{p}\bre_{\nu}{}^{\brq}V_{Ap}\brV_{B\brq}e^{2d}\frac{\delta S_{\rm m}}{\delta\cH_{AB}} 
	\, . \label{eq:Kmunu}
\end{equation}
This tensor is not necessarily symmetric, reflecting the generalized energy-momentum structure of DFT.

By expanding~\eqref{eq:Kmunu} explicitly and separating $K_{\mu\nu}$ into independent symmetric and skew-symmetric components, the first term of~\eqref{eq:deltaSDFT} can be expressed in spacetime variables as (see Appendix~\ref{sec:KgB})
\begin{equation}
	e^{2d}\frac{\delta S_{\rm m}}{\delta\cH_{AB}}\delta\cH_{AB} = -\frac{1}{2}K_{(\mu\nu)}\delta g^{\mu\nu} - \frac{1}{2}K_{[\mu\nu]}g^{\mu\lambda}g^{\nu\tau}\delta B_{\lambda\tau} \, . \label{eq:KgB}
\end{equation}
Similarly, the variation of the DFT dilaton $d$ can be expanded as
\begin{equation}
	\delta d  
	= \delta\phi + \frac{1}{4}g_{\mu\nu}\delta g^{\mu\nu} \, . \label{eq:dphig}
\end{equation} 
Inserting \eqref{eq:KgB} and \eqref{eq:dphig} into \eqref{eq:deltaSDFT}, we obtain the variation of the matter action in spacetime variables,
\begin{equation}
	\delta S_{\rm{m}} = \int\rd^{D}x\,\sqrt{-g}e^{-2\phi}\left[-\frac{1}{2}\left(K_{(\mu\nu)} - \frac{1}{2}g_{\mu\nu}\To\right)\delta g^{\mu\nu} - \frac{1}{2}K_{[\mu\nu]}g^{\mu\lambda}g^{\nu\tau}\delta B_{\lambda\tau} + \To\delta\phi\right] \, . \label{eq:deltaSgBphi}
\end{equation}

Comparing with \eqref{eq:deltaSGR}, we see that the standard `metric' energy-momentum tensor in string frame is
\begin{equation}
	T_{\mu\nu} = e^{-2\phi}\left(K_{(\mu\nu)} - \frac{1}{2}g_{\mu\nu}\To\right) \, . \label{eq:EMTmetric}
\end{equation}
It is also natural to define energy-momentum tensors for the $B$-field and dilaton, respectively, as
\begin{equation}
	\Theta^{\mu\nu} \equiv e^{-2\phi}g^{\mu\rho}g^{\nu\sigma}K_{[\rho\sigma]} \, , \qquad
	\sigma \equiv e^{-2\phi}\To \, . \label{eq:EMTBdil}
\end{equation}
These represent source terms in the $B$-field and dilaton equations of motion.  Note that in string frame, the dilaton decouples when $\To = 0$.

The DFT energy-momentum tensor conservation equation~\eqref{eq:DFTcon} can also be expanded in spacetime components, leading to generalized conservation laws which can be written as
\begin{align}
	\nabla^{\mu}T_{\mu\nu} + \frac{1}{2}H_{\nu\mu\lambda}\Theta^{\mu\lambda} - \nabla_{\nu}\phi\,\sigma &= 0 \, , \label{eq:conTThs1} \\
	\nabla^{\mu}\Theta_{\mu\nu} &= 0 \, . \label{eq:conTThs2}
\end{align}
Thus the standard conservation equation in GR is augmented by the presence of source terms depending on $\Theta_{\mu\nu}$ and $\sigma$, while $\Theta_{\mu\nu}$ itself is also conserved.

\subsubsection{Generalized fluids}
Let us apply this framework to a fluid.  Our immediate goal is to identify how the familiar description of a fluid in GR (which we will extend in section~\ref{sec:matpert} to include $B$-field and dilaton source terms) is related to the components of the DFT energy-momentum tensor.\footnote{See~\cite{Lescano:2021nju,Lescano:2023gge,Lescano:2024vrq} for a T-duality invariant construction of generalized fluids in DFT.}  We can define the energy density $\rho$ of a fluid 
as the timelike eigenvalue of the metric energy-momentum tensor,
\begin{equation}
	T^{\mu}{}_{\nu}U^{\nu} \equiv -\rho U^{\mu} \, , \label{rhoeigen}
\end{equation}
where $U^{\mu}$ is a timelike eigenvector 
that characterizes the $4$-velocity of the fluid and is normalized such that
\begin{equation}
	g_{\mu\nu}U^{\mu}U^{\nu} = -1 \, .
\end{equation}
Therefore
\begin{equation}
	\rho = T^{\mu}{}_{\nu}U_{\mu}U^{\nu} \, . \label{eq:rhodef}
\end{equation}

Furthermore, we can construct the projection operator
\begin{equation}
	h^{\mu}{}_{\nu} \equiv \delta^{\mu}{}_{\nu} + U^{\mu}U_{\nu} \, . \label{phproj}
\end{equation}
which satisfies $U_{\mu}h^{\mu}{}_{\nu} = 0$ and $h^{\mu}{}_{\nu}U^{\nu} = 0$.  Thus any covectors $h_{\mu\nu}V^{\nu}$ lie in spacelike hypersurfaces orthogonal to $U^{\mu}$.  In general $T^{\mu}{}_{\nu}$ may have a component proportional to $h^{\mu}{}_{\nu}$, so we can decompose
\begin{equation}
	T^{\mu}{}_{\nu} = \rho U^{\mu}U_{\nu} + p\left(h^{\mu}{}_{\nu} + \pi^{\mu}{}_{\nu}\right) \, , \label{eq:Tgeneral}
\end{equation}
where $p$ is the pressure
\begin{equation}
	p\equiv \frac{1}{3} h^{\nu}{}_{\mu}T^{\mu}{}_{\nu} \, \label{eq:pdef}
\end{equation}
and the anisotropic stress-energy $\pi^{\mu}{}_{\nu}$ satisfies
\begin{equation}
	U_{\mu}\pi^{\mu}{}_{\nu} = 0 \, , \qquad \pi^{\mu}{}_{\nu}U^{\nu} = 0 \, , \qquad h^{\nu}{}_{\mu}\pi^{\mu}{}_{\nu} = \pi^{\mu}{}_{\mu} = 0 \,  \label{eq:anistresscond}
\end{equation}
by construction.  Using the fact that $\pi^{\mu}{}_{\nu} = h^{\mu}{}_{\rho}\pi^{\rho}{}_{\nu}$, the anisotropic stress-energy can be obtained from~\eqref{eq:Tgeneral},
\begin{equation}
	p\pi^{\mu}{}_{\nu} \equiv h^{\mu}{}_{\rho}T^{\rho}{}_{\nu} - ph^{\mu}{}_{\nu}
	= h^{\mu}{}_{\rho}T^{\rho}{}_{\nu} - \frac{1}{3}h^{\mu}{}_{\nu}h^{\sigma}{}_{\rho}T^{\rho}{}_{\sigma} \, . \label{eq:Pidef}
\end{equation}
A perfect fluid is one whose energy-momentum tensor is completely defined by $\rho$ and $p$, such that
\begin{equation}
	T^{\mu}{}_{\nu} = \rho U^{\mu}U_{\nu} + ph^{\mu}{}_{\nu} = (\rho + p)U^{\mu}U_{\nu} + p\delta^{\mu}{}_{\nu} \, . \label{Tperfect}
\end{equation}

The fluid variables can thus be related to the components of the DFT energy-momentum tensor using \eqref{eq:EMTmetric}.  From the definitions \eqref{eq:rhodef},\eqref{eq:pdef}, and \eqref{eq:Pidef}, we find
\begin{align}
	\rho &= e^{-2\phi}\left(K_{(\mu\nu)}U^{\mu}U^{\nu} + \frac{1}{2}\To\right) \, , \label{eq:rhoDFT} \\
	p &= e^{-2\phi}\left(\frac{1}{3}K_{(\mu\nu)}h^{\mu\nu} - \frac{1}{2}\To\right) \, , \label{eq:pDFT} \\
	p\pi^{\mu}{}_{\nu} &= e^{-2\phi}\left(h^{\mu\rho}K_{(\rho\nu)} - \frac{1}{3}h^{\mu}{}_{\nu}h^{\rho\sigma}K_{(\rho\sigma)}\right) \, . \label{eq:PiDFT}
\end{align}
Contracting \eqref{eq:Tgeneral} with $U_{\mu}$ and $h^{\nu}{}_{\rho}$ and applying~\eqref{eq:PiDFT} 
leads to the constraint
\begin{equation}
	U^{\rho}K_{(\rho\sigma)}h^{\sigma}{}_{\nu} = 0 \, . \label{eq:Ki0constraint}
\end{equation}

For example, consider the case where $U_{\mu} = \left((-g_{00})^{1/2},\mathbf{0}\right)$ and $g_{0i} = 0$.  
In this case,~\eqref{eq:Ki0constraint} implies that $K_{(i0)} = 0$.  Therefore, on such backgrounds the energy density, pressure and anisotropic strain are given by
\begin{align}
	\rho &= e^{-2\phi}\left(-K^{0}{}_{0} + \frac{1}{2}\To\right) \, , \\
	p &= e^{-2\phi}\left(\frac{1}{3}K^{i}{}_{i} - \frac{1}{2}\To\right) \, , \\
	p\pi^{i}{}_{j} &= e^{-2\phi}\left(g^{ik}K_{(kj)} - \frac{1}{3}\delta^{i}{}_{j}g^{kl}K_{(kl)}\right) \, , 
\end{align}
while $\pi^{0}{}_{0} = 0$, $\pi^{i}{}_{0} = 0$ and $\pi^{0}{}_{j} = 0$.

\subsubsection{Einstein frame}
Einstein frame is defined by a Weyl rescaling of the metric,
\begin{equation}
	g_{\mu\nu} \equiv e^{2\phi}\tilde{g}_{\mu\nu} \, . \label{eq:Eframedef}
\end{equation}
Variations of the metric in string frame and Einstein frame are related as
\begin{equation}
	\delta g_{\mu\nu} 
	= e^{2\phi}(\delta\tilde{g}_{\mu\nu} + 2\tilde{g}_{\mu\nu}\delta\phi) \, , \qquad \delta g^{\mu\nu} = e^{-2\phi}\left(\delta\tg^{\mu\nu} - 2\tg^{\mu\nu}\delta\phi\right) \, . \label{eq:deltaEframe}
\end{equation}
Applying \eqref{eq:Eframedef} and \eqref{eq:deltaEframe} 
to \eqref{eq:deltaSgBphi} 
gives the variation of the matter action in Einstein frame,\footnote{Note that $\delta B_{\mu\nu}$ and $\delta\phi$ do not change with the frame.}
\begin{equation}
	\delta S_{\rm{m}}
	= \int\rd^{D}x\,\sqrt{-\tilde{g}}\bigg[-\frac{1}{2}e^{2\phi}T_{\mu\nu}\delta\tilde{g}^{\mu\nu} - \frac{1}{2}e^{4\phi}\Theta^{\mu\nu}\delta B_{\mu\nu} + e^{4\phi}(\sigma + T^{\mu}{}_{\mu})\delta\phi\bigg] \, , \label{eq:deltaSEframe}
\end{equation}
where $T^{\mu}{}_{\mu} \equiv g^{\mu\nu}T_{\mu\nu}$.  Thus the energy-momentum tensors in Einstein frame are defined as
\begin{align}
	\tilde{T}_{\mu\nu} &\equiv e^{2\phi}T_{\mu\nu} = K_{(\mu\nu)} - \frac{1}{2}g_{\mu\nu}\To \, , \label{eq:EMTmetricE} \\
	\tilde{\Theta}^{\mu\nu} &\equiv e^{4\phi}\Theta^{\mu\nu} = e^{2\phi}g^{\mu\rho}g^{\rho\sigma}K_{[\rho\sigma]} \, , \label{eq:EMTBE} \\
	\tilde{\sigma} &\equiv e^{4\phi}(\sigma + T^{\mu}{}_{\mu}) = e^{2\phi}(g^{\mu\nu}K_{\mu\nu} - \To) \, . \label{eq:EMTphiE}
\end{align}
Note that the dilaton source $\tsi$ vanishes when $g^{\mu\nu}K_{\mu\nu} - \To = 0$, which characterizes the `critical line' of GR-like solutions in \cite{Angus:2019bqs}.

When written in terms of Einstein frame variables, the DFT conservation laws take the same form as their string-frame presentations~\eqref{eq:conTThs1} and~\eqref{eq:conTThs2}, namely
\begin{align}
	\tilde{\nabla}^{\mu}\tilde{T}_{\mu\nu} + \frac{1}{2}H_{\nu\mu\lambda}\tilde{\Theta}^{\mu\lambda} - \tilde{\nabla}_{\nu}\phi\,\tilde{\sigma} &= 0 \, , \label{eq:conTThsE1} \\
	\tilde{\nabla}^{\mu}\tilde{\Theta}_{\mu\nu} &= 0 \, . \label{eq:conTThsE2}
\end{align}
The Einstein double field equations~\eqref{eq:EDFER1}--\eqref{eq:EDFER3} can also be converted to Einstein frame, yielding
\begin{align}
	8\pi G\tilde{T}_{\mu\nu} &= \tilde{R}_{\mu\nu} - \frac{1}{2}\tilde{g}_{\mu\nu}\tilde{R} - 2\left(\tilde{\nabla}_{\mu}\phi\tilde{\nabla}_{\nu}\phi - \frac{1}{2}\tilde{g}_{\mu\nu}\tilde{\nabla}^{\rho}\phi\tilde{\nabla}_{\rho}\phi\right) \nn \\
	&\quad - \frac{1}{4}e^{-4\phi}\left(\tilde{H}_{\mu}{}^{\rho\sigma}H_{\nu\rho\sigma} - \frac{1}{6}\tilde{g}_{\mu\nu}\tilde{H}^{\lambda\rho\sigma}H_{\lambda\rho\sigma}\right) \, , \label{eq:EDFEET} \\
	8\pi G\tilde{\sigma} &= -2\tilde{\nabla}^{\mu}\tilde{\nabla}_{\mu}\phi - \frac{1}{6}e^{-4\phi}\tilde{H}^{\mu\rho\sigma}H_{\mu\rho\sigma} \, , \label{eq:EDFEEs} \\
	16\pi G\tilde{\Theta}_{\mu\nu} &= \tilde{\nabla}^{\rho}\left(e^{-4\phi}H_{\rho\mu\nu}\right) \, . \label{eq:EDFEETh}
\end{align}
The conservation laws \eqref{eq:conTThsE1} and \eqref{eq:conTThsE2} can be verified by plugging in \eqref{eq:EDFEET}, \eqref{eq:EDFEEs} and \eqref{eq:EDFEETh}.  Note that \eqref{eq:EDFEET} can be interpreted in the standard GR sense as the Einstein curvature tensor being sourced by the energy-momentum tensors of the dilaton and $H$-flux, with $\tilde{T}_{\mu\nu}$ being the energy-momentum tensor for the additional matter only.

The energy-momentum tensor of a fluid in Einstein frame can be written as
\begin{equation}
	\tT^{\mu}{}_{\nu} = \trho\tU^{\mu}\tU_{\nu} + \tp\left(h^{\mu}{}_{\nu} + \pi^{\mu}{}_{\nu}\right) \, ,
\end{equation}
where the Einstein-frame fluid $4$-velocity $\tU^{\mu}$ is now normalized such that $\tg_{\mu\nu}\tU^{\mu}\tU^{\nu} = -1$, implying that $\tU^{\mu} = e^{\phi}U^{\mu}$ (and $\tU_{\nu} = e^{-\phi}U_{\nu}$).  Note that by construction, the mixed-index variables $h^{\mu}{}_{\nu}$ and $\pi^{\mu}{}_{\nu}$ do not change between frames; however, if indices are raised or lowered then relative factors of $e^{2\phi}$ will be introduced.  From~\eqref{eq:EMTmetricE} and~\eqref{eq:EMTphiE}, we see that the energy density, pressure and dilaton source in Einstein frame are related to those in string frame by
\begin{equation}
	\trho = e^{4\phi}\rho \, , \qquad \tp = e^{4\phi}p \, , \qquad \tsi = e^{4\phi}\left(\sigma - \rho + 3p\right) \, . \label{eq:rhopsigmaE}
\end{equation}

\subsection{Cosmological backgrounds} \label{sec:cosmobkgds}
If we restrict to solutions which are homogeneous and isotropic in $D=4$ spacetime dimensions, the most general gravitational ansatz can be written in the form
\begin{align}
	\rd s^{2}&=\brg_{\mu\nu}\rd x^{\mu}\rd x^{\nu} = -N^{2}(t)\rd t^{2} +a^{2}(t)\left[ \frac{1}{1-Kr^{2}} \rd r^2 +r^2 \left( \rd \vartheta^{2} + \sin^{2}\vartheta\,\rd\varphi^{2} \right)\right]\,, \label{eq:g0cosmo} \\
	\brB_{\scriptscriptstyle{(2)}} &=\frac{h r^{2}}{\sqrt{1-Kr^{2}}}\cos\vartheta\,\rd r\wedge\rd\varphi \,, \label{eq:B0cosmo} \\ \brphi &= \brphi(t) \, , \label{eq:phi0cosmo}
\end{align}
where $h$ and $K$ are constant.  Here and in the following, barred variables such as $\brg_{\mu\nu}$ refer to cosmological background quantities.  
The (magnetic) $H$-flux resulting from the $B$-field~\eqref{eq:B0cosmo} is proportional to the three-dimensional volume form,
\begin{equation}
	\brH_{\scriptscriptstyle{(3)}}= \rd_{4} \brB_{\scriptscriptstyle{(2)}} = \frac{hr^{2}}{\sqrt{1-Kr^{2}}}\sin\vartheta\, \rd r\wedge \rd\vartheta \wedge \rd\varphi = \frac{1}{3!}h\sqrt{\Omega}\,\epsilon_{ijk}\,\rd x^{i} \wedge \rd x^{j} \wedge \rd x^{k}\,, \label{eq:Hfluxbkgd}
\end{equation}
where $\Omega$ is the determinant of the three-dimensional spatial metric $\Omega_{ij}$ with constant curvature $K$.

A spatially homogeneous and isotropic DFT energy-momentum tensor is restricted to satisfy
\begin{equation}
	\brK^{\mu}{}_{\nu}=\left(\ba{cc}
	~\brK^{0}{}_{0}(t)~&~0~\\
	~0~&~\brK(t)\delta^{i}{}_{j}~
	\ea\right)\,,
	\qquad
	\brTo = \brTo(t) \, . \label{eq:CosmoEMT}
\end{equation}
Evaluating \eqref{eq:rhoDFT}, \eqref{eq:pDFT} and \eqref{eq:PiDFT} on a cosmological background, we find that the energy density, pressure and anisotropic strain are given by
\begin{equation}
	\brrho = e^{-2\brphi}\left(-\brK^{0}{}_{0}+\frac{1}{2}\brTo\right) \, , \qquad
	\brp = e^{-2\brphi}\left(\brK-\frac{1}{2}\brTo\right) \, , \qquad
	\brpi^{\mu}{}_{\nu} = 0 \, . \label{eq:rhopPi0cosmo}
\end{equation}
Thus from~\eqref{eq:EMTmetric} and~\eqref{eq:EMTBdil}, the most general DFT energy-momentum tensor on a cosmological background takes the form
\begin{equation}
	\brT^{\mu}{}_{\nu} = \left(\ba{cc}
	~-\brrho(t)~&~0~\\
	~0~&~\brp(t)\delta^{i}{}_{j}~
	\ea\right) \, ,
	\qquad \brTheta_{\mu\nu} = 0 \, , 
	\qquad \brsigma = \brsigma(t) \, .
\end{equation}

Plugging \eqref{eq:g0cosmo}, \eqref{eq:B0cosmo}, \eqref{eq:phi0cosmo}, \eqref{eq:CosmoEMT} and \eqref{eq:rhopPi0cosmo} into the gravitational equations of motion~\eqref{eq:EDFER1}--\eqref{eq:EDFER3} gives the $\ODD$-completion of the Friedmann equations~\cite{Angus:2019bqs},
\begin{eqnarray}
	\frac{8\pi G}{3}\brrho e^{2\brphi} + \frac{h^2}{12a^6} &=& H^{2} - 2\left(\frac{\brphi^{\prime}}{N}\right)H + \frac{2}{3}\left(\frac{\brphi^{\prime}}{N}\right)^2 +  \frac{K}{a^2} \, , \label{eq:OFE1}
	\\ \frac{4\pi G}{3} (\brrho+3\brp)e^{2\brphi} + \frac{h^2}{6a^6} &=& - H^{2} - \frac{H^{\prime}}{N} + \left(\frac{\brphi^{\prime}}{N}\right)H - \frac{2}{3}\left(\frac{\brphi^{\prime}}{N}\right)^2 + \frac{1}{N}\left(\frac{\brphi^{\prime}}{N}\right)' \, , \label{eq:OFE2}
	\\ \frac{4\pi G}{3}\left(2\brrho - \brsigma\right)e^{2\brphi} &=& - H^{2} - \frac{H^{\prime}}{N} + \frac{2}{3N}\left(\frac{\brphi^{\prime}}{N}\right)' \, , \label{eq:OFE3}
\end{eqnarray}
where the prime denotes differentiation with respect to $t$, and the Hubble parameter $H \equiv a'/(Na)$.  In addition, from~\eqref{eq:conTThs1} there is one non-trivial conservation law,
\begin{equation}
	\brrho^{\prime}+3NH (\brrho + \brp) + \brphi^{\prime} \brsigma = 0 \, , \label{eq:Conservation}
\end{equation}
which differs from that of standard GR cosmology due to the presence of the final term depending on the dilaton source $\brsigma$.

Since the background evolution of matter in DFT cosmology depends on three functions, $\brrho$, $\brp$ and $\brsigma$, we may introduce 
two equation-of-state parameters,
\begin{equation}
	w \equiv \frac{\brp}{\brrho} \, , \qquad \lambda \equiv \frac{\brsigma}{\brrho} \, . \label{eq:wlambda}
\end{equation}
Here $w$ is the standard equation-of-state parameter of GR cosmology, while $\lambda$ characterizes the magnitude of the dilaton stress-energy relative to the energy density.  When the dilaton is constant and the $H$-flux vanishes, some solutions in GR cosmology can be recovered.  From equations~\eqref{eq:OFE2} and~\eqref{eq:OFE3}, we see that this implies the constraint
\begin{equation}
	\brsigma = \brrho - 3\brp \qquad \implies \qquad \lambda = 1 - 3w \, . \label{eq:critline}
\end{equation}
This means solutions with constant dilaton and vanishing $H$-flux lie on a `critical line' in the $(w,\lambda)$-plane defined by~\eqref{eq:critline}.\footnote{However, the converse does not hold: the critical line also admits solutions with non-trivial $H$-flux and dilaton.}  If we define the corresponding Einstein-frame parameters
\begin{equation}
	\tw \equiv \frac{\brtp}{\brtrho} \, , \qquad \tlambda \equiv \frac{\brtsigma}{\brtrho} \, , \label{eq:wlambdaE}
\end{equation}
which from~\eqref{eq:rhopsigmaE} are related to their string-frame counterparts as
\begin{equation}
	\tw = w \, , \qquad \tlambda = \lambda - 1 + 3w \, ,
\end{equation}
we see that the critical line corresponds to $\tlambda = 0$ in Einstein frame.  Most of our examples of analytic solutions, whose perturbations we will study in section~\ref{sec:ex}, lie on this line.

\section{Cosmological perturbations} \label{sec:pert}
Now we turn to linear perturbations in DFT cosmology, which is the main focus of this work.  After first defining our perturbation variables for both the DFT gravitational fields and a generic matter sector in sections~\ref{sec:gravpert} and~\ref{sec:matpert}, respectively, we present the explicit equations of motion for the independent perturbation variables to linear order in section~\ref{sec:pertOFE}.  We separate the degrees of freedom according to a scalar-vector-tensor (SVT) decomposition, which remains consistent at linear order.  Finally, in section~\ref{sec:adiabatic} we obtain the perturbed conservation equations and discuss conditions under which adiabatic superhorizon perturbations are conserved.

\subsection{Gravitational sector} \label{sec:gravpert}
For the gravitational fields, we consider perturbations around the homogeneous and isotropic cosmological background~\eqref{eq:g0cosmo},~\eqref{eq:B0cosmo} and~\eqref{eq:phi0cosmo}, 
\begin{equation}
	g_{\mu\nu} \equiv \brg_{\mu\nu} + \delta g_{\mu\nu} \,, \qquad B_{\mu\nu} \equiv \brB_{\mu\nu} + \delta B_{\mu\nu} \,, \qquad
	\phi \equiv \brphi + \delta\phi \,. \label{eq:pertgBphi}
\end{equation}
Note that while $\brB_{\mu\nu}$ itself is not homogeneous and isotropic, it is defined up to a $U(1)$ gauge transformation such that the resulting $H$-flux 
respects the Killing symmetries of homogeneity and isotropy, as in~\eqref{eq:Hfluxbkgd}.  Let us consider each field in turn.

\subsubsection{Metric perturbations}
We consider a perturbed metric in string frame of the form
\begin{equation}
	\rd s^2 = g_{\mu\nu}\rd x^{\mu}\rd x^{\nu} = -N(t)^2(1+2A)\rd t^2 + 2 N(t)a(t)B_i\rd t\rd x^i + a(t)^2(\Omega_{ij}+h_{ij})\rd x^i\rd x^j\, \, , 
	\label{eq:perturbedmetric}
\end{equation}
where $A(t,\mathbf{x})$, $B_{i}(t,\mathbf{x})$ and $h_{ij}(t,\mathbf{x})$ are linear perturbations, and $\Omega_{ij}\rd x^i\rd x^j$ is the metric of a three-dimensional space with constant curvature $K$,
\begin{equation}
	\Omega_{ij}\rd x^i\rd x^j = \frac{1}{1-Kr^{2}} \rd r^2 +r^2 \left( \rd \vartheta^{2} + \sin^{2}\vartheta\,\rd\varphi^{2} \right) \, .
\end{equation}
Under the linearized gauge (diffeomorphism) transformation $x^{\mu}\to x^{\mu}-\xi^{\mu}$, the metric perturbations transform to linear order as 
\begin{align}
	A & \to A + \frac{N'}{N}\xi^{0} + \xi^{0}{}'\,, \nn \\
	B_{i} & \to B_{i} - \frac{N}{a}D_{i}\xi^{0} + \frac{a}{N}\xi_{i}'\, \, , \nn \\
	h_{ij} & \to h_{ij} + D_{i}\xi_{j} + D_{j}\xi_{i} + 2\cH\xi^{0}\Omega_{ij}\, \, , 
\end{align}
where 
$' \equiv \rd/\rd t$, $\cH \equiv a'/a = NH$ is the conformal Hubble parameter,
and $D_i$ is the three-dimensional covariant derivative compatible with $\Omega_{ij}$.

Under a scalar-vector-tensor (SVT) decomposition, we may expand $B_i$ as 
\begin{equation}
	B_i = D_iB + \hat{B}_i\, \, , 
\end{equation}
where $\hat{B}_i$ satisfies the transverse condition
\begin{equation}
	D^i\hat{B}_i=0 \, , \qquad D^i\equiv \Omega^{ij}D_j \, ,
\end{equation}
and decompose $h_{ij}$ as 
\begin{equation}
	h_{ij} = 2C\Omega_{ij} + 2\left(D_iD_j-\frac{1}{3}\Omega_{ij}D^kD_k\right)E
	+ \left(D_i\hat{E}_j+D_j\hat{E}_i\right) + \hat{E}_{ij} \, ,
\end{equation}
where $\hat{E}_i$ and $\hat{E}_{ij}$ are transverse and transverse traceless, respectively,
\begin{equation}
	D^i\hat{E}_i=0 \, ,  \qquad
	D^i\hat{E}_{ij}=0 \, , \qquad
	\Omega^{ij}\hat{E}_{ij} = 0 \, .
\end{equation}
Under the gauge transformation $x^{\mu}\to x^{\mu}-\xi^{\mu}$, each component then transforms as 
\begin{align}
	A & \to A + \frac{N'}{N}\xi^{0} + \xi^{0}{}' \, , \qquad
	B \to B - \frac{N}{a}\xi^{0} + \frac{a}{N}\xi' \, , \nn \\
	C & \to C + \cH\xi^{0} + \frac{1}{3}D^{k}D_{k}\xi \, , \qquad
	E \to E + \xi \, , \nn \\
	\hat{B}_{i} & \to  \hat{B}_{i} + \frac{a}{N}\hat{\xi}_i' \, , \qquad
	\hat{E}_{i} \to  \hat{E}_{i} + \hat{\xi}_{i} \, , \qquad
	\hat{E}_{ij} \to \hat{E}_{ij} \, ,
\end{align}
where 
the spatial diffeomorphism parameter $\xi_{i} \equiv \Omega_{ij}\xi^{j}$ is decomposed as 
\begin{equation}
	\xi_{i} = D_{i}\xi + \hat{\xi}_{i} \, , \qquad D^{i}\hat{\xi}_{i} = 0 \, .
\end{equation}
To linear order, the inverse and determinant of the spacetime metric are given by\footnote{Here and in the following discussion, `$=$' is assumed to hold only to linear order in perturbations.}
\begin{equation}
	g^{\mu\nu} = \left(
	\begin{array}{cc}
		- N^{-2}\left(1 - 2A\right) & \left(Na\right)^{-1}B^{j} \\
		\left(Na\right)^{-1}B^{i} & a^{-2}\left(\Omega^{ij} - h^{ij}\right)
	\end{array}
	\right) \, , \qquad g = -N^{2}a^{6}\Omega\left(1 + 2A + 6C\right) \, ,
\end{equation}
where $\Omega^{ij}$ is the inverse of $\Omega_{ij}$, $\Omega \equiv \det(\Omega_{ij})$, $B^{i}\equiv \Omega^{ik}B_{k}$ and $h^{ij} \equiv \Omega^{ik}h_{kl}\Omega^{lj}$.

For practical applications it is often convenient to study perturbations in terms of the Einstein frame metric, $\tg_{\mu\nu} \equiv e^{-2\phi}g_{\mu\nu}$.  Explicitly, the perturbed Einstein frame metric takes the form
\begin{equation}
	d\tilde{s}^{2} = \tilde{g}_{\mu\nu}\rd x^{\mu}\rd x^{\nu} = -\tilde{N}(t)^{2}\left(1 + 2\tA\right)\rd t^{2} + 2 \tilde{N}(t)\tilde{a}(t)B_{i}\rd t\rd x^{i} + \tilde{a}(t)^{2}\left(\Omega_{ij}+\tilde{h}_{ij}\right)\rd x^{i}\rd x^{j} \, ,
\end{equation}
where $\tilde{N}=e^{-\bar{\phi}}N$, $\tilde{a}=e^{-\bar{\phi}}a$ and
\begin{equation}
	\tilde{h}_{ij} = h_{ij} - 2\delta\phi \Omega_{ij} = 
	2\tilde{C}\Omega_{ij} + 2\left(D_iD_j-\frac{1}{3}\Omega_{ij}D^kD_k\right)E
	+ \left(D_i\hat{E}_j+D_j\hat{E}_i\right) + \hat{E}_{ij}\,. 
\end{equation}
Thus the metric perturbations in string frame and Einstein frame are related as
\begin{equation}
	A = \tA + \delta\phi \, , \qquad C = \tC + \delta\phi \, . \label{eq:EinsteinAC}
\end{equation}
Note that as defined above, $B$, $E$, $\hat{B}_{i}$, $\hat{E}_{i}$ and $\hat{E}_{ij}$ are all frame independent (to linear order).  Similarly, the inverse metric and determinant take the form
\begin{equation}
	\tg^{\mu\nu} \simeq \left(
	\begin{array}{cc}
		- \tN^{-2}\big(1 - 2\tA\big) & \big(\tN\ta\big)^{-1}B^{j} \\
		\big(\tN\ta\big)^{-1}B^{i} & \ta^{-2}\left(\Omega^{ij} - \th^{ij}\right)
	\end{array}
	\right) \, , \qquad \tg = -\tN^{2}\ta^{6}\Omega\left(1 + 2\tA + 6\tC\right) \, .
\end{equation}
This structure reflects the general rule: spacetime metric-dependent quantities --- connection coefficients, Ricci curvature, etc. --- take the same algebraic form in Einstein frame as in string frame, with appropriate variables replaced by their `tilde' ($\,\tilde{\cdot}\,$) variants.

\subsubsection{Perturbations of the dilaton and $B$-field}
In DFT, the gravitational degrees of freedom include not just the usual metric but also the dilaton and $B$-field.  Thus in order to study cosmological perturbations we must also consider an SVT decomposition of these additional fields.  For the dilaton we set
\begin{equation}
	\phi = \bar{\phi}(t) + \delta\phi\,, 
\end{equation}
where the background dilaton $\brphi$ is time dependent only while the perturbation $\delta\phi$ may have a more general coordinate dependence.  As a scalar field, the dilaton transforms under linearized diffeomorphisms as
\begin{equation}
	\delta\phi \rightarrow \delta\phi + \brphi'\xi^{0} \, .
\end{equation}

Up to a gauge transformation, the background $B$-field is expressed most simply in polar coordinates as
\begin{equation}
	\brB_{(2)} = \frac{h r^2}{\sqrt{1-Kr^2}}\cos\vartheta\, \rd r \wedge \rd \varphi \, ,
\end{equation}
with $h$ a constant.  It follows that the corresponding background $H$-flux is homogeneous and isotropic,
\begin{equation}
	\brH_{(3)} \equiv \rd\brB_{(2)} =  \frac{hr^{2}}{\sqrt{1 - Kr^{2}}}\sin\vartheta\, \rd r \wedge \rd \vartheta \wedge \rd \varphi = \frac{1}{3!}h\sqrt{\Omega}\,\epsilon_{ijk}\,\rd x^{i}\wedge \rd x^{j} \wedge \rd x^{k} \, .
\end{equation}
For the $B$-field perturbations, it is sufficient to consider an ansatz of the form
\begin{equation}
	B_{(2)} = \brB_{(2)} + f_i \, \rd x^i\wedge \rd t + m_{ij} \, \rd x^i\wedge \rd x^j\,.
\end{equation}
The perturbations $f_{i}$ and $m_{ij}$ may be decomposed as
\begin{equation}
	f_{i} = \hat{f}_{i} + D_{i}f \, , \qquad m_{ij} = D_{i}\hat{m}_{j} - D_{j}\hat{m}_{i} + \sqrt{\Omega}\,\epsilon_{ijk}D^{k}m \, , 
\end{equation}
where without loss of generality we may define $\hat{f}_{i}$ and $\hat{m}_{i}$ to be transverse,
\begin{equation}
	D^{i}\hat{f}_{i} = 0 \, , \qquad D^{i}\hat{m}_{i} = 0 \, .
\end{equation}

It is instructive to consider how the various perturbations transform under $B$-field gauge transformations, 
\begin{equation}
	B_{(2)} \rightarrow B_{(2)} + \rd\zeta_{(1)} \, .
\end{equation} 
Decomposing the gauge one-form $\zeta_{(1)}$ into SVT components,
\begin{equation}
	\zeta_{(1)} = \zeta_{0}\rd t + \zeta_{i}\rd x^{i} \, , \qquad \zeta_{i} = \hat{\zeta}_{i} + D_{i}\zeta \, , \qquad D^{i}\hat{\zeta}_{i} = 0 \, ,
\end{equation}
the individual perturbation variables transform as
\begin{equation}
	f \rightarrow f + \zeta_{0} - \p_{0}\zeta \, , \qquad \hat{f}_{i} \rightarrow \hat{f}_{i} - \p_{0}\hat{\zeta}_{i} \, , \qquad \hat{m}_{i} \rightarrow \hat{m}_{i} + \hat{\zeta}_{i} \, , \qquad m \rightarrow m \, .
\end{equation}
With this decomposition, the total $H$-flux becomes
\begin{align}
	H_{(3)} \equiv \rd B_{(2)} &= \frac{1}{2!}\left(D_{i}\hat{F}_{j} - D_{j}\hat{F}_{i} + \sqrt{\Omega}\,\epsilon_{ijk}D^{k}\p_{0}m\right)\rd t \wedge \rd x^{i} \wedge \rd x^{j} \nn \\
	&\quad + \frac{1}{3!}\big(h + D^{2}m\big) \sqrt{\Omega}\,\epsilon_{ijk}\,\rd x^{i} \wedge \rd x^{j} \wedge \rd x^{k} \, ,
\end{align}
where
$\hat{F}_{i} \equiv \hat{f}_{i} + \p_{0}\hat{m}_{i}$
is invariant under $B$-field gauge transformations.  In addition, under the linearized diffeomorphism $x^{\mu} \rightarrow x^{\mu} - \xi^{\mu}$ the $B$-field perturbations transform as
\begin{equation}
	\hat{F}_{k} \rightarrow \hat{F}_{k} + h\p_{0}\check{\xi}_{k} \, , \qquad m \rightarrow m + h\xi \, ,
\end{equation}
where have expanded $\hat{\xi}^{i}$ as the `curl' of a covector $\check{\xi}_{k}$,\footnote{Note that in the equations of motion, $\hat{F}_{i}$ always appears under a derivative in the skew-symmetric combination $D_{[i}\hat{F}_{j]}$.  Thus there is no ambiguity arising from the particular choice of $\check{\xi}_{k}$ satisfying~\eqref{eq:checkxi}.}
\begin{equation}
	\hat{\xi}^{i} \equiv \frac{1}{\sqrt{\Omega}}\,\epsilon^{ijk}D_{j}\check{\xi}_{k} \, . \label{eq:checkxi}
\end{equation}

\subsection{Matter sector} \label{sec:matpert}
As discussed in section~\ref{sec:rev}, the extended gravitational sector in DFT is coupled $\ODD$-covariantly to a generalized energy-momentum tensor, whose components consist of stress-energy sources for the $B$-field and dilaton in addition to the metric.  In a cosmological setting, the homogeneous and isotropic ansatz places constraints on the background fields.  However, at the level of perturbations all components receive contributions, so let us examine each in turn.

First of all, consider perturbations of the `metric' energy-momentum tensor~\eqref{eq:EMTmetric}, decomposed according to~\eqref{eq:Tgeneral} as
\begin{equation}
	T^{\mu}{}_{\nu} = \left(\rho + p\right)U^{\mu}U_{\nu} + p\left(\delta^{\mu}{}_{\nu} + \pi^{\mu}{}_{\nu}\right) \, . \label{eq:Tmnfluid}
\end{equation}
Expanding around a homogeneous and isotropic background, we may write
\begin{align}
	\rho & = \brrho + \delta\rho = \bar{\rho}(1+\delta)\,, \nonumber\\
	p & = \brp + \delta p = \brp(1+\pi_{\mathrm{L}})\,, \nonumber\\
	U^{i} & = U^{0}\frac{N}{a}v^{i}\,,  \nonumber\\
	\pi_{ij} & = \Omega_{ik}\pi^k_{\ j}\,, \label{eq:EMTpertdecomp}
\end{align}
where the background quantities $\bar{\rho}$ and $\brp$ are functions of $t$ only. Requiring the four-velocity normalization, $g_{\mu\nu}U^{\mu}U^{\nu} = -1$, to be preserved under perturbations yields the first-order expressions 
\begin{align}
	&U^0 = \frac{1}{N}(1-A)\,, \qquad \,\,
	U^i = \frac{v^i}{a}\,, \nonumber\\
	&U_0 = -N(1+A)\,, \qquad
	U_i = a(v_i+B_i)\,,
\end{align}
while the conditions in~\eqref{eq:anistresscond} imply that 
\begin{equation}
	\pi^{0}{}_{0} = 0 \, , \qquad \pi^{0}{}_{j} = 0 \, , \qquad \pi^{i}{}_{0} = 0 \, , \qquad \pi^{i}{}_{i} = 0 \, .
\end{equation}
Therefore the components of $T^{\mu}{}_{\nu}$ 
are, to linear order,
\begin{align}
	T^0{}_{0} & = -\left(\brrho+\delta\rho\right)\,, \qquad\quad\,\,\,
	T^0{}_{j} = \left(\brrho+\brp\right)\frac{a}{N}\left(v_i+B_i\right)\,, \nn \\
	T^i{}_{0} & = -\left(\brrho+\brp\right)\frac{N}{a}v^i\,, \qquad
	T^i{}_{j} = \left[\left(\brp+\delta p\right)\delta^i_j+\brp\pi^i_{\ j}\right]\,,
\end{align}
where $v_i\equiv \Omega_{ij}v^j$.
Under the gauge transformation $x^{\mu}\to x^{\mu}-\xi^{\mu}$, each component of the stress-energy 
transforms as
\begin{align}
	\delta\rho & \to \delta\rho + \brrho'\xi^{0} \, , \qquad \delta p \to \delta p + \brp'\xi^{0} \, , \nn \\ 
	v_i & \to v_i-\frac{a}{N}\xi_i' \, , \qquad
	\pi_{ij} \to \pi_{ij} \, ,
\end{align}
where $\xi_i \equiv \Omega_{ij}\xi^j$. 

Under an SVT decomposition, the components of $v_i\equiv\Omega_{ij}v^j$ and $\pi_{ij}$ can be expanded as
\begin{equation}
	v_i = D_{i}v + \hat{v}_i \, , \qquad
	\pi_{ij} = \left(D_iD_j-\frac{1}{3}\Omega_{ij}D^kD_k\right)\pi_{\mathrm{T}} + \frac{1}{2}(D_i\hat{\pi}_j+D_j\hat{\pi}_i) + \pi^{\mathrm{TT}}_{ij} \, ,
\end{equation}
where $\hat{v}_i$ and $\hat{\pi}_i$ are transverse, 
\begin{equation}
	D^i\hat{v}_i = D^i\hat{\pi}_i = 0 \, , 
\end{equation}
and $\pi^{\mathrm{TT}}_{ij}$ satisfies the transverse traceless condition 
\begin{equation}
	D^i\pi^{\mathrm{TT}}_{ij}=0\,, \quad
	\Omega^{ij}\pi^{\mathrm{TT}}_{ij} = 0\,.
\end{equation}
Each component thus transforms under linear diffeomorphisms as 
\begin{align}
	v & \to v -\frac{a}{N}\xi' \, , \qquad 
	\hat{v}_i \to \hat{v}_i-\frac{a}{N}\hat{\xi}_i' \, , \nn \\
	\pi_{\mathrm{T}} & \to \pi_{\mathrm{T}}\, , \qquad
	\hat{\pi}_i \to \hat{\pi}_i \, , \qquad
	\pi^{\mathrm{TT}}_{ij} \to \pi^{\mathrm{TT}}_{ij} \, .
\end{align}

Next consider the additional gravitational modes.  For the dilaton source, expand
\begin{equation}
	\sigma = \brsigma(t) + \delta\sigma \, , \label{eq:sigmapertdecomp}
\end{equation}
where the background term $\brsigma$ is a function of $t$ only.  Meanwhile, for the $B$-field source, write
\begin{equation}
	\Theta_{(2)} = \cJ_{i} \rd x^{i}\wedge\rd t + \frac{1}{2}\sqrt{\Omega}\,\epsilon_{ijk}\cI^{k}\rd x^{i} \wedge \rd x^{j} \, ,
\end{equation}
where $\cJ_{i}$ and $\cI_{i} \equiv \Omega_{ij}\cI^{j}$ have SVT decompositions
\begin{equation}
	\cJ_{i} = \hat{\cJ}_{i} + D_{i}\cJ \, , \qquad \cI_{i} = \hat{\cI}_{i} + D_{i}\cI \, ; \qquad D^{i}\hat{\cJ}_{i} = 0 \, , \qquad D^{i}\hat{\cI}_{i} = 0 \, .
\end{equation}
Under linearized diffeomorphisms $x^{\mu}\rightarrow x^{\mu} + \xi^{\mu}$, the dilaton source transforms as a scalar, while the $B$-field sources transform at second order due to the absence of a background term.  Thus at linear order,
\begin{equation}
	\delta\sigma \rightarrow \delta\sigma + \brsigma'\xi^{0} \, , \qquad \cJ \rightarrow \cJ \, , \qquad \cI \rightarrow \cI \, , \qquad \hat{\cJ}_{i} \rightarrow \hat{\cJ}_{i} \, , \qquad \hat{\cI}_{i} \rightarrow \hat{\cI}_{i} \, .
\end{equation}

Finally, note that the above discussion, while constructed in string frame, does not depend explicitly on the choice of frame.  Thus equivalent Einstein-frame expressions can be read off by inserting `tilde' ($\,\tilde{\cdot}\,$) superscripts in appropriate places.  
From equation~\eqref{eq:rhopsigmaE} we see that the stress-energy components of the background in string and Einstein frame are related as 
\begin{equation}
	\brtrho = e^{4\brphi}\brrho \, , \qquad \brtp = e^{4\brphi}\brp \, , \qquad \brtsigma = e^{4\brphi}\left(\brsigma - \brrho + 3\brp\right) \, . \label{eq:rhopsigmaEbkgd}
\end{equation}
Correspondingly, the first-order matter perturbations in Einstein frame are related to those in string frame as
\begin{align}
	&\delta\trho = e^{4\brphi}\left(\delta\rho + 4\brrho\delta\phi\right) \, , \qquad \delta\tp = e^{4\brphi}\left(\delta p + 4\brp\delta\phi\right) \, , \qquad \tv_{i} = v_{i} \, , \qquad \tpi_{ij} = \pi_{ij} \, , \nn \\
	&\tcJ_{i} = \cJ_{i} \, , \qquad \tcI_{i} = \cI_{i} \, , \qquad
	\delta\tsi = e^{4\brphi}\left[\delta\sigma - \delta\rho + 3\delta p  + 4\left(\brsigma - \brrho + 3\brp\right)\delta\phi\right] \, . \label{eq:Einsteinmatterpert}
\end{align}
In particular, note that with the above definitions, $v_{i}$, $\pi_{ij}$, $\cJ_{i}$ and $\cI_{i}$ are frame-independent at linear order.

\subsection{Gauge-invariant variables} \label{sec:gaugeinv}
As seen above, scalar and vector perturbations often transform non-trivially under linearized diffeomorphism (`gauge') transformations.  Thus it is sometimes useful to introduce gauge-invariant variables.

First of all, consider the scalar degrees of freedom.  For the metric perturbations, it is convenient to introduce the variables\footnote{These are equivalent to Bardeen variables~\cite{Bardeen:1980kt}, with $\Psi = \cA$ and $\Phi = -\cC$.  The notation is chosen to avoid confusion with e.g. the dilaton $\phi$ or the scalar field $\Phi$ (see section~\ref{sec:scalarfield}).}
\begin{align}
	\cA &\equiv A + \frac{N'}{N}\left[\frac{a}{N}\left(B - \frac{a}{N}E'\right)\right] + \left[\frac{a}{N}\left(B - \frac{a}{N}E'\right)\right]' \, , \\
	\cC &\equiv C + \cH\left[\frac{a}{N}\left(B - \frac{a}{N}E'\right)\right] - \frac{1}{3}D^{2}E \, .
\end{align}
Meanwhile, the dilaton and $B$-field scalar perturbations can also be written in gauge-invariant combinations, such as
\begin{equation}
	\widehat{\delta\phi} \equiv \delta\phi + \brphi'\left[\frac{a}{N}\left(B - \frac{a}{N}E'\right)\right] \, , \qquad \cM \equiv m - hE \, .
\end{equation}
Turning to the matter sector, perturbations of the metric energy-momentum tensor can be written as
\begin{equation}
	\brrho\Delta \equiv \delta\rho + \brrho'\frac{a}{N}\left(v + B\right) \, , \qquad \brp\Pi_{\mathrm{L}} \equiv \delta p + \brp'\frac{a}{N}\left(v + B\right) \, , \qquad V \equiv v + \frac{a}{N}E' \, .
\end{equation}
Similarly, the dilaton source term can be dressed as
\begin{equation}
	\brsigma\Sigma \equiv \delta\sigma + \brsigma'\frac{a}{N}\left(v + B\right) \, .
\end{equation}
The other source terms, such as the $B$-field sources, are automatically gauge invariant at linear order.

Since it is often convenient to work in Einstein frame, we may also define corresponding variables
\begin{align}
	\tcA &\equiv \tA + \frac{\tN'}{\tN}\left[\frac{\ta}{\tN}\left(B - \frac{\ta}{\tN}E'\right)\right] + \left[\frac{\ta}{\tN}\left(B - \frac{\ta}{\tN}E'\right)\right]' = \cA - \widehat{\delta\phi} \, , \\
	\tcC &\equiv \tC + \tcH\left[\frac{\ta}{\tN}\left(B - \frac{\ta}{\tN}E'\right)\right] - \frac{1}{3}D^{2}E = \cC - \widehat{\delta\phi}  \, , \\
	\brtrho\tDelta &\equiv \delta\trho + \brtrho'\frac{\ta}{\tN}\left(v + B\right) \, , \qquad \brtp\tPi_{\mathrm{L}} \equiv \delta \tp + \brtp'\frac{\ta}{\tN}\left(v + B\right) \, , \qquad
	\brtsigma\tSigma \equiv  \delta\tsi + \brtsigma'\frac{\ta}{\tN}\left(v + B\right) \, . \label{eq:gaugeinvEmatter}
\end{align}
The string and Einstein-frame matter perturbations are related via
\begin{equation}
	\tDelta = \Delta + 4\widehat{\delta\phi}_{V} \, , \qquad
	\tPi_{\mathrm{L}} = \Pi_{\mathrm{L}} + 4\widehat{\delta\phi}_{V} \, , \qquad
	\tSigma = \frac{\lambda\Sigma - \Delta + 3w\Pi_{\mathrm{L}}}{\lambda - 1 + 3w} + 4\widehat{\delta\phi}_{V} \, ,
\end{equation}
where for simplicity we have introduced the shorthand notation
\begin{equation}
	\widehat{\delta\phi}_{V} \equiv \delta\phi + \brphi'\frac{a}{N}\left(v + B\right) = \widehat{\delta\phi} + \brphi'\frac{a}{N}V \, .
\end{equation}

Meanwhile, for the vector perturbations, some useful gauge-invariant variables are
\begin{equation}
	\hcB_{i} \equiv \hat{B}_{i} - \frac{a}{N}\hat{E}_{i}' \, , \qquad \hcG_{i} \equiv h\hat{B}_{i} - \frac{a}{N}\sqrt{\Omega}\epsilon_{ijk}D^{j}\hat{F}^{k} \, , \qquad \hat{V}_{i} \equiv \hat{v}_{i} + \frac{a}{N}\hat{E}_{i}' \, , \label{eq:vectorGIvariables}
\end{equation}
which take the same form in string and Einstein frame.

\subsection{Perturbed $\ODD$ Friedmann equations} \label{sec:pertOFE}
Having defined our cosmological ansatz and perturbation variables, we may now evaluate the DFT gravitational equations of motion on cosmological backgrounds up to linear order in perturbations.  Having performed an SVT decomposition, we present in turn the equations for tensor, vector and scalar modes.  For technical convenience we will work in Einstein frame and apply our ansatz to the equations of motion~\eqref{eq:EDFEET}, \eqref{eq:EDFEEs} and \eqref{eq:EDFEETh}.  Note that as the results are mathematically equivalent in any frame, we may easily convert to string frame using the transformation laws of sections~\ref{sec:gravpert} and~\ref{sec:matpert}.  For some alternative presentations of the perturbation equations, see Appendix~\ref{sec:stringframe}.

\subsubsection{Tensor perturbations} \label{sec:pertOFEtensor}
The tensor perturbations $\hat{E}_{ij}$ satisfy the equation
\begin{equation}
	16\pi G \brtp \pi^{\mathrm{TT}}_{ij} = \frac{1}{\tN^{2}}\left[\hat{E}_{ij}'' + \left(3\tcH - \frac{\tN'}{\tN}\right)\hat{E}_{ij}'\right] - \frac{1}{\ta^{2}}D^{2}\hat{E}_{ij} \, ,
\end{equation}
where $\tcH \equiv \ta'/\ta$ is the conformal Hubble parameter in Einstein frame.  For simplicity, let us focus on the case of flat space ($K = 0$).  Decomposing into Fourier modes and expanding in a circular polarization basis,\footnote{Here $e^{\lambda}_{ij} = \varepsilon^{\lambda}_{i}\varepsilon^{\lambda}_{j}$, where $\varepsilon^{\lambda}_{i}$ is the corresponding circular polarization basis for vector modes, appearing in~\eqref{eq:BTcirc}.}
\begin{equation}
	\hat{E}_{ij}\left(t,\mathbf{x}\right) \equiv \sum_{\lambda = \pm}\int\frac{\rd^{3}\mathbf{k}}{\left(2\pi\right)^{3/2}}\,e^{\lambda}_{ij}(\mathbf{k})\hat{E}_{\lambda}(t,\mathbf{k})e^{i\mathbf{k}\cdot\mathbf{x}} \, , \label{eq:ETTcirc} 
\end{equation}
the equations of motion for the two polarization modes may be written as
\begin{equation}
	16\pi G \brtp \pi^{\mathrm{TT}}_{\pm} = \frac{1}{\tN^{2}}\left[\hat{E}_{\pm}'' + \left(3\tcH - \frac{\tN'}{\tN}\right)\hat{E}_{\pm}'\right] + \frac{k^{2}}{\ta^{2}}\hat{E}_{\pm} \, , \label{eq:EDFEtensork}
\end{equation}
where $k \equiv |\mathbf{k}|$ and $\pi^{\mathrm{TT}}_{\pm}(t,\mathbf{k})$ is defined in terms of $\pi^{\mathrm{TT}}_{ij}(t,\mathbf{x})$ 
analogously to~\eqref{eq:ETTcirc}.  Equation~\eqref{eq:EDFEtensork} characterizes the propagation of tensor gravitational waves on DFT cosmological backgrounds, whose evolution in Einstein frame resembles that of GR~\cite{Caprini:2018mtu}.

\subsubsection{Vector perturbations} \label{sec:pertOFEvector}
Next we turn to the vector perturbations.  Upon expanding the equations of motion, it turns out that the vector modes always appear in gauge-invariant combinations.  Simplifying and using the gauge-invariant variables~\eqref{eq:vectorGIvariables} yields the following equations:
\begin{align}
	8\pi G\left(\brtrho + \brtp\right)\left(\hat{V}_{i} + \hcB_{i}\right) &= \frac{1}{2\ta^{2}}\left(D^{2} + 2K\right)\hcB_{i} - \frac{e^{-4\brphi}h}{2\ta^{6}}\hcG_{i} \, ; \label{eq:EDFEvecE1} \\
	8\pi G\brtp D_{(i}\hat{\pi}_{j)} &= - \frac{1}{\tN\ta}\left(D_{(i}\hcB_{j)}' + 2\tcH D_{(i}\hcB_{j)}\right) \, ; \label{eq:EDFEvecE2} \\
	16\pi G\sqrt{\Omega}\epsilon_{ijk}\hcJ^{k} &= \frac{e^{-4\brphi}}{\tN\ta^{5}}D_{[i}\hcG_{j]} \, ; \label{eq:EDFEvecE3} \\
	16\pi G\hcI_{i} &= \frac{e^{-4\brphi}}{\tN\ta}\left[\hcG_{i}' - \left(2\tcH + 4\brphi'\right)\hcG_{i}\right] \, . \label{eq:EDFEvecE4}
\end{align}

Again restricting the following discussion to $K = 0$, the vector degrees of freedom can also be expanded in Fourier modes with a circular polarization basis, e.g.
\begin{equation}
	\hcB_{i}(t,\mathbf{x}) \equiv \sum_{\lambda = \pm}\int\frac{\rd^{3}\mathbf{k}}{(2\pi)^{3/2}}\,\varepsilon^{\lambda}_{i}(\mathbf{k})\hcB_{\lambda}(t,\mathbf{k})e^{i\mathbf{k}\cdot\mathbf{x}} \, . \label{eq:BTcirc}
\end{equation}
Applying similar expansions to all vector modes, the equations of motion~\eqref{eq:EDFEvecE1}--\eqref{eq:EDFEvecE4} can be written as
\begin{align}
	8\pi G\left(\brtrho + \brtp\right)\left(\hat{V}_{\pm} + \hcB_{\pm}\right) &= -\frac{k^{2}}{2\ta^{2}}\hcB_{\pm} - \frac{e^{-4\brphi}h}{2\ta^{6}}\hcG_{\pm} \, , 
	\label{eq:EDFEvecE1k} \\
	8\pi G\brtp k^{2}\hat{\pi}_{\pm} &= - \frac{k^{2}}{\tN\ta}\left(\hcB_{\pm}' + 2\tcH\hcB_{\pm}\right) \, , \label{eq:EDFEvecE2k} \\
	16\pi G\hcJ_{\pm} &= \pm\frac{e^{-4\brphi}}{\tN\ta^{5}} k\hcG_{\pm} \, , \label{eq:EDFEvecE3k} \\
	16\pi G\hcI_{\pm} &= \frac{e^{-4\brphi}}{\tN\ta}\left[\hcG_{\pm}' - \left(2\tcH + 4\brphi'\right)\hcG_{\pm}\right] \, . \label{eq:EDFEvecE4k}
\end{align}

An important class of solutions is those without vector-like sources, i.e. with $\hat{V}_{\pm} = \hat{\pi}_{\pm} = \hcJ_{\pm} = \hcI_{\pm} = 0$.  This encompasses all examples of DFT cosmological backgrounds studied in the next section.  In such cases, assuming nonvanishing (but possibly very small) $k$, equation~\eqref{eq:EDFEvecE3k} implies  
\begin{equation}
	\hcG_{\pm} = 0 \, ,
\end{equation}
such that equation~\eqref{eq:EDFEvecE1k} becomes
\begin{equation}
	\left[8\pi G\left(\brtrho + \brtp\right) + \frac{k^{2}}{2\ta^{2}}\right]\hcB_{\pm} = 0 \, .
\end{equation}
Thus, provided $\brtrho + \brtp \geq 0$ (or equivalently, $\brrho + \brp \geq 0$), it follows that $\hcB_{\pm} = 0$, implying that in the absence of vector-like sources the vector perturbations are trivial.

\subsubsection{Scalar perturbations} \label{pertOFEscalar}
Next we turn to the scalar perturbations.  From the metric energy-momentum tensor equation~\eqref{eq:EDFEET} we obtain four equations,
\begin{align}
	&8\pi G\left(\delta\trho + 2\brtrho\tA\right)
	= -\frac{2}{\tN\ta}\tcH D^{2}B + \frac{6}{\tN^{2}}\tcH\tC' - \frac{2}{\ta^{2}}D^{2}\left(\tC - \frac{1}{3}D^{2}E\right) + \frac{6K}{\ta^{2}}\left(\tA - \tC + \frac{1}{3}D^{2}E\right) \nn \\
	&\qquad\qquad\qquad\qquad\,\, - \frac{2}{\tN^{2}}\brphi'\delta\phi' - \frac{e^{-4\brphi}h^{2}}{2\ta^{6}}\left(\tA - 3\tC - 2\delta\phi\right) - \frac{e^{-4\brphi}h}{2\ta^{6}}D^{2}m \, , \label{eq:EDFEscalarE1} \\
	&D_{i}\left[8\pi G\left(\brtrho + \brtp\right)\left(v + B\right)\right] = D_{i}\Bigg[-\frac{2}{\tN\ta}\tcH\tA + \left(\frac{2K}{\ta^{2}} - \frac{e^{-4\brphi}h^{2}}{2\ta^{6}}\right)B + \frac{2}{\tN\ta}\left(\tC - \frac{1}{3}D^{2}E\right)' \nn \\
	&\qquad\qquad\qquad\qquad\qquad\qquad\quad\,\,\, - \frac{2K}{\tN\ta}E' + \frac{2}{\tN\ta}\brphi'\delta\phi + \frac{e^{-4\brphi}h}{2\tN\ta^{5}}m'\Bigg] \, , \label{eq:EDFEscalarE2} \\
	&8\pi G\left(\delta\tp + 2\brtp\tA\right) = \frac{2}{\tN^{2}}\tcH\tA' - \frac{2}{\tN^{2}}\tC'' - \frac{2}{\tN^{2}}\left(3\tcH - \frac{\tN'}{\tN}\right)\tC' - \frac{2K}{\ta^{2}}\left(\tA - \tC + \frac{1}{3}D^{2}E\right)
	\nn \\
	&\qquad\qquad\qquad\qquad\,\, + \frac{2}{3\ta^{2}}D^{2}\bigg[\tA + \frac{\ta}{\tN}\left(B' + 2\tcH B\right) + \tC - \frac{1}{3}D^{2}E\bigg] - \frac{2}{\tN^{2}}\brphi'\delta\phi' \nn \\
	&\qquad\qquad\qquad\qquad\,\, - \frac{e^{-4\brphi}h^{2}}{2\ta^{6}}\left(\tA - 3\tC - 2\delta\phi\right) - \frac{e^{-4\brphi}h}{2\ta^{6}}D^{2}m \, , \label{eq:EDFEscalarE3} \\
	&0 = \left(D_{i}D_{j} - \frac{1}{3}\Omega_{ij}D^{2}\right)\Bigg\{8\pi G\ta^{2}\brtp\pi_{\mathrm{T}} + \tA + \frac{\ta}{\tN}\left(B' + 2\tcH B\right) + \tC - \frac{1}{3}D^{2}E \nn \\
	&\qquad\qquad\qquad\qquad\qquad\quad - \frac{\ta^{2}}{\tN^{2}}\left[E'' + \left(3\tcH - \frac{\tN'}{\tN}\right)E'\right]\Bigg\} \, . \label{eq:EDFEscalarE4}
\end{align}
The dilaton source equation~\eqref{eq:EDFEEs} yields one scalar perturbation equation,
\begin{align}
	4\pi G\left(\delta\tsi + 2\brtsigma\tA\right) &= \frac{1}{\tN^{2}}\delta\phi'' + \frac{1}{\tN^{2}}\left(3\tcH - \frac{\tN'}{\tN}\right)\delta\phi' - \frac{1}{\ta^{2}}D^{2}\delta\phi + \frac{\brphi'}{\tN^{2}}\left(-\tA' + 3\tC' - \frac{\tN}{\ta}D^{2}B\right) \nn \\
	&\quad - \frac{e^{-4\brphi}h^{2}}{\ta^{6}}\left(\tA - 3\tC - 2\delta\phi\right) - \frac{e^{-4\brphi}h}{\ta^{6}}D^{2}m \, . \label{eq:EDFEscalarE7}
\end{align}
Finally, equation~\eqref{eq:EDFEETh} for the $B$-field source provides an additional two equations,
\begin{align}
	16\pi Ge^{4\brphi}D_{i}\cJ &= 0 \, , \label{eq:EDFEscalarE5} \\
	16\pi Ge^{4\brphi}D_{i}\cI &= D_{i}\Bigg\{-\frac{1}{\tN^{2}}m'' + \frac{1}{\tN^{2}}\left(\tcH + \frac{\tN'}{\tN} + 4\brphi'\right)m' + \frac{1}{\ta^{2}}D^{2}m \nn \\
	&\qquad\quad\,\, + \frac{h}{\ta^{2}}\left[\tA + \frac{\ta}{\tN}\left(B' - 2\tcH B - 4\brphi'B\right) - 3\tC - 4\delta\phi\right]\Bigg\} \, . \label{eq:EDFEscalarE6}
\end{align}
In section~\ref{sec:ex} we will study solutions of these equations in various backgrounds.

\subsection{Fluid conservation laws} \label{sec:adiabatic} 
We can study the evolution of fluctuations by expanding the DFT conservation equations to linear order, using the parametrizations of sections~\ref{sec:gravpert} and~\ref{sec:matpert}.  As seen in section~\ref{sec:DFTemtensor}, the conservation laws take the same algebraic form in string frame and Einstein frame.  Therefore the analysis can be performed in either frame, with conversion to the other frame amounting to adding or removing `tilde's ($\,\tilde{\cdot}\,$) as appropriate.  Note however that the physical implications will differ, in particular in relation to the inequivalent meanings of $\sigma$ and $\tsi$.
Thus in keeping with the previous subsection, for now we proceed in Einstein frame, taking equations~\eqref{eq:conTThsE1} and~\eqref{eq:conTThsE2} as our starting point, and include relevant commentary on string frame as required.

The $\nu = 0$ component of~\eqref{eq:conTThsE1} includes the background conservation law in Einstein frame (\textit{c.f.}~\eqref{eq:Conservation}),
\begin{equation}
	0 = \bar{\tilde{\rho}}' + 3\tilde{\cH}\left(\bar{\tilde{\rho}} + \bar{\tilde{p}}\right) + \bar{\tilde{\phi}}'\bar{\tilde{\sigma}} \, . \label{eq:conservation_background}
\end{equation}
Employing an SVT decomposition for the linear perturbations, we find four scalar conservation equations.  The $\nu = 0$ component of~\eqref{eq:conTThsE1} yields the continuity equation,
\begin{equation}
	0 = \delta\trho' + 3\tilde{\cH}\left(\delta\trho + \delta\tp\right) + \left(\brtrho + \brtp\right)\left(3\tC' + \frac{\tN}{\ta}D^{2}v\right) + \brphi'\delta\tsi + \delta\phi'\brtsigma \, . \label{eq:continuity}
\end{equation}
The $\nu = j$ component of \eqref{eq:conTThsE1} leads to an Euler-like equation with an $H$-flux contribution.
Its SVT decomposition includes one scalar equation,
\begin{align}
	0 &= D_{i}\Bigg\{\frac{\ta}{\tN}\left[\left(\brtrho + \brtp\right)\left(v + B\right)\right]' + 4\tcH\frac{\ta}{\tN}\left(\brtrho + \brtp\right)\left(v + B\right) + \delta\tp + \left(\brtrho + \brtp\right)\tA \nn \\ &\quad + 2\brtp\left(\frac{1}{3}D^{2} + K\right)\pi_{\mathrm{T}} + \frac{h}{\ta^{4}}\cI - \brtsigma\delta\phi\Bigg\} \, . \label{eq:Euler}
\end{align}
In addition, the conservation law for $\tTh_{\mu\nu}$~\eqref{eq:conTThsE2} includes two more scalar components,
\begin{equation}
	0 = D^{2}\cJ \, , \qquad 0 = D_{i}\left[\cJ' + \left(\tcH - \frac{\tN'}{\tN}\right)\cJ\right] \, . \label{eq:Thcons12}
\end{equation}
However, from~\eqref{eq:EDFEscalarE5} these turn out to be trivial.
The associated vector-mode conservation equations corresponding to~\eqref{eq:Euler} and~\eqref{eq:Thcons12} are
\begin{align}
	0 &= \frac{\ta}{\tN}\left\{\left[\left(\brtrho + \brtp\right)\left(\hat{V}_{i} + \hcB_{i}\right)\right]' + 4\tcH\left(\brtrho + \brtp\right)\left(\hat{V}_{i} + \hcB_{i}\right)\right\} + \brtp\left(\frac{1}{2}D^{2} + K\right)\hat{\pi}_{i} + \frac{h}{\ta^{4}}\hcI_{i} \, , \label{eq:veccons1} \\
	0 &= \left[\hcJ_{i}' + \left(\tcH - \frac{\tN'}{\tN}\right)\hcJ_{i}\right] - \frac{\tN^{2}}{\ta^{2}}\sqrt{\Omega}\epsilon_{ijk}D^{j}\hcI^{k} \, . \label{eq:veccons2}
\end{align}
These can also be derived straightforwardly from the vector equations of motion~\eqref{eq:EDFEvecE1}--\eqref{eq:EDFEvecE4}.

Let us now study the scalar equations in more detail.  For the following 
we restrict our attention to flat space ($K = 0$) and expand in Fourier modes. 
Meanwhile, 
in order to extract physical predictions for the scalar modes, we should fix the gauge: 
a convenient choice is spatially flat gauge, $\tC = E = 0$, in which 
the perturbed scalar conservation equations become
\begin{equation}
	0 = \delta\trho' + 3\tcH\left(\delta\trho + \delta\tp\right) + \brphi'\delta\tsi + \delta\phi'\brtsigma - \frac{\tN}{\ta}k^{2}\left(\brtrho + \brtp\right)v \, , \label{eq:continuityk}
\end{equation}
and
\begin{align}
	0 &= \frac{\ta}{\tN}\left\{\left[\left(\brtrho + \brtp\right)\left(v + B\right)\right]' + 4\tilde{\cH}\left[\left(\brtrho + \brtp\right)\left(v + B\right)\right]\right\} + \delta\tp + \left(\brtrho + \brtp\right)\tA - \frac{2}{3}k^{2}\brtp\pi_{\mathrm{T}} + \frac{h}{\ta^{4}}\cI - \brtsigma\delta\phi \, . \label{eq:Eulerk}
\end{align}
Taking the superhorizon limit of~\eqref{eq:continuityk}, $k^{2}/\ta^{2}\rightarrow 0$, the continuity equation in this limit can be obtained from a shift of the background equation~\eqref{eq:conservation_background}.\footnote{Note that in spatially flat gauge, $\tcH$ is invariant under shifts of the background, since its variation would be proportional to $\tC'$.}

We may define the density contrast in Einstein frame,
\begin{equation}
	\tdelta \equiv \frac{\delta\trho}{\brtrho} \, .
\end{equation}
It is also useful to introduce the isotropic pressure perturbation $\tpi_{\mathrm{L}}$ and the entropy perturbation $\tGamma_{p}$ as
\begin{equation}
	\tpi_{\mathrm{L}} \equiv \frac{\delta\tp}{\brtp} \, , \qquad \tGamma_{p} \equiv \tpi_{\mathrm{L}} - \frac{\tc_{\rm{s}}^{2}}{\tw}\tdelta \, , \label{eq:deltapiso}
\end{equation}
where the sound speed $\tc_{\rm{s}}^{2} \equiv \brtp'/\brtrho'$.  Moreover, to account for the additional dilaton source terms, we define analogous variables\footnote{These are related to corresponding variables in string frame,
\begin{equation}
	\delta \equiv \frac{\delta\rho}{\brrho} \, , \qquad \pi_{\mathrm{L}} \equiv \frac{\delta p}{\brp} \, , \qquad \pi_{\sigma} \equiv \frac{\delta\sigma}{\brsigma} \, , 
\end{equation}
via the first-order relations
\begin{equation}
	\tdelta = \delta + 4\delta\phi \, , \qquad \tpi_{\mathrm{L}} = \pi_{\mathrm{L}} + 4\delta\phi \, , \qquad \tpi_{\sigma} = \frac{\lambda\pi_{\sigma} - \delta + 3w\pi_{\mathrm{L}}}{\lambda - 1 + 3w} + 4\delta\phi \, .
\end{equation}}
\begin{equation}
	\tpi_{\sigma} \equiv \frac{\delta\tsi}{\brtsigma} \, , \qquad \tc_{\lambda}^{2} \equiv \frac{\brtsigma'}{\brtrho'} \, , \qquad \tGamma_{\sigma} \equiv \tpi_{\sigma} - \frac{\tc_{\lambda}^{2}}{\tlambda}\tdelta \, . \label{eq:deltasigmaiso}
\end{equation}
The continuity equation can then be written as
\begin{equation}
	0 = \tdelta' + \left[3\tcH\left(\tc_{\rm{s}}^{2} - \tw\right) + \brphi'\left(\tc_{\lambda}^{2} - \tlambda\right)\right]\tdelta + 3\tcH\tw\tGamma_{p} + \tlambda\left(\brphi'\tGamma_{\sigma} + \delta\phi'\right) - \frac{\tN}{\ta}k^{2}\left(1 + \tw\right)v \, . \label{eq:continuitydeltak}
\end{equation}

\subsubsection{Conservation of adiabatic superhorizon perturbations}
The modified conservation laws can have non-trivial consequences for the cosmological evolution.  In particular, we should re-examine the conditions under which adiabatic superhorizon perturbations are conserved.

First let us consider the case where the dilaton source and its fluctuations vanish identically.  Setting $\brtsigma =\delta\tsi=0$ (and thus $\tlambda = \tc_{\lambda}^{2} = 0$) in the continuity equations~\eqref{eq:conservation_background} and~\eqref{eq:continuitydeltak}, one obtains
\begin{equation}
	0 = \brtrho' + 3\tcH\left(\brtrho + \brtp\right) \,
\end{equation}
and 
\begin{equation}
	0 = \tdelta'
	+ 3\tcH\left(\tc_{\rm{s}}^{2} - \tw\right)\tdelta
	+3\tcH\tw\tGamma_{p}
	-\frac{\tN}{\ta}k^{2}\left(1 + \tw\right)v \, . 
\end{equation}
By introducing the new variable
\begin{equation}
	\tzeta \equiv \frac{\tdelta}{3\left(1+\tw\right)} \, , \label{eq:tzetadef}
\end{equation}
we may write this as
\begin{equation}
	\tzeta' = \frac{\tN}{3\ta}k^{2}v - \frac{\tw\tcH}{1+\tw}\tGamma_{p} \, . \label{eq:dzetadt-criticalline}
\end{equation}
In the superhorizon limit, $k^2/\tilde{a}^2\rightarrow 0$, and for adiabatic perturbations, i.e. $\tGamma_{p}=0$, equation~\eqref{eq:dzetadt-criticalline} implies that $\tzeta$ is conserved.  This result holds in the presence of an $H$-flux background ($h\neq 0$), $B$-field sources ($\cI\neq 0$), and even an extra scalar field $\Phi$ with an arbitrary potential $V(\Phi)$ (see section~\ref{sec:scalarfield}). 

Now consider the general case with $\brtsigma\neq 0$ and $\delta\tsi\neq 0$.  In the superhorizon limit, $k^{2}/\ta^{2}\rightarrow 0$, the continuity equation~\eqref{eq:continuitydeltak} can be written as
\begin{equation}
	0 = \tdelta' + \left[3\tcH\left(\tc_{\rm{s}}^{2} - \tw\right) + \brphi'\left(\tc_{\lambda}^{2} - \tlambda\right)\right]\tdelta + 3\tw\tcH\tGamma_{p} + \tlambda\brphi'\hat\tGamma_{\sigma} 
	\, , \label{eq:ddeltadt-generalk->0}
\end{equation}
where we have defined
\begin{equation}
	\hat\tGamma_{\sigma} \equiv \tGamma_{\sigma} + \frac{\delta\phi'}{\brphi'} \, .
\end{equation}
When $\brtsigma \neq 0$ and $\delta\tsi \neq 0$, in general it does not seem easy to find physically reasonable conditions under which $\tzeta$ is conserved.  However, the form of~\eqref{eq:ddeltadt-generalk->0} suggests that a conserved variable may nevertheless be identified under certain conditions.  
In particular, we can define a new variable 
\begin{equation}
	\tZ \equiv e^{\tf_{\delta}}\tdelta \, , \qquad \tf_{\delta} \equiv 
	\int\rd t\,\left[3\tcH\left(\tc_{\rm{s}}^{2} - \tw\right) + \brphi'\left(\tc_{\lambda}^{2} - \tlambda\right)\right] \, ,
\end{equation}
which is conserved whenever a generalized adiabaticity condition is satisfied, 
\begin{equation}
	3\tw\tcH\tGamma_{p} + \tlambda\brphi'\hat\tGamma_{\sigma} = 0 \, .
\end{equation}
In the case where $\tc_{\lambda}^{2} = \tlambda$, which corresponds to $\tlambda$ being constant, the resulting conserved variable $\tZ$ is equivalent to the $\tzeta$ of~\eqref{eq:tzetadef} (up to an integration constant).

Finally, we remark that the entire analysis above translates directly to string frame.  The sound speed and entropy perturbation in string frame are
\begin{equation}
	c_{\rm{s}}^{2} \equiv \frac{\brp'}{\brrho'} \, , \qquad \Gamma_{p} \equiv \pi_{\mathrm{L}} - \frac{c_{\rm{s}}^{2}}{w}\delta \, .
\end{equation}
Note that these are in general not equivalent to $\tc_{\rm{s}}^{2}$ and $\tGamma_{p}$, so the conditions for adiabaticity in string and Einstein frames carry different physical meanings.  Moreover, in situations where $\brsigma = \delta\sigma = 0$, superhorizon perturbations of
\begin{equation}
	\zeta \equiv \frac{\delta}{3(1 + w)} \,
\end{equation}
that are adiabatic with respect to string frame are conserved.

\section{Examples} \label{sec:ex}
In this section we study scalar perturbations around various solutions of the $\ODD$ Friedmann equations.  For convenience we will mostly work in Einstein frame and conformal gauge, $\tN = \ta$, and for simplicity we will restrict the following discussion to flat space, $K = 0$.  Moreover, we fix the gauge of metric perturbations to spatially flat gauge in Einstein frame,
\begin{equation}
	\tC = E = 0 \, , \qquad \hat{E}_{i} = 0 \, . \label{eq:spatiallyflatEgauge}
\end{equation}
The backgrounds we will consider constitute various examples of matter for which analytic solutions are known~\cite{Angus:2019bqs}, and our goal is study linear fluctuations around these known background solutions.  The perturbations are best studied in terms of Fourier modes with spatial wavevector $k_i$: in many cases explicit solutions for the fluctuations can be obtained in the superhorizon limit, $k \equiv \sqrt{|k_{i}k^{i}|} \rightarrow 0$.

In Einstein frame, the background equations of motion in the presence of matter can be written as
\begin{align}
	8\pi G\ta^{2}\brtrho &= 3\tcH^{2} - \brphi'{}^{2} - \frac{e^{-4\brphi}}{4\ta^{4}}h^{2} \, , \label{eq:Ebkgd1} \\
	4\pi G\ta^{2}\left(\brtrho -\brtp\right) &= \tcH' + 2\tcH^{2} \, , \label{eq:Ebkgd2} \\
	4\pi G\ta^{2}\brtsigma &= \brphi'' + 2\tcH\brphi' - \frac{e^{-4\brphi}}{2\ta^{4}}h^{2} \, . \label{eq:Ebkgd3}
\end{align}
For the scalar perturbation equations, we expand~\eqref{eq:EDFEscalarE1}--\eqref{eq:EDFEscalarE6} in Fourier modes with $K = 0$ and fix the gauge to~\eqref{eq:spatiallyflatEgauge}.  Note that the $H$-flux scalar perturbation $m$ always appears accompanied by a factor of $h$, so in order to obtain finite perturbations in the $h\rightarrow 0$ limit, we define
\begin{equation}
	\widehat{m} \equiv hm \, .
\end{equation}
After rearranging and simplifying, the scalar perturbation equations of motion in Einstein frame and conformal time ($\tN = \ta$) are\footnote{Note that we have dropped some overall factors of $k$ from equations~\eqref{eq:EDFEscalarE2k}, \eqref{eq:EDFEscalarE4k}, 
and~\eqref{eq:EDFEscalarE6k}.  Although we will consider the case where $k\rightarrow 0$, for consistency this should be taken as a limit with $k$ small but finite; we do not set $k = 0$ exactly.}
\begin{align}
	&8\pi G\ta^{2}\left(\delta\trho + 2\brtrho\tA\right) = 2\tcH k^{2}B - 2\brphi'\delta\phi' - \frac{e^{-4\brphi}h^{2}}{2\ta^{4}}\left(\tA - 2\delta\phi\right) + \frac{e^{-4\brphi}}{2\ta^{4}}k^{2}\widehat{m} \, , \label{eq:EDFEscalarE1k} \\
	&0 = 8\pi G\ta^{2}\left(\brtrho + \brtp\right)\left(v + B\right) + 2\tcH\tA + \frac{e^{-4\brphi}h^{2}}{2\ta^{4}}B - 2\brphi'\delta\phi - \frac{e^{-4\brphi}}{2\ta^{4}}\widehat{m}' \, , \label{eq:EDFEscalarE2k} \\
	&8\pi G\ta^{2}\left[\delta\trho - \delta\tp + 2\left(\brtrho - \brtp\right)\tA - \frac{2}{3}k^{2}\brtp\pi_{\mathrm{T}}\right] = 2\tcH\left(-\tA' + k^{2}B\right) \, , \label{eq:EDFEscalarE3k} \\
	&0= 8\pi G\ta^{2}\brtp\pi_{\mathrm{T}} + \tA + B' + 2\tcH B \, , \qquad 0 = 16\pi G\ta^{2}e^{4\brphi}\cJ \, , \label{eq:EDFEscalarE4k} \\
	&0 = 8\pi G\ta^{2}\left(2e^{4\brphi}h\cI - \brtp h^{2}\pi_{\mathrm{T}}\right) + \widehat{m}'' - \left(2\tcH + 4\brphi'\right)\widehat{m}' + k^{2}\widehat{m} + 4h^{2}\left[\left(\tcH + \brphi'\right)B + \delta\phi\right] \, , \label{eq:EDFEscalarE6k} \\
	&4\pi G\ta^{2}\left(\delta\tsi + 2\brtsigma\tA\right) = \delta\phi'' + 2\tcH\delta\phi' + k^{2}\delta\phi + \brphi'\left(-\tA' + k^{2}B\right) - \frac{e^{-4\brphi}h^{2}}{\ta^{4}}\left(\tA - 2\delta\phi\right) + \frac{e^{-4\brphi}}{\ta^{4}}k^{2}\widehat{m} \, . \label{eq:EDFEscalarE7k}
\end{align}

As an initial observation, the second equation of~\eqref{eq:EDFEscalarE4k} implies that to linear order $\cJ$ should vanish on shell,
\begin{equation}
	\cJ = 0 \, .
\end{equation}
This leaves six non-trivial perturbation equations, which we will now analyse for some specific examples.  We reiterate that although we will obtain solutions in Einstein frame, it is straightforward to recover the string-frame perturbations using~\eqref{eq:EinsteinAC} and~\eqref{eq:Einsteinmatterpert}.  Moreover, the gauge-invariant variables of section~\ref{sec:gaugeinv} can be reconstructed in our gauge choice~\eqref{eq:spatiallyflatEgauge}: for the gravity sector, we find
\begin{equation}
	\tcA = \tA + B' + \tcH B \, , \qquad \tcC = \tcH B \, , \qquad \widehat{\delta\phi} = \delta\phi + \brphi'B \, , \qquad \widehat{\cM} \equiv h\cM = \widehat{m} \, ,
\end{equation}
while the matter-sector variables can be obtained from~\eqref{eq:gaugeinvEmatter}.

\subsection{Generalized perfect fluid with barotropic equation of state} \label{sec:gpf}
In many applications we are interested in perfect fluids, and this concept can be generalized to the DFT case.  For a barotropic fluid $\tp = \tp(\trho)$, and for simplicity we also make the assumption that $\tsi = \tsi(\trho)$.  In particular, we may write
\begin{equation}
	\brtp = \tw\brtrho \, , \qquad \brtsigma = \tlambda\brtrho \, ,
\end{equation}
as in~\eqref{eq:wlambdaE}. 
If $\tw$ and $\tlambda$ are constant, then the background conservation equation~\eqref{eq:conservation_background},
\begin{equation}
	0 = \brtrho' + 3\tcH\left(\brtrho + \brtp\right) + \brphi'\brtsigma \, , \label{eq:GPFbkgdcons}
\end{equation}
can be integrated to yield the solution
\begin{equation}
	\brtrho = \brrho_{0}\ta^{-3(1+\tw)}e^{-\tlambda\brphi} \, . \label{eq:GPFbkgdrho}
\end{equation}
Note that we may drop the `$\sim$' in $\brrho_{0}$, since it is equivalent in string and Einstein frames, as seen below.

\subsubsection{Power law ansatz}
To get a feel for the structure of the equations, we consider an ansatz where the scale factor and dilaton follow a power law,
\begin{equation}
	\ta = \left(\frac{\eta - \eta_{0}}{\eta_{\ast}}\right)^{\tn} \, , \qquad e^{\brphi} = \left(\frac{\eta - \eta_{0}}{\eta_{\ast}}\right)^{-\ts} \, . \label{eq:powerlawaphiE}
\end{equation}
It follows that
\begin{equation}
	\tcH = \frac{\tn}{(\eta - \eta_{0})} \, , \qquad \brphi' = -\frac{\ts}{(\eta - \eta_{0})} \, , \qquad \tcH' = -\frac{\tn}{(\eta - \eta_{0})^{2}} \, , \qquad \brphi'' = \frac{\ts}{(\eta - \eta_{0})^{2}} \, . \label{eq:GPFhubblephi'E}
\end{equation}
At this point we can make two observations.  The first is that with this ansatz, the $H$-flux terms in the background equations of motion~\eqref{eq:Ebkgd1}--\eqref{eq:Ebkgd3} already take a power-law form, corresponding to the case $\tw = 1$ and $\tlambda = 4$.  Moving this contribution to the left-hand side and treating it as an energy density, it can be subsumed into the general analysis of power-law solutions (insofar as we are considering a single power-law contribution to the energy density).  Thus without loss of generality, in the following we set $h = 0$.

The second observation is that by counting powers of $(\eta - \eta_{0})$ in~\eqref{eq:Ebkgd1}--\eqref{eq:Ebkgd3}, since every term on the right-hand side has degree $-2$, non-trivial solutions with $\brtrho \neq 0$ are only possible if $\brtrho \propto (\eta - \eta_{0})^{-2}$.  Comparing with~\eqref{eq:GPFbkgdrho}, this implies the constraint
\begin{equation}
	(1 + 3\tw)\tn - \tlambda\ts = 2 \, . \label{eq:GPFbkgd-nsconstraint}
\end{equation}
Furthermore, the background equations of motion reduce to the algebraic equations
\begin{align}
	2\ckrho_{0} &= 3\tn^{2} - \ts^{2} \, , \label{eq:GPFbkgd1alg} \\
	(\tw - 1)\ckrho_{0} &= \tn(1 - 2\tn) \, , \label{eq:GPFbkgd2alg} \\
	\tlambda\ckrho_{0} &= \ts(1 - 2\tn) \, , \label{eq:GPFbkgd3alg}
\end{align}
where we have defined $\ckrho_{0} \equiv 4\pi G\brrho_{0}\eta_{\ast}^{2}$.  Equations~\eqref{eq:GPFbkgd2alg} and~\eqref{eq:GPFbkgd3alg} imply the constraint
\begin{equation}
	(\tw - 1)\ts - \tlambda\tn = 0 \, , \label{eq:GPFbkgd-nsconstraint2}
\end{equation}
which together with~\eqref{eq:GPFbkgd-nsconstraint} can be solved for $\tn$ and $\ts$, yielding
\begin{equation}
	\tn = \frac{2(1 - \tw)}{\tlambda^{2} - 3\tw^{2} + 2\tw + 1} \, , \qquad \ts = \frac{-2\tlambda}{\tlambda^{2} - 3\tw^{2} + 2\tw + 1} \, . \label{eq:GPFbkgd-ns}
\end{equation}
Using these expressions in~\eqref{eq:GPFbkgd1alg}, we also obtain
\begin{equation}
	\ckrho_{0} = \frac{1}{2}\left(3\tn^{2} - \ts^{2}\right) = \frac{6(1 - \tw)^{2} - 2\tlambda^{2}}{\left(\tlambda^{2} - 3\tw^{2} + 2\tw + 1\right)^{2}} \, . \label{eq:GPFbkgd-ckrho}
\end{equation}

In~\cite{Angus:2019bqs} a power-law ansatz was studied in terms of cosmic time in string frame,
\begin{equation}
	a = \left(\frac{t}{t_{0}}\right)^{n} \, , \qquad e^{\brphi} = \left(\frac{t}{t_{0}}\right)^{-s} \, .
\end{equation}
This turns out to be equivalent to the above construction: switching to conformal time, $\rd t = a\rd\eta$, 
and converting to Einstein frame, we recover~\eqref{eq:powerlawaphiE} with the identification
\begin{equation}
	\tn = \frac{n + s}{1 - n} \, , \qquad \ts = \frac{s}{1 - n} \, .
\end{equation}
Note that the solution in conformal time also extends to include the case of a GR-like cosmological constant in Einstein frame ($\tw = -1$, $\tlambda = 0$), for which $\tn = -1$.  However, a DFT cosmological constant ($\tw = -1$, $\tlambda = -2$) is still not included, as can be seen from the fact that the expressions for~\eqref{eq:GPFbkgd-ns} and~\eqref{eq:GPFbkgd-ckrho} become singular in this limit (also c.f.~\eqref{eq:DFTLa} and~\eqref{eq:DFTLbrphi}).\footnote{Note also that the $n = 1$ linear ansatz, with $a\propto t$ in string frame, becomes exponential in conformal time.  This corresponds to
\begin{equation}
	2\ckrho_0 = 3 + 6s + 2s^{2} \, , \qquad (1 + 3w)\ckrho_{0} = -2s^{2} \, , \qquad (2 - \lambda)\ckrho_{0} = 2s \, ,
\end{equation}
which includes the GR-like solution with $w = -\frac{1}{3}$ and $\lambda = 2$.}

Next we turn to the perturbations.  Our goal is to solve the scalar perturbation equations~\eqref{eq:EDFEscalarE1k}--\eqref{eq:EDFEscalarE7k}. Since we are considering a generic fluid without specifying the underlying Lagrangian description, we must make some assumptions in order to obtain analytic solutions.  First of all, since we are interested in perfect fluids, we may assume that the anisotropic stress vanishes, $\pi_{\mathrm{T}} = 0$.  Moreover, since the $B$-field stress energy source $\cI$ is absent for every explicit model we consider in this work, we also set $\cI = 0$ in this discussion.  
As above we set $h = 0$, and in order to obtain explicit solutions we will focus on the superhorizon limit, $k\rightarrow 0$.  Finally, we assume that the perturbations are adiabatic in Einstein frame, such that the entropy perturbation~\eqref{eq:deltapiso} vanishes,
\begin{equation}
	\brtp\tGamma_{p} = \delta\tp - \tc_{\rm{s}}^{2}\delta\trho = 0 \, , 
\end{equation}
where $\tc_{\rm{s}}^{2} \equiv \brtp'/\brtrho'$ is the sound speed (in this case, since $\tw$ is constant, we have $\tc_{\rm{s}}^{2} = \tw$).  It follows that $\delta\tp = \tw\delta\trho$, and by similar reasoning we assume that $\delta\tsi = \tlambda\delta\trho$.  In the end, this leaves six independent perturbation variables, $\tA$, $\tB$, $\delta\phi$, $m$, $\delta\trho$ and $\tv$, in six equations,
\begin{align}
	8\pi G \ta^{2} \left[\delta\trho + 2\brtrho\tA\right] &= -2\brphi'\delta\phi' \, , \qquad
	8\pi G\ta^{2}\left(1 - \tw\right)\left[\delta\trho + 2\brtrho\tA\right] = -2\tcH\tA' \, , \label{eq:GPFpertE1} \\
	8\pi G\ta^{2}\left(1 + \tw\right)\brtrho\left(v + B\right) &= -2\tcH\tA + 2\brphi'\delta\phi + \frac{e^{-4\brphi}}{2\ta^{4}}\widehat{m}' \, , \qquad 0 = \tA + B' + 2\tcH B \, , \label{eq:GPFpertE2} \\
	4\pi G\ta^{2}\tlambda\left[\delta\trho + 2\brtrho\tA\right] &= \delta\phi'' + 2\tcH\delta\phi' - \brphi'\tA' \, , \qquad 0 = \widehat{m}'' - \left(2\tcH + 4\brphi'\right)\widehat{m}' \, . \label{eq:GPFpertE6}
\end{align}

Now we would like to solve these equations.  Applying the power-law ansatz~\eqref{eq:powerlawaphiE}, and for now assuming $\tn \neq \frac{1}{2}$, the two equations in~\eqref{eq:GPFpertE1} 
can be combined to obtain a relation between $\delta\phi'$ and $\tA'$,
\begin{equation}
	(1-\tw)\brphi'\delta\phi' = \tcH\tA' \, . \label{eq:GPFpertE13}
\end{equation}
Using this 
to eliminate $\tA'$ from the first of~\eqref{eq:GPFpertE6}, 
we can integrate and find general solutions for $\delta\phi$ and $\tA$, 
\begin{equation}
	\delta\phi = \frac{-\alpha_{\delta\phi}}{(2\tn - 1)\left(\eta - \eta_{0}\right)^{2\tn - 1}} + \beta_{\delta\phi} \, , \qquad
	\tA = \frac{-\tlambda\alpha_{\delta\phi}}{(2\tn - 1)\left(\eta - \eta_{0}\right)^{2\tn - 1}} + \beta_{\tA} \, , \label{eq:GPFdeltaphiA}
\end{equation}
where $\alpha_{\delta\phi}$, $\beta_{\delta\phi}$ and $\beta_{\tA}$ are integration constants.
Then 
the second of~\eqref{eq:GPFpertE2} can be solved to obtain
\begin{equation}
	B = \frac{\tlambda\alpha_{\delta\phi}}{2\left(2\tn - 1\right)\left(\eta - \eta_{0}\right)^{2\tn - 2}} - \frac{\beta_{\tA}\left(\eta - \eta_{0}\right)}{\left(2\tn + 1\right)} + \gamma_{B}\left(\frac{\eta_{\ast}}{\eta - \eta_{0}}\right)^{2\tn} \, ,
\end{equation}
where $\gamma_{B}$ is another integration constant.
Furthermore, 
the second of~\eqref{eq:GPFpertE6} yields the general solution
\begin{equation}
	\widehat{m} = \left\{
	\begin{array}{ll}
		\frac{\eta_{\ast}\alpha_{\widehat{m}}}{\left(2\tn - 4\ts + 1\right)}\left(\frac{\eta - \eta_{0}}{\eta_{\ast}}\right)^{2\tn - 4\ts + 1} + \beta_{\widehat{m}} \, , \qquad &2\tn - 4\ts \neq -1 \, , \\
		\eta_{\ast}\alpha_{\widehat{m}}\ln\left(\eta - \eta_{0}\right) + \beta_{\widehat{m}} \, , \qquad &2\tn - 4\ts = -1 \, ,
	\end{array}
	\right. \label{eq:GPFpertm}
\end{equation}
with two further integration constants $\alpha_{\widehat{m}}$ and $\beta_{\widehat{m}}$.
Applying ~\eqref{eq:GPFdeltaphiA} in~\eqref{eq:GPFpertE1} gives the density perturbation,
\begin{equation}
	\delta\trho = \frac{\tlambda\ckrho_{0}\eta_{\ast}^{2\tn}\alpha_{\delta\phi}}{4\pi G\left(2\tn - 1\right)\left(\eta - \eta_{0}\right)^{4\tn + 1}} - \frac{\ckrho_{0}\eta_{\ast}^{2\tn}\beta_{\tA}}{2\pi G\left(\eta - \eta_{0}\right)^{2\tn + 2}} \, .
\end{equation}
Combining the above ingredients with the first of~\eqref{eq:GPFpertE2}, we find
\begin{equation}
	\left(1 + \tw\right)\ckrho_{0}\left(v + B\right) = \frac{4\tw\ts\alpha_{\delta\phi} + \left(2\tn - 1\right)\alpha_{\widehat{m}}\eta_{\ast}^{2\tn}}{4\left(2\tn - 1\right)\left(\eta - \eta_{0}\right)^{2\tn - 2}} - \left(\tn\beta_{\tA} + \ts\beta_{\delta\phi}\right)\left(\eta - \eta_{0}\right) \, .
\end{equation}

Finally, consider the special case where $\tn = \frac{1}{2}$.  From the background equations~\eqref{eq:GPFbkgd2alg} and~\eqref{eq:GPFbkgd3alg}, we see that this corresponds to $w = 1$ and $\tlambda = 0$.  In terms of cosmic time in string frame, this corresponds to the continuous family of solutions satisfying $3n + 2s = 1$, which were argued in~\cite{Angus:2019bqs} to describe massless scalar field solutions.  In this case $\ts$ is unconstrained; however, equation~\eqref{eq:GPFbkgd1alg} becomes $2\ckrho_{0} = \frac{3}{4} - \ts^{2}$, which implies that for non-negative energy densities $\ts$ should lie in the range $-\frac{\sqrt{3}}{2} \leq \ts \leq \frac{\sqrt{3}}{2}$.  Analysing the perturbation equations in this case, following the same procedure and assumptions as above, we find the solutions
\begin{align}
	&\tA = \beta_{\tA} \, , \quad \delta\phi = \alpha_{\delta\phi}\ln\left(\eta - \eta_{0}\right) + \beta_{\delta\phi} \, , \quad B = -\frac{1}{2}\beta_{\tA}\left(\eta - \eta_{0}\right) + \frac{\gamma_{B}}{\left(\eta - \eta_{0}\right)} \, , \quad \delta\trho = \frac{\eta_{\ast}\left(\ts\alpha_{\delta\phi} - 2\ckrho_{0}\beta_{\tA}\right)}{4\pi G\left(\eta - \eta_{0}\right)^{3}} \, , \nn \\
	&4\ckrho_{0}\left(v + B\right) =  - 2\ts\alpha_{\delta\phi}\left(\eta - \eta_{0}\right)\ln\left(\eta - \eta_{0}\right) - \left(\beta_{\tA} + 2\ts\beta_{\delta\phi} - \frac{\eta_{\ast}\alpha_{\widehat{m}}}{2}\right)\left(\eta - \eta_{0}\right) \, , \label{eq:GPFspecial}
\end{align}
while $\widehat{m}$ takes the same form as in~\eqref{eq:GPFpertm}.  We will see in section~\ref{sec:scalarfield} that these indeed correspond to the perturbation equations for a massless scalar field with $h = 0$ in the superhorizon limit.

\subsection{DFT vacuum}
We now turn to explicit analytic solutions with non-trivial background $H$-flux, $h\neq 0$.  As a simplest example, consider the case without additional matter beyond the metric, $B$-field and dilaton, which in this context we refer to as the `DFT vacuum' (although the solution itself is well-known in the string cosmology literature~\cite{Copeland:1994vi,Lidsey:1999mc}).  
Let us briefly review the background solution. 
When $\brtrho - \brtp = 0$, equation~\eqref{eq:Ebkgd2} can be written as $(\ta^{2})'' = 0$, which can be integrated easily to give
\begin{equation}
	\ta^{2} = C_{1}\left(\eta - \eta_{0}\right) \, , \qquad \tcH = \frac{1}{2\left(\eta - \eta_{0}\right)} \, . \label{eq:Evacta}
\end{equation}
Here $C_{1}$ and $\eta_{0}$ are integration constants, the latter chosen such that $\ta(\eta_{0}) = 0$.  Then for $\brtsigma = 0$, multiplying~\eqref{eq:Ebkgd3} by $2\ta^{4}\brphi'$ and integrating gives
\begin{equation}
	\brphi'{}^{2}\ta^{4} + \frac{e^{-4\brphi}h^{2}}{4} = \frac{h_{\circ}^{2}}{4} \, , \label{eq:Evaccrit}
\end{equation}
where $h_{\circ}$ is another integration constant.  This equation holds for any solution on the critical line, 
i.e. $\brtsigma = 0$.  In this context, $h_{\circ}^{2}$ can be interpreted as the total energy density of the dilaton and $H$-flux.  Plugging~\eqref{eq:Evacta} and~\eqref{eq:Evaccrit} into the Friedmann equation~\eqref{eq:Ebkgd1} reveals the constraint
\begin{equation}
	h_{\circ}^{2} = 3C_{1}^{2} \, . \label{eq:EvacbachoC1}
\end{equation}
Finally, integrating~\eqref{eq:Evaccrit} yields a solution for the dilaton,
\begin{equation}
	e^{2\brphi} = \left(\frac{\eta - \eta_{0}}{\eta_{\ast}}\right)^{\pm\sqrt{3}} + \frac{h^{2}}{12C_{1}^{2}}\left(\frac{\eta - \eta_{0}}{\eta_{\ast}}\right)^{\mp\sqrt{3}} \, , \label{eq:Evacbrphi}
\end{equation}
where $\eta_{\ast}$ is another constant of integration.  This has a minimum at $e^{2\brphi} = |h|/(\sqrt{3}C_{1})$, corresponding to 
\begin{equation}
	\left(\frac{\eta - \eta_{0}}{\eta_{\ast}}\right)^{\pm\sqrt{3}} = \frac{|h|}{2\sqrt{3}C_{1}} \, .
\end{equation}

We now turn to the perturbations.  
In the absence of $H$-flux and external matter, equations~\eqref{eq:EDFEscalarE1k}--\eqref{eq:EDFEscalarE7k} reduce to a set of independent damped oscillator equations for the perturbation modes~\cite{Mueller:1989in}.  However, for $h \neq 0$ the perturbations mix non-trivially and source each other.
Nevertheless, analytic solutions can be obtained in the superhorizon limit, $k\rightarrow 0$.  First of all, equation~\eqref{eq:EDFEscalarE3k} implies that in this limit $\tA' = 0$ and thus
\begin{equation}
	\tA = \beta_{\tA} \, , \label{eq:EvactA}
\end{equation}
where $\beta_{\tA}$ is a constant of integration.  From this, equation~\eqref{eq:EDFEscalarE4k} 
can be integrated to yield
\begin{equation}
	B = \frac{1}{2}\left[-\beta_{\tA}\left(\eta - \eta_{0}\right) + \frac{\beta_{B}}{C_{1}\left(\eta - \eta_{0}\right)}\right] \, , \label{eq:EvactB}
\end{equation}
with $\beta_{B}$ another constant.
Next, $\delta\phi$ can be obtained from~\eqref{eq:EDFEscalarE1k} in this limit by substituting the background equation~\eqref{eq:Ebkgd3} and writing the terms containing $\delta\phi$ as a total derivative. 
Integrating gives, for constant $\beta_{\delta\phi}$,
\begin{equation}
	\delta\phi = \beta_{\delta\phi}\left(\frac{x - \frac{h^{2}}{12C_{1}^{2}x}}{x + \frac{h^{2}}{12C_{1}^{2}x}}\right) + \frac{\beta_{\tA}}{2} \, ,
	\qquad
	x \equiv \left(\frac{\eta - \eta_{0}}{\eta_{\ast}}\right)^{\pm\sqrt{3}} \, .
	\label{eq:Evacdeltaphi}
\end{equation}
Finally, the magnetic $H$-flux perturbation $\widehat{m}$ is most easily obtained from~\eqref{eq:EDFEscalarE2k}. 
Applying the previous results and integrating yields the solution
\begin{align}
	\widehat{m} &= \widehat{m}_{0} + \left[\left(2\mp\sqrt{3}\right)\beta_{\tA} \mp 2\sqrt{3}\beta_{\delta\phi}\right]\frac{C_{1}^{2}\left(\eta - \eta_{0}\right)^{2}x^{2}}{2\left(1\pm\sqrt{3}\right)} 
	- \left(\beta_{\tA} \mp 2\sqrt{3}\beta_{\delta\phi} \right)\frac{h^{2}\left(\eta - \eta_{0}\right)^{2}}{12} \nn \\ 
	&\quad + \frac{h^{2}\beta_{\tB}}{2C_{1}}\ln\left(\eta - \eta_{0}\right) + \left[\left(2\pm\sqrt{3}\right)\beta_{\tA} \mp 2\sqrt{3}\beta_{\delta\phi}\right]\frac{h^{4}\left(\eta - \eta_{0}\right)^{2}}{288\left(1\mp\sqrt{3}\right)C_{1}^{2}x^{2}} \, .
\end{align}
Note that when $h = 0$ and up to integration constant definitions, these solutions match the power-law solutions in~\eqref{eq:GPFspecial} for $\ts = \mp \frac{\sqrt{3}}{2}$ (with $\alpha_{\delta\phi} = 0$, as required 
by the first of~\eqref{eq:GPFpertE1}
when $\brtrho = \delta\trho = 0$).

\subsection{Scalar field} \label{sec:scalarfield}
A scalar field $\Phi$ in DFT has energy-momentum tensor components~\cite{Angus:2018mep}
\begin{equation}
	K_{\mu\nu} = \p_{\mu}\Phi\p_{\nu}\Phi \, , \qquad \To = g^{\mu\nu}\p_{\mu}\Phi\p_{\nu}\Phi + 2V(\Phi) = -2L_{\Phi} \, . \label{KTscalar}
\end{equation}
We expand the scalar field around a cosmological solution in conformal time $\eta$ as
\begin{equation}
	\Phi(\eta,\mathbf{x}) = \brPhi(\eta) + \delta\Phi(\eta,\mathbf{x}) \, . \label{pertscalar}
\end{equation} 
From~\eqref{eq:EMTBdil},~\eqref{eq:rhopsigmaE} and~\eqref{eq:rhopPi0cosmo} we obtain the background energy-momentum tensor components in Einstein frame, 
\begin{align}
	\brtrho = \frac{1}{2\ta^{2}}\brPhi'{}^{2} + e^{2\brphi}V(\brPhi) \, , \qquad
	\brtp = \frac{1}{2\ta^{2}}\brPhi'{}^{2} - e^{2\brphi}V(\brPhi) \, , \qquad
	\brtsigma = -2e^{2\brphi}V(\brPhi) \, .
\end{align}
Meanwhile, from~\eqref{eq:EMTpertdecomp},~\eqref{eq:sigmapertdecomp} and~\eqref{eq:Einsteinmatterpert} the scalar perturbations at linear order can be read off as
\begin{align}
	\delta\trho &= \frac{1}{\ta^{2}}\left[\brPhi'\delta\Phi' - \brPhi'{}^{2}\tA\right] + e^{2\brphi}\left[\frac{\rd V}{\rd\Phi}(\brPhi)\delta\Phi + 2V(\brPhi)\delta\phi\right] \, , \\
	\delta\tp &= \frac{1}{\ta^{2}}\left[\brPhi'\delta\Phi' - \brPhi'{}^{2}\tA\right] - e^{2\brphi}\left[\frac{\rd V}{\rd\Phi}(\brPhi)\delta\Phi + 2V(\brPhi)\delta\phi\right] \, , \\
	\delta\tsi &= -2e^{2\brphi}\left[\frac{\rd V}{\rd\Phi}(\brPhi)\delta\Phi + 2V(\brPhi)\delta\phi\right] \, , \\
	v + B &= - \frac{\delta\Phi}{\brPhi'} \, , \qquad	\pi_{\mathrm{T}} = 0 \, , \qquad \cI = 0 \, , \qquad \cJ = 0 \, . \label{eq:EscalarEMtensoretc}
\end{align}
For the velocity perturbation in~\eqref{eq:EscalarEMtensoretc}, we have made use of the fact that
\begin{equation}
	\delta K_{i0} - \frac{1}{2}e^{2\brphi}\tg_{i0}\brTo = \tT_{i0} = -\ta^{2}\left[\left(\brtrho + \brtp\right)v_{i} + \brtrho B_{i}\right] \, 
\end{equation}
to linear order and extracted the scalar component.

In addition to the background equations of motion~\eqref{eq:Ebkgd1}, \eqref{eq:Ebkgd2} and~\eqref{eq:Ebkgd3},
the conservation law~\eqref{eq:conservation_background} gives the equation of motion for the background evolution of the scalar field,
\begin{equation}
	0 = \brPhi'' + 2\tcH\brPhi' + e^{2\brphi}\frac{\rd V}{\rd\brPhi}(\brPhi) \, . \label{eq:scalarbkgdPhi}
\end{equation}
Likewise, supplementing the scalar perturbation equations of motion~\eqref{eq:EDFEscalarE1k}--\eqref{eq:EDFEscalarE7k}, one perturbed conservation equation is non-trivial and gives the equation of motion for $\delta\Phi$,
\begin{equation}
	0 = \delta\Phi'' + 2\tcH\delta\Phi' + \left(k^{2} + \ta^{2}e^{2\brphi}\frac{\rd^{2}V(\brPhi)}{\rd\Phi^{2}}\right)\delta\Phi + 2\ta^{2}e^{2\brphi}\frac{\rd V(\brPhi)}{\rd\Phi}\left(\tA + \delta\phi\right) + \brPhi'\left(-\tA' + k^{2} B\right) \, . \label{eq:eomdeltaPhi}
\end{equation}

\subsubsection{Massless scalar (modulus)}
Consider the simple case of $V(\Phi) = 0$, corresponding to a massless scalar field without a potential (e.g. a modulus field).  In this case $\brtrho = \brtp$ and $\brtsigma = 0$, so this solution lies on the critical line at $\tw = 1$.  Such solutions may be relevant for studying `kination' epochs, which can arise in the moduli-dominated early universe in some string cosmology scenarios~\cite{Apers:2024ffe}.  
Let us briefly review the background solution.  Equations~\eqref{eq:Ebkgd2} and~\eqref{eq:scalarbkgdPhi} can be integrated to obtain the scale factor and background scalar field solutions, respectively,
\begin{equation}
	\ta^{2} = C_{1}\left(\eta - \eta_{0}\right) \, , \qquad \brPhi = \frac{\alpha_{\brPhi}}{C_{1}}\ln\left(\eta - \eta_{0}\right) + \brPhi_{0} \, , \label{eq:scalarbkgdaPhisolutions}
\end{equation}
where $C_{1}$, $\eta_{0}$, $\alpha_{\brPhi}$ and $\brPhi_{0}$ are integration constants, and we have fixed $\ta(\eta_{0}) = 0$.  
Since we are on the critical line, equation~\eqref{eq:Ebkgd3} integrates to~\eqref{eq:Evaccrit}, as in the DFT vacuum solution. 
However, in this case equation~\eqref{eq:Ebkgd1} yields a more general relation between the integration constants,
\begin{equation}
	16\pi G\alpha_{\brPhi}^{2} = 3C_{1}^{2} - h_{\rm{o}}^{2} \, ,
	\label{eq:hoC1scalar}
\end{equation}
implying the bound $|h_{\rm{o}}| \leq \sqrt{3}C_{1}$.\footnote{Note that $|h_{\rm{o}}| = \sqrt{3}C_{1}$ corresponds to the DFT vacuum solution~\eqref{eq:Evacbrphi}.}
The dilaton background can be solved by integrating~\eqref{eq:Evaccrit}, yielding~\cite{Copeland:1994vi,Lidsey:1999mc}
\begin{equation}
	e^{2\brphi} = \left(\frac{\eta - \eta_{0}}{\eta_{\ast}}\right)^{\pm\frac{h_{\rm{o}}}{C_1}} + \frac{h^{2}}{4h_{\rm{o}}^{2}}\left(\frac{\eta - \eta_{0}}{\eta_{\ast}}\right)^{\mp\frac{h_{\rm{o}}}{C_1}} \, , \label{eq:scalarbkgdphisolution}
\end{equation}
where $\eta_{\ast}$ is another integration constant.
Without loss of generality, in any given region we can absorb the $\pm$ sign in the exponent into the definition of $h_{\rm{o}}$.  
In the case of vanishing background $H$-flux, $h=0$, this becomes
\begin{equation}
	\brphi = \brphi_{0} \pm\frac{h_{\rm{o}}}{2C_{1}}\ln\left(\eta - \eta_{0}\right) \, , \label{eq:scalarbkgdphih0}
\end{equation}
where $\brphi_{0} \equiv \mp(h_{\rm{o}}/2C_{1})\ln\eta_{\ast}$.  Furthermore, when $h_{\rm{o}} = 0$ this reduces to $\brphi = \brphi_{0}$, a constant dilaton.

Now we turn to the perturbations.  
For a scalar field with $V(\brPhi) = 0$, the system of equations~\eqref{eq:EDFEscalarE1k}--\eqref{eq:EDFEscalarE7k} and~\eqref{eq:eomdeltaPhi} admits analytic solutions in the superhorizon limit, $k\rightarrow 0$.  Much of the calculation proceeds analogously to the DFT vacuum case: first of all, equations~\eqref{eq:EDFEscalarE3k} and~\eqref{eq:EDFEscalarE4k} imply
\begin{equation}
	\tA = \beta_{\tA} \, 
	, \qquad
	B = \frac{1}{2}\left[-\beta_{\tA}\left(\eta - \eta_{0}\right) + \frac{\beta_{B}}{C_{1}\left(\eta - \eta_{0}\right)}\right] 
	\, , \label{eq:VlessscalartAB}
\end{equation}
where $\beta_{\tA}$ 
and $\beta_{\tB}$ are integration constants.  
Moreover, equation~\eqref{eq:eomdeltaPhi} yields, for constant $\alpha_{\delta\Phi}$ and $\delta\Phi_{0}$, 
\begin{equation}
	\delta\Phi = \frac{\alpha_{\delta\Phi}}{C_{1}}\ln\left(\eta - \eta_{0}\right) + \delta\Phi_{0} \, , \label{eq:VlessscalardeltaPhi}
\end{equation}
which clearly 
corresponds to a shift of the integration constants of the corresponding background solution~\eqref{eq:scalarbkgdaPhisolutions}, as expected for adiabatic modes.

Next, the solution for the dilaton perturbation $\delta\phi$ in the superhorizon limit can be obtained by integrating equation~\eqref{eq:EDFEscalarE1k} (see Appendix~\ref{sec:ana} for a detailed derivation),
\begin{equation}
	\delta\phi = \left(\frac{x - \frac{h^{2}}{4h_{\circ}^{2}x}}{x + \frac{h^{2}}{4h_{\circ}^{2}x}}\right)\left[\beta_{\delta\phi} - \frac{8\pi G\alpha_{\brPhi}\alpha_{\delta\Phi}}{h_{\circ}C_{1}}\ln\left(\eta - \eta_{0}\right)\right] + \frac{\beta_{\tA}}{2} + \frac{8\pi G\alpha_{\brPhi}\alpha_{\delta\Phi}}{h_{\circ}^{2}} \, , \quad 
	x \equiv \left(\frac{\eta - \eta_{0}}{\eta_{\ast}}\right)^{\frac{h_{\circ}}{C_{1}}} \, , 
	\label{eq:Vlessscalardeltaphi}
\end{equation}
where $\beta_{\delta\phi}$ is another integration constant. 
Note that for $h = 0$, the solution~\eqref{eq:Vlessscalardeltaphi} reduces to the form
\begin{equation}
	\delta\phi = -\frac{8\pi G\alpha_{\brPhi}\alpha_{\delta\Phi}}{h_{\circ}C_{1}}\ln\left(\eta - \eta_{0}\right) + \delta\phi_{0} \, , \qquad \delta\phi_{0} \equiv \beta_{\delta\phi} + \frac{\beta_{\tA}}{2} + \frac{8\pi G\alpha_{\brPhi}\alpha_{\delta\Phi}}{h_{\circ}^{2}} \, .
\end{equation} 
It is straightforward to show that the most general adiabatic variation of the background solution~\eqref{eq:scalarbkgdphisolution} at linear order takes the functional form
\begin{equation}
	\delta\phi = \alpha + \brphi'\left[\beta + \gamma\ln\left(\eta - \eta_{0}\right)\right] \, ,
\end{equation}
in agreement with the form of~\eqref{eq:Vlessscalardeltaphi}, implying that the superhorizon dilaton perturbation is indeed adiabatic.  Moreover, when the scalar field energy density vanishes, corresponding to $\alpha_{\brPhi} = 0 = \alpha_{\delta\Phi}$ and $|h_{\circ}| = \sqrt{3}C_{1}$,  
the dilaton perturbation reduces to that of the DFT vacuum solution~\eqref{eq:Evacdeltaphi}.

Finally, to solve for $\widehat{m}$ we may use~\eqref{eq:EDFEscalarE2k} 
to obtain 
an expression for $\widehat{m}'$ and then integrate the result.  
In the simple case where $h = 0$, this gives
\begin{equation}
	\widehat{m} = \left\{\begin{array}{lll}
		\frac{\alpha_{m}}{2(C_{1} + h_{\circ})}C_{1}^{2}\left(\eta - \eta_{0}\right)^{2}x^{2} + \widehat{m}_{0}
		= \frac{\alpha_{m}}{2(C_{1} + h_{\circ})}\ta^{4}e^{4\brphi} + \widehat{m}_{0} & , & \frac{h_{\circ}}{C_{1}} \neq -1 \, ; \\
		\alpha_{m}C_{1}\eta_{\ast}^{2}\ln\left(\eta - \eta_{0}\right) + \widehat{m}_{0} & , & \frac{h_{\circ}}{C_{1}} = -1 \, ,
	\end{array}\right. \label{eq:Vlessscalarmh0}
\end{equation}
where
$\alpha_{m} \equiv (2C_{1} - h_{\circ})\beta_{\tA} - 16\pi G\alpha_{\brPhi}(\delta\Phi_{0} + \alpha_{\delta\Phi}/h_{\circ}) - 2h_{\circ}\beta_{\delta\phi}$. For non-zero $h$ the solution can take three different forms, depending on whether $h_{\circ} = \pm C_{1}$ or neither.  The general result is given in Appendix~\ref{sec:m}.

It is also useful to study the limit $h \rightarrow 0$ as a consistency check.  In this case, the dependence of $e^{\brphi}$ becomes a power law, so we ought to recover one of the generalized perfect fluid solutions discussed in section~\ref{sec:gpf}.  
From the background solution~\eqref{eq:scalarbkgdaPhisolutions} and~\eqref{eq:scalarbkgdphih0}, we see that consistency with~\eqref{eq:powerlawaphiE} requires
\begin{equation}
	\tn = \frac{1}{2} \, , \qquad \ts = -\frac{h_{\rm{o}}}{2C_{1}} \, , \qquad C_{1} = \frac{1}{\eta_{\ast}} \, .
\end{equation}
Note that in general we may independently set the overall constant in $\ta$ and $e^{\brphi}$, and the final relation is an artefact of our particular choice in the power-law ansatz.  Meanwhile, the background energy density of the scalar field is
identified with the generalized perfect fluid solution~\eqref{eq:GPFbkgdrho} for $\tw = 1$ and $\tlambda = 0$ when
\begin{equation}
	\brrho_{0} = \frac{\alpha_{\brPhi}^{2}}{2} \, .
\end{equation}
With these identifications, the solutions for perturbations in the $h\rightarrow 0$ limit also agree with~\eqref{eq:GPFpertm} and~\eqref{eq:GPFspecial} up to integration constant definitions.

\subsubsection{DFT cosmological constant}
Another interesting special case is the DFT `cosmological constant'~\cite{Jeon:2011cn}, which due to the weight factor $e^{-2d}$ effectively corresponds to an exponential potential for the dilaton.
It can be recovered from the general scalar field case by setting $\brPhi' = 0$ and $V = \Lambda/(8\pi G)$.  
The background is most conveniently studied in string frame, where
$a = \ta e^{\brphi}$ and
$\cH = \tcH + \brphi'$, 
and cosmic time $t$, where $\rd t  = a\rd\eta$.  Denoting $\dot{} \equiv \rd/\rd t$, we find that $\cH = aH$ and $\brphi' = a\dot\brphi$, where
$H\equiv \dot a / a = \rd\ln a/\rd t$. 
Moreover, it is convenient to exchange $\brphi$ for the background DFT dilaton $\brd$ using the relation $e^{-2\brd} = e^{-2\brphi}\sqrt{-\brg}$, which 
implies $2\brd = 2\brphi - 3\ln a$.
Thus from~\eqref{eq:Ebkgd1}, \eqref{eq:Ebkgd2} and~\eqref{eq:Ebkgd3}, the background equations of motion can be expressed as
\begin{equation}
	2\Lambda = 4\dot\brd^{2} - 3H^{2} - \frac{h^{2}}{4a^{6}} \, , \qquad 
	0 = \dot H - 2\dot\brd H \, , \qquad 
	-2\Lambda = 2\ddot\brd - 4\dot\brd^{2} + \frac{h^{2}}{2a^{6}} \, . 
	\label{eq:DFTLd}
\end{equation}

When $h = 0$, exact solutions can be found. 
For $\Lambda > 0$, the third of~\eqref{eq:DFTLd} has solutions of the form
\begin{equation}
	e^{-2\brd} = C_{\brd}\sinh\left(2M(t-t_{0})\right) \, , \label{eq:DFTLdbar}
\end{equation}
where $C_{\brd}$ and $t_{0}$ are integration constants and $M \equiv \sqrt{\Lambda/2}$.  For $C_{\brd} > 0$ this is well-defined for $t > t_{0}$~\cite{Mueller:1989in}; alternatively, if $C_{\brd} < 0$ we have
\begin{equation}
	e^{-2\brd} = |C_{\brd}|\sinh\left(2M(t_{0} - t)\right) \, , \label{eq:DFTLdbaralt}
\end{equation}
which is well-defined for $t < t_{0}$.  The latter solution can be obtained from the former via a combined T-duality and time reversal transformation~\cite{Angus:2019bqs}.
The
scale factor can then be obtained from the first of~\eqref{eq:DFTLd},
\begin{equation}
	a = a_{0}\tanh^{\frac{1}{\sqrt{3}}}\left(M(t - t_{0})\right) \, , \qquad a = a_{0}\coth^{\frac{1}{\sqrt{3}}}\left(M(t_{0} - t)\right) \, , \label{eq:DFTLa}
\end{equation}
where the left- and right-hand expressions correspond to choosing \eqref{eq:DFTLdbar} or \eqref{eq:DFTLdbaralt}, respectively.  Finally, the spacetime dilaton $\brphi$ can be obtained from the prior results using $e^{2\brphi} = e^{2\brd}a^{3}$, giving 
\begin{equation}
	e^{2\brphi} = C_{\brphi}\frac{\tanh^{\sqrt{3}}\left(M(t - t_{0})\right)}{\sinh\left(2M(t - t_{0})\right)} \, , \qquad e^{2\brphi} = C_{\brphi}\frac{\coth^{\sqrt{3}}\left(M(t_{0} - t)\right)}{\sinh\left(2M(t_{0} - t)\right)} \, , \label{eq:DFTLbrphi}
\end{equation}
where $C_{\brphi} \equiv a_{0}^{3}/|C_{\rmd}|$.

We now turn to the perturbations.  We can study the equations of motion in conformal time, spatially flat gauge and Einstein frame by setting $\brPhi' = 0$ and $V = \Lambda/(8\pi G)$ in \eqref{eq:EDFEscalarE1k}--\eqref{eq:EDFEscalarE7k}. 
In the superhorizon limit, $k\rightarrow 0$, and for the $h = 0$ background discussed above, we may apply~\eqref{eq:DFTLa} and~\eqref{eq:DFTLbrphi} in order to look for exact solutions.  For $\widehat{m}$, integrating equation~\eqref{eq:EDFEscalarE6k} twice yields
\begin{equation}
	\widehat{m} 
	= \widehat{m}_{0} + \frac{\sqrt{3}\alpha_{\widehat{m}}C_{\brphi}}{8Ma_{0}^{3}}\ta^4e^{4\brphi} \, , \label{eq:DFTLEm}
\end{equation}
where $\alpha_{\widehat{m}}$ and $\widehat{m}_{0}$ are integration constants.
Next, combining equations~\eqref{eq:EDFEscalarE3k} and~\eqref{eq:EDFEscalarE7k} 
then 
using~\eqref{eq:EDFEscalarE1k} to eliminate $\tA$ yields an equation for $\delta\phi$,
\begin{equation}
	0 = \tcH\delta\phi'' + \left(2\tcH^{2} - 2\tcH\brphi' - 2\brphi'{}^{2}\right)\delta\phi' \, .
\end{equation}
Substituting the background equations, 
this can be integrated to obtain
\begin{equation}
	C_{\delta\phi} = \frac{\tcH}{\ta^{2}e^{2\brphi}}\delta\phi' + \Lambda\delta\phi \, ,
\end{equation}
where $C_{\delta\phi}$ is another integration constant.  Using~\eqref{eq:EDFEscalarE1k}--\eqref{eq:EDFEscalarE3k} and applying the solution for $\widehat{m}$ in~\eqref{eq:DFTLEm}, we may solve for $\delta\phi$ and $\tA$ without further integration.  This gives
\begin{equation}
	\delta\phi = \alpha_{\delta\phi}\frac{\brphi'}{\tcH} - \frac{\alpha_{\widehat{m}}}{4\ta^{2}\tcH} \, , \qquad \tA = \alpha_{\delta\phi}\frac{\brphi'^{2}}{\tcH^{2}} + \frac{\alpha_{\widehat{m}}}{4\ta^{2}\tcH^{2}}\left(\tcH - \brphi'\right) \, , \label{eq:DFTLEdeltaphiA}
\end{equation}
where we have defined $\alpha_{\delta\phi} \equiv -C_{\delta\phi}{\Lambda}$.  
Finally, the shift perturbation $B$ can be obtained by integrating~\eqref{eq:EDFEscalarE4k}.  Since the explicit expression appears non-trivial, this may be done numerically.

\subsection{Radiation}
Finally, we consider a perfect fluid with 
$\tw = w = \frac{1}{3}$ and $\tlambda = \lambda = 0$, corresponding to radiation in both string and Einstein frames.  The background conservation equation~\eqref{eq:conservation_background} gives
\begin{equation}
	\brrho = \brrho_{0}a^{-4} \qquad \Longleftrightarrow \qquad \brtrho = \brrho_{0}\ta^{-4} \, , \label{eq:rhorad}
\end{equation}
where the equivalence follows from the fact that $\brtrho = e^{4\brphi}\brrho$.  
Defining $\cE_{0}\equiv 8\pi G\brrho_{0}/3$, we can 
solve~\eqref{eq:Ebkgd2} to obtain the scale factor,
\begin{equation}
	\ta^{2} = C_{1}\left(\eta - \eta_{0}\right) + \cE_{0}\left(\eta - \eta_{0}\right)^{2} \, . \label{eq:radbkgdEasolution}
\end{equation}
Since we are on the critical line, \eqref{eq:Ebkgd1} can be integrated 
to give~\eqref{eq:Evaccrit}.
Equation~\eqref{eq:Ebkgd1} then implies the same constraint as the DFT vacuum solution~\eqref{eq:EvacbachoC1},
$h_{\rm{o}}^{2} = 3C_{1}^{2}$. 
Solving~\eqref{eq:Evaccrit} with the scale factor~\eqref{eq:radbkgdEasolution},
the resulting expression for the background dilaton is
\begin{equation}
	e^{2\brphi} = \left(\frac{\eta - \eta_{0}}{\eta_{\star}\left(1 + \frac{\cE_{0}}{C_{1}}\left(\eta - \eta_{0}\right)\right)}\right)^{\pm\sqrt{3}}
	+ \frac{h^{2}}{12C_{1}^{2}}\left(\frac{\eta - \eta_{0}}{\eta_{\star}\left(1 + \frac{\cE_{0}}{C_{1}}\left(\eta - \eta_{0}\right)\right)}\right)^{\mp\sqrt{3}} \, . \label{eq:radbkgdphisolution}
\end{equation}
The solution given by~\eqref{eq:radbkgdEasolution} and~\eqref{eq:radbkgdphisolution} may provide a candidate for a bouncing universe in string frame~\cite{Angus:2019bqs}.

In order to solve the perturbation equations~\eqref{eq:EDFEscalarE1k}--\eqref{eq:EDFEscalarE7k} for this radiation background, as we do not specify the underlying Lagrangian it is necessary to make some reasonable assumptions.  To this end, 
consider solutions where only $\delta\trho$, $\delta\tp$ and $\tv$ are non-vanishing, with $\tGamma_{p} = 0$.  If we also take the superhorizon limit, $k\rightarrow 0$, 
the continuity equation~\eqref{eq:continuityk} becomes
\begin{align}
	0 = \delta\trho' + 3\tcH\left(1 + \tc_{\rm{s}}^{2}\right)\delta\trho \, . \label{eq:radscalarEC1sup} 
\end{align}
For constant $\tc_{\rm{s}}^{2}$ we can easily integrate this to obtain the density perturbation.
Since we are interested in a fluid with a constant equation of state, we have $\tc_{\rm{s}}^{2} = \tw = \frac{1}{3}$.  Thus
\begin{equation}
	\delta\trho = \delta\rho_{0}\ta^{-4} \, . \label{eq:raddeltatrho}
\end{equation}

Now we solve for the gravitational fluctuations. 
For the following 
we relegate the calculation details to Appendix~\ref{sec:ana}. 
First of all, the solution for $\tA$ can be obtained from~\eqref{eq:EDFEscalarE3k}. 
The result is
\begin{equation}
	\tA = 
	\frac{\widehat{\beta}_{\tA}C_{1}^{2}}{4\ta^{4}\tcH^{2}} - \frac{\delta_{0}}{2} \, , \label{eq:radtA}
\end{equation}
where we have defined $\delta_{0} \equiv \delta\rho_{0}/\brrho_{0}$.  This matches the DFT vacuum solution~\eqref{eq:EvactA} in the limit $\cE_{0} \rightarrow 0$\footnote{When $\cE_{0} \rightarrow 0$, for consistency the energy density perturbation $\delta\rho_{0}$ should also vanish.  However, $\delta_{0} = \delta\rho_{0}/\brrho_{0}$ may remain finite provided it is sufficiently small, $\delta_{0}\ll 1$, so that the validity of the perturbative expansion is maintained while taking the limit.} if we identify the integration constant $\widehat{\beta}_{\tA} \equiv \beta_{\tA} + \frac{\delta_{0}}{2}$.
Solving~\eqref{eq:EDFEscalarE4k} for $B$,
a somewhat lengthy calculation yields
\begin{equation}
	B = - \frac{\beta_{\tA}C_{1}^{2}\left(\eta - \eta_{0}\right)^{2}}{4\ta^{4}\tcH} + \frac{\beta_{B}}{2\ta^{2}} + \frac{\delta_{0}\cE_{0}\left(\eta - \eta_{0}\right)^{2}\left(C_{1}\left(\eta - \eta_{0}\right) + \ta^{2}\right)}{6\ta^{4}\tcH} \, , \label{eq:radtB}
\end{equation}
where $\beta_{B}$ is an integration constant.  As a consistency check, when $\cE_{0} \rightarrow 0$ this reduces to the DFT vacuum result~\eqref{eq:EvactB} (or, equivalently,~\eqref{eq:VlessscalartAB} for the massless scalar).
From~\eqref{eq:EDFEscalarE1k} we find the dilaton perturbation,
\begin{equation}
	\delta\phi = \left(\frac{\tx - \frac{h^{2}}{12C_{1}^{2}\tx}}{\tx + \frac{h^{2}}{12C_{1}^{2}\tx}}\right)\left[\widehat{\beta}_{\delta\phi} \pm \frac{\sqrt{3}C_{1}\widehat{\beta}_{\tA}}{C_{1} + 2\cE_{0}\left(\eta - \eta_{0}\right)}\right] + \frac{\beta_{\tA}}{2} \, , 
	\quad \tx \equiv \left(\frac{\eta - \eta_{0}}{\eta_{\ast}\left(1 + \frac{\cE_{0}}{C_{1}}\left(\eta - \eta_{0}\right)\right)}\right)^{\pm\sqrt{3}} \, ,
	\label{eq:raddeltaphi}
\end{equation}
which reduces to the dilaton perturbation of the DFT vacuum solution~\eqref{eq:Evacdeltaphi} for $\cE_{0} \rightarrow 0$ if we identify 
$\beta_{\delta\phi} \equiv \widehat{\beta}_{\delta\phi} \pm \sqrt{3}\widehat{\beta}_{\tA}$. 
From the above expressions, one may pursue this further to solve~\eqref{eq:Eulerk} and~\eqref{eq:EDFEscalarE2k} for $v$ and $\widehat{m}$, respectively.  However, as the solution appears somewhat involved, we leave this as an exercise to the interested reader.

\subsubsection{Radiation and scalar field}
As a final example, it is also possible to combine the previous ingredients and write down an analytic solution for radiation plus a scalar field with vanishing potential, in the presence of a non-trivial dilaton and $H$-flux~\cite{Angus:2019bqs}.  
Conservation of the total energy-momentum tensor combined with the scalar field equation of motion~\eqref{eq:scalarbkgdPhi} implies that the radiation energy density is independently conserved,
\begin{equation}
	\brtrho_{\rm r}' + 3\tcH(1 + w_{\rm r})\brtrho_{\rm r} = 0 \quad \Longrightarrow \quad \brtrho_{\rm r} = \brrho_{{\rm r}0}\ta^{-4} \, ,
\end{equation}
where $\brtrho_{\rm r}$ is the energy density of a fluid with $\tw_{\rm r} = \frac{1}{3}$ and $\tlambda_{\rm r} = 0$.

The solution follows readily by recycling much of the previous analysis.  
Equations~\eqref{eq:Ebkgd2} and~\eqref{eq:Ebkgd3} behave identically to the case of pure radiation, yielding solutions
\begin{equation}
	\ta^{2} = C_{1}\left(\eta - \eta_{0}\right) + \cE_{0}\left(\eta - \eta_{0}\right)^{2} \, , \qquad \cE_{0} \equiv \frac{8\pi G\brrho_{{\rm r}0}}{3} \, , \label{eq:scalaradta}
\end{equation}
\begin{equation}
	e^{2\brphi} = \left(\frac{\eta - \eta_{0}}{\eta_{\star}\left(1 + \frac{\cE_{0}}{C_{1}}\left(\eta - \eta_{0}\right)\right)}\right)^{\pm\frac{h_{\rm{o}}}{C_{1}}}
	+ \frac{h^{2}}{4h_{\rm{o}}^{2}}\left(\frac{\eta - \eta_{0}}{\eta_{\star}\left(1 + \frac{\cE_{0}}{C_{1}}\left(\eta - \eta_{0}\right)\right)}\right)^{\mp\frac{h_{\rm{o}}}{C_{1}}} \label{eq:scalaradbrphi} \, ,
\end{equation}
respectively.  Meanwhile,~\eqref{eq:scalarbkgdPhi} can be solved 
analogously to the pure massless scalar field case, giving  
\begin{equation}
	\brPhi = \frac{\alpha_{\brPhi}}{C_{1}}\ln\left(\frac{\eta - \eta_{0}}{1 + \frac{\cE_{0}}{C_{1}}\left(\eta - \eta_{0}\right)}\right) + \brPhi_{0} \, , \label{eq:scalaradbrPhi}
\end{equation}
while equation~\eqref{eq:Ebkgd1} implies the same relation~\eqref{eq:hoC1scalar} between integration constants as for the massless scalar,
\begin{equation}
	16\pi G\alpha_{\brPhi}^{2} = 3C_{1}^{2} - h_{\circ}^{2} \, . \label{eq:hoC1scalarad}
\end{equation}

The generalization also extends to linear perturbations in the superhorizon limit (see Appendix~\ref{sec:ana} for details).  
For a fluid with $\tc_{\rm{s,r}}^{2} = \tw_{\rm r} = \frac{1}{3}$, the 
continuity equation~\eqref{eq:continuityk}
again implies that
\begin{equation}
	\delta\trho_{\rm r} = \delta\rho_{{\rm r}0}\ta^{-4} \, .
\end{equation}
Thus from~\eqref{eq:EDFEscalarE3k} and~\eqref{eq:EDFEscalarE4k}, $\tA$ and $\tB$ have the same superhorizon solutions~\eqref{eq:radtA} and~\eqref{eq:radtB} as for radiation,
\begin{equation}
	\tA = \frac{\widehat{\beta}_{\tA}C_{1}^{2}}{4\ta^{4}\tcH^{2}} - \frac{\delta_{{\rm r}0}}{2} \, , \qquad \delta_{{\rm r}0} \equiv \frac{\delta\rho_{{\rm r}0}}{\brrho_{{\rm r}0}} \, , \label{eq:scalaradtA}
\end{equation}
\begin{equation}
	B = - \frac{\beta_{\tA}C_{1}^{2}\left(\eta - \eta_{0}\right)^{2}}{4\ta^{4}\tcH} + \frac{\beta_{B}}{2\ta^{2}} + \frac{\delta_{{\rm r}0}\cE_{0}\left(\eta - \eta_{0}\right)^{2}\left(C_{1}\left(\eta - \eta_{0}\right) + \ta^{2}\right)}{6\ta^{4}\tcH} \, . \label{eq:scalaradtB}
\end{equation}
However, 
in this case $\tA'$ contributes non-trivially to the $\delta\Phi$ equation of motion~\eqref{eq:eomdeltaPhi}, giving the solution 
\begin{equation}
	\delta\Phi = \frac{\alpha_{\delta\Phi}}{C_1}\ln\left(\frac{\eta - \eta_{0}}{1 + \frac{\cE_{0}}{C_{1}}\left(\eta - \eta_{0}\right)}\right) - \frac{2\alpha_{\brPhi}\widehat{\beta}_{\tA}\cE_{0}\left(\eta - \eta_{0}\right)}{C_{1}\ta^{2}\tcH} + \delta\Phi_{0} \, , \label{eq:scalaraddeltaPhi}
\end{equation}
where the integration constants $\alpha_{\delta\Phi}$ and $\delta\Phi_{0}$ have been chosen to match~\eqref{eq:VlessscalardeltaPhi} 
in the limit $\cE_{0} \rightarrow 0$.
Next, as in previous cases, the dilaton perturbation $\delta\phi$ can be obtained from equation~\eqref{eq:EDFEscalarE1k}, yielding 
\begin{equation}
	\delta\phi = \frac{2\ta^{2}\brphi'}{h_{\circ}}\left[\widehat{\beta}_{\delta\phi} - \frac{8\pi G\alpha_{\brPhi}\alpha_{\delta\Phi}}{h_{\circ}C_{1}}\ln\left(\frac{\eta - \eta_{0}}{1 + \frac{\cE_{0}}{C_{1}}\left(\eta - \eta_{0}\right)}\right) + \frac{h_{\circ}\widehat{\beta}_{\tA}}{2\ta^{2}\tcH}\right] + \frac{\beta_{\tA}}{2} + \frac{8\pi G\alpha_{\brPhi}\alpha_{\delta\Phi}}{h_{\circ}^{2}} \, ,
\end{equation}
where $\widehat{\beta}_{\delta\phi} \equiv \beta_{\delta\phi} - \frac{h_{\circ}}{C_{1}}\widehat{\beta}_{\tA}$.  When the massless scalar field coefficients vanish ($\alpha_{\brPhi} \rightarrow 0$, $\alpha_{\delta\Phi} \rightarrow 0$), this reduces to the expression~\eqref{eq:raddeltaphi} from the case of pure radiation; similarly, when the radiation contribution is set to zero ($\cE_{0} \rightarrow 0$, $\delta\rho_{{\rm r}0}\rightarrow 0$), the result converges to that of the massless scalar field solution~\eqref{eq:Vlessscalardeltaphi}.

\section{Discussion} \label{sec:discuss}
If string theory is the true theory of quantum gravity, it is natural to take the full closed-string massless sector, consisting of the spacetime metric, $B$-field, and dilaton, as the low-energy effective theory of gravity.  The corresponding action enjoys an $\ODD$ symmetry, which is most simply described using double field theory (DFT).  Extending this description to include additional matter leads to a DFT generalization of Einstein's equations, with the allowed couplings to matter constrained accordingly.

In this work we have applied this string-inspired modified gravity framework to cosmology and studied perturbations around homogeneous and isotropic background solutions.  We have derived the equations of motion for the perturbation variables, we have obtained solutions in the superhorizon limit for fluctuations around various DFT cosmological backgrounds, and we have identified the generalized condition for the conservation of adiabatic superhorizon perturbations.

Having developed the formalism of cosmological perturbations in DFT, there are a number of potential applications.  A natural extension would be to study perturbations for nonzero spatial curvature $K$, for which explicit background solutions are already known, and may be relevant for late-time applications~\cite{Choi:2022srv,Lee:2023boi}.  Furthermore, it would be worthwhile to go beyond the superhorizon limit and examine the $k_{i}$-dependence of the perturbations, including the corresponding analysis for spatially curved backgrounds.  Connecting to late-time cosmology and analysing perturbations can also provide an important consistency check and further test the viability of DFT cosmological scenarios.

There are other aspects of DFT cosmology that deserve further study.  Fermions in DFT couple to two separate local Lorentz groups, and an embedding of the Standard Model was considered in~\cite{Choi:2015bga}.  It is worth considering fermions in more detail and assessing possible cosmological signatures of such a model.  Moreover, the presence of non-trivial background $H$-flux also opens new avenues for exploration.  A coupling between DFT fermions and $H$-flux may generate new signals from astrophysical sources, while $\ODD$ covariance may constrain the allowed parity-violating interactions involving $H$-flux, which may be relevant to non-Gaussianity~\cite{Bartolo:2004if} or cosmic birefringence~\cite{Minami:2020odp}.  A dilaton coupling to electromagnetism could affect radiation signals, while gravitational wave signatures may also be modified at higher orders.

There are also more speculative directions to consider.  In addition to usual Riemannian spacetime, double field theory also admits non-Riemannian backgrounds~\cite{Morand:2017fnv}, on which the usual spacetime metric is degenerate but the DFT metric is regular.  This has been argued to provide a smooth description of various gravitational singularities~\cite{Morand:2021xeq}.  Particular cases of non-Riemannian geometry include Newton--Cartan `non-relativistic' and Carrollian `ultra-relativistic' gravities.  In particular, the latter was proposed as an alternative framework for inflation~\cite{deBoer:2021jej}.  It would be interesting to explore such scenarios for the early universe in a DFT context.

\section*{Acknowledgements}
The authors would like to thank Stephen Appleby, Eric Lescano and Jeong-Hyuck Park for helpful discussions and comments.  SA is supported by the National Research Foundation of Korea (NRF) via the Grant NRF-2022R1F1A1070999 as well as through the Center for Quantum Spacetime (CQUeST) of Sogang University (NRF-2020R1A6A1A03047877), funded by the Korean Government (MSIT).  SA also wishes to thank the Asia Pacific Center for Theoretical Physics and the Yukawa Institute for Theoretical Physics for their support and hospitality while this work was developed.  SM is supported by the Japan Society for the Promotion of
Science (JSPS) Grants-in-Aid for Scientific Research (KAKENHI) No. JP24K07017, and by the World Premier International Research
Center Initiative (WPI), MEXT, Japan.

~\\
\noindent This version of the article has been
accepted for publication after peer review, but is not the Version of Record and does not
reflect post-acceptance improvements or corrections. The Version of Record is available online at:
\href{https://doi.org/10.1140/epjc/s10052-025-13841-7}{https://doi.org/10.1140/epjc/s10052-025-13841-7}.

\appendix

\section{DFT energy-momentum tensor on spacetime backgrounds} \label{sec:KgB}
In this Appendix we relate in detail the DFT gravitational fields to the usual supergravity variables and derive the spacetime variation of the matter action given in~\eqref{eq:KgB} and~\eqref{eq:dphig}.  
With the conventional section choice, $\p/\p\tilde{x}_{\mu} = 0$, the DFT metric can be expanded as
\begin{equation}
	\cH_{AB}=\left(\ba{cc}
	\quad g^{\mu\nu}\quad&\quad -g^{\mu\lambda}B_{\lambda\sigma}\quad\\
	\quad B_{\rho\kappa}g^{\kappa\nu}\quad&\quad
	g_{\rho\sigma}-B_{\rho\kappa}g^{\kappa\lambda}B_{\lambda\sigma}\quad
	\ea\right) \, , \label{HAB}
\end{equation}
while the vielbeins can similarly be expanded as
\begin{equation}
	\ba{ll}
	V_{Ap}=\frac{1}{\sqrt{2}}\left(\ba{c}e_{p}{}^{\mu}\\
	e_{\nu}{}^{q}\eta_{qp}+B_{\nu\sigma}e_{p}{}^{\sigma}\ea\right)\,,\quad&\quad
	\brV_{A\brp}=\frac{1}{\sqrt{2}}\left(\ba{c}\bre_{\brp}{}^{\mu}\\
	\bre_{\nu}{}^{\brq}\breta_{\brq\brp}+B_{\nu\sigma}\bre_{\brp}{}^{\sigma}\ea\right)\,. \label{VbrV}
	\ea
\end{equation}
Here we have introduced a pair of Riemannian vielbeins $e_{\mu}{}^{p}$ and $\bre_{\mu}{}^{\brp}$, which satisfy the defining relations
\begin{equation}
	e_{\mu}{}^{p}e_{\nu}{}^{q}\eta_{pq} = g_{\mu\nu} = -\bre_{\mu}{}^{\brp}\bre_{\nu}{}^{\brq}\breta_{\brp\brq} \, . \label{ebre}
\end{equation}
The local $D$-dimensional Lorentz metrics take the standard form, $\eta = \breta = \textrm{diag}(-1,1\ldots 1)$.  Note that it follows from \eqref{ebre} that the inverse vielbeins,
\begin{equation}
	e_{p}{}^{\mu} \equiv (e_{\mu}{}^{p})^{-1} = g^{\mu\nu}\eta_{pq}e_{\nu}{}^{q} \, , \qquad \bre_{\brp}{}^{\mu} \equiv (\bre_{\mu}{}^{p})^{-1} = -g^{\mu\nu}\breta_{\brp\brq}\bre_{\nu}{}^{\brq} \, ,
\end{equation}
similarly satisfy
\begin{equation}
	e_{p}{}^{\mu}e_{q}{}^{\nu}\eta^{pq} = g^{\mu\nu} = - \bre_{\brp}{}^{\mu}\bre_{\brq}{}^{\nu}\breta^{\brp\brq} \, .
\end{equation}
In addition, the DFT dilaton $d$ is related to the supergravity dilaton $\phi$ and the determinant of the spacetime metric via
\begin{equation}
	e^{-2d} = \sqrt{-g}e^{-2\phi} \, . \label{eq:d}
\end{equation}

Using \eqref{HAB}, the variation of $\cH_{AB}$ can be written explicitly as
\begin{align}
	\delta\cH^{\mu\nu} &= \delta g^{\mu\nu} = -g^{\mu\rho}g^{\nu\sigma}\delta g_{\rho\sigma} \, ; \label{deltaHAB1} \\
	\delta\cH^{\mu}{}_{\sigma} &= -\delta g^{\mu\lambda}B_{\lambda\sigma} - g^{\mu\lambda}\delta B_{\lambda\sigma} \, ; \label{deltaHAB2} \\
	\delta\cH_{\rho}{}^{\nu} &= \delta B_{\rho\kappa}g^{\kappa\nu} + B_{\rho\kappa}\delta g^{\kappa\nu} \, ; \label{deltaHAB3} \\
	\delta\cH_{\rho\sigma} &= \delta g_{\rho\sigma} - \delta B_{\rho\kappa} g^{\kappa\lambda}B_{\lambda\sigma} - B_{\rho\kappa}\delta g^{\kappa\lambda}B_{\lambda\sigma} - B_{\rho\kappa}g^{\kappa\lambda}\delta B_{\lambda\sigma} \, , \label{deltaHAB4}
\end{align}
where we have distinguished the different components of $\delta\cH_{AB}$ by the positions of the spacetime indices.  
From these we can expand
\begin{align}
	\frac{\delta S_{\rm m}}{\delta\cH_{AB}}\delta\cH_{AB} &= \frac{\delta S_{\rm m}}{\delta\cH^{\mu\nu}}\delta\cH^{\mu\nu} + \frac{\delta S_{\rm m}}{\delta\cH^{\mu}{}_{\sigma}}\delta\cH^{\mu}{}_{\sigma} + \frac{\delta S_{\rm m}}{\delta\cH_{\rho}{}^{\nu}}\delta\cH_{\rho}{}^{\nu} + \frac{\delta S_{\rm m}}{\delta\cH_{\rho\sigma}}\delta\cH_{\rho\sigma} \nn \\
	&=\left[\frac{\delta S_{\rm m}}{\delta\cH^{\mu\nu}} - \frac{\delta S_{\rm m}}{\delta\cH^{\mu}{}_{\sigma}}B_{\nu\sigma} - \frac{\delta S_{\rm m}}{\delta\cH_{\rho}{}^{\nu}}B_{\mu\rho} - \frac{\delta S_{\rm m}}{\delta\cH_{\rho\sigma}}(g_{\mu\rho}g_{\nu\sigma} - B_{\mu\rho}B_{\nu\sigma})\right]\delta g^{\mu\nu} \nn \\
	&\quad + \left[- \frac{\delta S_{\rm m}}{\delta\cH^{\mu}{}_{\sigma}}g_{\nu\sigma} + \frac{\delta S_{\rm m}}{\delta\cH_{\rho}{}^{\nu}}g_{\mu\rho} - \frac{\delta S_{\rm m}}{\delta\cH_{\rho\sigma}}(g_{\mu\rho}B_{\nu\sigma} - B_{\mu\rho}g_{\nu\sigma})\right]g^{\mu\lambda}g^{\nu\tau}\delta B_{\lambda\tau} \, . \label{eq:deltaLmdeltaHAB}
\end{align}

We can apply the same treatment to \eqref{eq:Kmunu} to obtain the spacetime kinetic tensor
\begin{equation}
	K_{\mu\nu} \equiv 2e_{\mu}{}^{p}\bre_{\nu}{}^{\brq} K_{p\brq} = -4e_{\mu}{}^{p}\bre_{\nu}{}^{\brq}V_{Ap}\brV_{B\brq}e^{2d}\frac{\delta S_{\rm m}}{\delta\cH_{AB}} \, .
\end{equation}
Expanding this in terms of Riemannian variables yields
\begin{align}
	K_{\mu\nu} &= -2e^{2d}\bigg[\frac{\delta S_{\rm m}}{\delta\cH^{\mu\nu}} - \frac{\delta S_{\rm m}}{\delta\cH^{\mu}{}_{\sigma}}(g_{\nu\sigma} + B_{\nu\sigma}) + \frac{\delta S_{\rm m}}{\delta\cH_{\rho}{}^{\nu}}(g_{\mu\rho} - B_{\mu\rho}) \nn \\
	&\qquad\quad - \frac{\delta S_{\rm m}}{\delta\cH_{\rho\sigma}}(g_{\mu\rho} - B_{\mu\rho})(g_{\nu\sigma} + B_{\nu\sigma})\bigg] \, .
\end{align}
Separating into symmetric and skew-symmetric pieces,~\eqref{eq:deltaLmdeltaHAB} can thus be expressed as
\begin{equation}
	e^{2d}\frac{\delta S_{\rm m}}{\delta\cH_{AB}}\delta\cH_{AB} = -\frac{1}{2}K_{(\mu\nu)}\delta g^{\mu\nu} - \frac{1}{2}K_{[\mu\nu]}g^{\mu\lambda}g^{\nu\tau}\delta B_{\lambda\tau} \, . \label{KgB}
\end{equation}
Similarly, in the case of the dilaton, expanding~\eqref{eq:d} gives
\begin{equation}
	\delta d = \delta\phi - \frac{1}{4}\delta\ln(-g) = \delta\phi + \frac{1}{4}g_{\mu\nu}\delta g^{\mu\nu} \, , \label{dphig}
\end{equation}
where we have used $g^{-1}\delta g = g^{\mu\nu}\delta g_{\mu\nu}$.

\section{String frame and gauge-invariant variables} \label{sec:stringframe}
In section~\ref{pertOFEscalar} the equations of motion for scalar perturbations were presented in Einstein frame in an arbitrary gauge.  While this choice simplifies many calculations and conversion to other frames and variables is relatively straightforward, in this Appendix we include some other relevant presentations for completeness.

First of all, the scalar perturbation equations of motion take the following form in string frame:
\begin{align}
	&8\pi Ge^{2\brphi}N^{2}\left[\delta\rho + 2\brrho\left(A + \delta\phi\right)\right] \nn \\
	&= -\frac{2N}{a}\left(\cH - \brphi'\right)D^{2}B + 6\left(\cH - \brphi'\right)C' - \frac{2N^{2}}{a^{2}}D^{2}\left(C - \delta\phi - \frac{1}{3}D^{2}E\right) + \frac{6KN^{2}}{a^{2}}\left(A - C + \frac{1}{3}D^{2}E\right) \nn \\
	&\quad - 2\left(3\cH - 2\brphi'\right)\delta\phi' - \frac{N^{2}h^{2}}{2a^{6}}\left(A - 3C\right) - \frac{N^{2}h}{2a^{6}}D^{2}m \, ; \label{eq:EDFEscalar1} \\
	&D_{i}\left[8\pi Ge^{2\brphi}\left(\brrho + \brp\right)\left(v + B\right)\right] \nn \\
	&= D_{i}\Bigg[-2\left(\cH - \brphi'\right)A + \left(2K - \frac{h^{2}}{2a^{4}}\right)\frac{N}{a}B + 2\left(C - \delta\phi - \frac{1}{3}D^{2}E\right)'
	- 2KE' + 2\cH\delta\phi + \frac{h}{2a^{4}}m'\Bigg] \, ; \label{eq:EDFEscalar2} \\
	&8\pi Ge^{2\brphi}N^{2}\left[\delta p + 2\brp\left(A + \delta\phi\right)\right] \nn \\
	&= 2\left(\cH - \brphi'\right)A' - 2C'' - 2\left(3\cH - \frac{N'}{N} - 2\brphi'\right)C' - \frac{2KN^{2}}{a^{2}}\left(A - C + \frac{1}{3}D^{2}E\right)
	\nn \\
	&\quad + \frac{2N^{2}}{3a^{2}}D^{2}\bigg[A + \frac{a}{N}\left(B' + 2\left(\cH - \brphi'\right)B\right) + C - 2\delta\phi - \frac{1}{3}D^{2}E\bigg] + 2\delta\phi'' + 2\left(2\cH - \frac{N'}{N} - 2\brphi'\right)\delta\phi' \nn \\
	&\quad - \frac{N^{2}h^{2}}{2a^{6}}\left(A - 3C\right) - \frac{N^{2}h}{2a^{6}}D^{2}m \, ; \label{eq:EDFEscalar3} \\
	&0 = \left(D_{i}D_{j} - \frac{1}{3}\Omega_{ij}D^{2}\right)\Bigg\{8\pi Ge^{2\brphi}a^{2}\brp\pi_{\mathrm{T}} + A + \frac{a}{N}\left(B' + 2\left(\cH - \brphi'\right)B\right) + C - \frac{1}{3}D^{2}E - 2\delta\phi \nn \\
	&\qquad\qquad\qquad\qquad\qquad\quad - \frac{a^{2}}{N^{2}}\left[E'' + \left(3\cH - \frac{N'}{N} - 2\brphi'\right)E'\right]\Bigg\} \, ; \label{eq:EDFEscalar4} \\
	&4\pi Ge^{2\brphi}N^{2}\left[\delta\sigma - \delta\rho + 3\delta p + 2\left(\brsigma - \brrho + 3\brp\right)\left(A + \delta\phi\right)\right] \nn \\ &= \delta\phi'' + \left(3\cH - \frac{N'}{N} - 4\brphi'\right)\delta\phi' - \frac{N^{2}}{a^{2}}D^{2}\delta\phi + \brphi'\left(- A' + 3C' - \frac{N}{a}D^{2}B\right)
	- \frac{N^{2}h^{2}}{a^{6}}\left(A - 3C\right) \nn \\ &\quad - \frac{N^{2}h}{a^{6}}D^{2}m \, ; \label{eq:EDFEscalar7} 
\end{align}
\begin{align}
	&16\pi Ge^{4\brphi}D_{i}\cJ = 0 \, ; \label{eq:EDFEscalar5} \\
	&16\pi Ge^{4\brphi}N^{2}D_{i}\cI = D_{i}\Bigg\{- m'' + \left(\cH + \frac{N'}{N} + 2\brphi'\right)m' + \frac{N^{2}}{a^{2}}D^{2}m \nn \\
	&\qquad\qquad\qquad\qquad\quad\quad\,\, 
	+ \frac{N^{2}h}{a^{2}}\left[A + \frac{a}{N}\left(B' - 2\left(\cH + \brphi'\right)B\right) - 3C - 2\delta\phi\right]\Bigg\} \, . \label{eq:EDFEscalar6}
\end{align}

Moreover, equations~\eqref{eq:EDFEscalarE1}--\eqref{eq:EDFEscalarE7} may be expressed using the gauge-invariant variables of section~\ref{sec:gaugeinv}.  Explicitly, we find the following set of equations:
\begin{align}
	&0 = 8\pi G\left[\brtrho\left(\tDelta + 2\tcA\right) - \brtrho'\frac{\ta}{\tN}V\right] - \frac{6}{\tN^{2}}\tcH\tcC' + \frac{2}{\tN^{2}}\brphi'\widehat{\delta\phi}' + \frac{2}{\ta^{2}}D^{2}\tcC - \frac{6K}{\ta^{2}}\left(\tcA - \tcC\right) \nn \\
	&\quad\,\,\, + \frac{e^{-4\brphi}h^{2}}{2\ta^{6}}\left(\tcA - 3\tcC - 2\widehat{\delta\phi}\right) + \frac{e^{-4\brphi}h}{2\ta^{6}}D^{2}\cM \, ; \\
	&0 = D_{i}\left\{8\pi G\left(\brtrho + \brtp\right)V + \frac{2}{\tN\ta}\tcH\tcA - \frac{2}{\tN\ta}\tcC' - \frac{2}{\tN\ta}\brphi'\widehat{\delta\phi} - \frac{e^{-4\brphi}h}{2\tN\ta^{5}}\cM'\right\} \, ; \\
	&0 = 8\pi G\left[\brtp\left(\tPi_{\mathrm{L}} + 2\tcA\right) - \brtp'\frac{\ta}{\tN}V\right] - \frac{2}{\tN^{2}}\tcH\tcA' + \frac{2}{\tN^{2}}\tcC'' + \frac{2}{\tN^{2}}\left(3\tcH - \frac{\tN'}{\tN}\right)\tcC' + \frac{2}{\tN^{2}}\brphi'\widehat{\delta\phi}' \nn \\
	&\quad\,\,\, - \frac{2}{3\ta^{2}}D^{2}\left(\tcA + \tcC\right) + \frac{2K}{\ta^{2}}\left(\tcA - \tcC\right) + \frac{e^{-4\brphi}h^{2}}{2\ta^{6}}\left(\tcA - 3\tcC - 2\widehat{\delta\phi}\right) + \frac{e^{-4\brphi}h}{2\ta^{6}}D^{2}\cM \, ; \\
	&0 = \left(D_{i}D_{j} - \frac{1}{3}\Omega_{ij}D^{2}\right)\left\{8\pi G\ta^{2}\brtp\pi_{\mathrm{T}} + \tcA + \tcC\right\} \, ; \\
	&0 = 4\pi G\left[\brtsigma\left(\tSigma + 2\tcA\right) - \brtsigma'\frac{\ta}{\tN}V\right] - \frac{1}{\tN^{2}}\widehat{\delta\phi}'' - \frac{1}{\tN^{2}}\left(3\tcH - \frac{\tN'}{\tN}\right)\widehat{\delta\phi}' + \frac{1}{\ta^{2}}D^{2}\widehat{\delta\phi} \nn \\
	&\quad\,\,\, + \frac{\brphi'}{\tN^{2}}\left(\tcA' - 3\tcC'\right) + \frac{e^{-4\brphi}h^{2}}{\ta^{6}}\left(\tcA - 3\tcC - 2\widehat{\delta\phi}\right) + \frac{e^{-4\brphi}h}{\ta^{6}}D^{2}\cM \, ; \\
	&0 = 16\pi Ge^{4\brphi}D_{i}\cJ \, ; \\
	&0 = D_{i}\left\{16\pi Ge^{4\brphi}\cI + \frac{1}{\tN^{2}}\left[\cM'' - \left(\tcH + \frac{\tN'}{\tN} + 4\brphi'\right)\cM'\right] - \frac{1}{\ta^{2}}D^{2}\cM - \frac{h}{\ta^{2}}\left(\tcA - 3\tcC - 4\widehat{\delta\phi}\right)\right\} \, .
\end{align}
Furthermore, the conservation equations~\eqref{eq:continuity} and~\eqref{eq:Euler} for the scalar perturbations can be written  in terms of these variables as
\begin{align}
	0 &= \left(\brtrho\tDelta - \brtrho'\frac{\ta}{\tN}V\right)' + 3\tcH\left(\brtrho\tDelta - \brtrho'\frac{\ta}{\tN}V + \brtp\tPi_{\mathrm{L}} - \brtp\frac{\ta}{\tN}V\right) + \left(\brtrho + \brtp\right)\left(3\tcC' + \frac{\tN}{\ta}D^{2}V\right) \nn \\
	&\quad + \brphi'\left(\brtsigma\tSigma - \brtsigma'\frac{\ta}{\tN}V\right) + \brtsigma\widehat{\delta\phi} \, , \\
	0 &= D_{i}\left\{\brtp\tPi_{\mathrm{L}} - \brtsigma\widehat{\delta\phi}_{V} + \left(\brtrho + \brtp\right)\left[\tcA + \frac{\tN'}{\tN}\left(\frac{\ta}{\tN}V\right) + \left(\frac{\ta}{\tN}V\right)'\right] + 2\brtp\left(\frac{1}{3}D^{2} + K\right)\pi_{\mathrm{T}} + \frac{h}{\ta^{4}}\cI\right\} \, .
\end{align}

\section{Analytic solutions} \label{sec:ana}
In this Appendix we present derivations of some analytic solutions discussed in Section~\ref{sec:ex} for perturbation variables in the superhorizon limit, $k \rightarrow 0$.  
Specifically, we consider the most general case studied therein, consisting of a background of radiation plus a massless scalar field in the presence of a non-trivial dilaton and background $H$-flux, from which other examples can be obtained by taking various limits.  The background evolution is given by~\eqref{eq:scalaradta}, \eqref{eq:scalaradbrphi} and~\eqref{eq:scalaradbrPhi}, where the constants of integration are constrained to satisfy~\eqref{eq:hoC1scalarad}.

The full set of perturbation equations is
\begin{align}
	8\pi G\left[\ta^{2}\left(\delta\trho_{\rm r} + 2\tA\brtrho_{\rm r}\right) + \brPhi'\delta\Phi'\right] &= - \frac{e^{-4\brphi}h^{2}}{2\ta^{4}}\tA - 2\brphi'\delta\phi' + \frac{e^{-4\brphi}h^{2}}{\ta^{4}}\delta\phi \, , \label{eq:scalaradscalarReE1} \\
	8\pi G\left[\ta^{2}\brtrho_{\rm r}\left(1 + \tw_{\rm r}\right)\left(v_{\rm r} + B\right) - \brPhi'\delta\Phi\right] &= - 2\tcH\tA - \frac{e^{-4\brphi}h^{2}}{2\ta^{4}}\tB + 2\brphi'\delta\phi + \frac{e^{-4\brphi}}{2\ta^{4}}\widehat{m}' \, , \label{eq:scalaradscalarReE2} \\
	8\pi G\ta^{2}\left[\left(1-\tc_{\rm{s,r}}^{2}\right)\delta\trho_{\rm r} + 2\tA\left(1 - \tw_{\rm r}\right)\brtrho_{\rm r}\right] &= -2\tcH\tA' \, , \label{eq:scalaradscalarReE3} \\
	0 &= \tA + \tB' + 2\tcH\tB \, , \label{eq:scalaradscalarReE4} \\
	0 &= \widehat{m}'' - \left(2\tcH + 4\brphi'\right)\widehat{m}' + 4h^{2}\left(\left(\tcH + \brphi'\right)\tB + \delta\phi\right) \, , \label{eq:scalaradscalarReE5} \\
	0 &= \delta\phi'' + 2\tcH\delta\phi' + \frac{2e^{-4\brphi}h^{2}}{\ta^{4}}\delta\phi - \brphi'\tA' - \frac{e^{-4\brphi}h^{2}}{\ta^{4}}\tA \, , \label{eq:scalaradscalarReE6} \\
	0 &= \delta\Phi'' + 2\tcH\delta\Phi' - \brPhi'\tA' \, , \label{eq:scalaradscalarReEPhi}
\end{align}
and the conservation equations are
\begin{align}
	0 &= \delta\trho_{\rm r}' + 3\tcH\left(1 + \tc_{\rm{s,r}}^{2}\right)\delta\trho_{\rm r} \, , \label{eq:scalaradscalarEC1sup} \\
	0 &= \left[\left(1 + \tw_{\rm r}\right)\brtrho_{\rm r}\left(v_{\rm r} + B\right)\right]' + 4\tcH\left[\left(1 + \tw_{\rm r}\right)\brtrho_{\rm r}\left(v_{\rm r} + B\right)\right] + \tc_{\rm{s,r}}^{2}\delta\trho + \left(1 + \tw_{\rm r}\right)\brtrho_{\rm r}\tA \, . \label{eq:scalaradscalarEC2sup}
\end{align}
For an adiabatic fluid corresponding to radiation, $\tc_{\rm{s,r}}^{2} = \tw_{\rm r} = \frac{1}{3}$, and thus
\begin{equation}
	\delta\trho_{\rm r} = \delta\rho_{{\rm r}0}\ta^{-4} \, . \label{eq:scalaraddeltatrho}
\end{equation}

Now we can solve for the gravitational perturbations.  One necessary ingredient is the explicit form of the Hubble parameter,
\begin{equation}
	\tcH = \frac{\ta'}{\ta} = \frac{C_{1} + 2\cE_{0}\left(\eta - \eta_{0}\right)}{2\ta^{2}} \, . \label{eq:scalaradbkgdtcHsolution}
\end{equation}
First of all, the solution for $\tA$ can be obtained from~\eqref{eq:scalaradscalarReE3}.  Using $\trho_{\rm r} = \brrho_{{\rm r}0}\ta^{-4}$, $\cE_{0} = 8\pi G\brrho_{{\rm r}0}/3$ as well as~\eqref{eq:scalaraddeltatrho} and~\eqref{eq:scalaradbkgdtcHsolution}, and setting $\tc_{\rm{s,r}}^{2} = \tw_{\rm r} = \frac{1}{3}$, we can express~\eqref{eq:scalaradscalarReE3} as
\begin{equation}
	\frac{1}{2}\left[C_{1} + 2\cE_{0}\left(\eta - \eta_{0}\right)\right]\tA' + 2\cE_{0}\tA = -\frac{8\pi G}{3}\delta\rho_{{\rm r}0} \, .
\end{equation}
We can integrate this by multiplying throughout by another factor of $\frac{1}{2}\left[C_{1} + 2\cE_{0}\left(\eta - \eta_{0}\right)\right]$.  The result is
\begin{equation}
	\tA = \frac{\beta_{\tA}C_{1}^{2} - 2\delta_{{\rm r}0}\cE_{0}\left[C_{1}\left(\eta - \eta_{0}\right) + \cE_{0}\left(\eta - \eta_{0}\right)^{2}\right]}{\left[C_{1} + 2\cE_{0}\left(\eta - \eta_{0}\right)\right]^{2}} = \frac{\beta_{\tA}C_{1}^{2}}{4\ta^{4}\tcH^{2}} - \frac{\delta_{{\rm r}0}\cE_{0}}{2\ta^{2}\tcH^{2}} \, , \label{eq:AscalaradtA}
\end{equation}
where we have defined $\delta_{{\rm r}0} = \delta\rho_{{\rm r}0}/\brrho_{{\rm r}0}$, and $\beta_{A}$ is an integration constant.
If we define $\widehat{\beta}_{\tA} \equiv \beta_{\tA} + \frac{\delta_{{\rm r}0}}{2}$, this can be written simply as
\begin{equation}
	\tA = \frac{\widehat{\beta}_{\tA}C_{1}^{2}}{\left[C_{1} + 2\cE_{0}\left(\eta - \eta_{0}\right)\right]^{2}} - \frac{\delta_{{\rm r}0}}{2} = \frac{\widehat{\beta}_{\tA}C_{1}^{2}}{4\ta^{4}\tcH^{2}} - \frac{\delta_{{\rm r}0}}{2} \, .
\end{equation}

Next we solve for $B$ using equation~\eqref{eq:scalaradscalarReE4}.  Inserting~\eqref{eq:AscalaradtA} and multiplying through by $\ta^{2}$, this becomes
\begin{equation}
	\left(\ta^{2}B\right)' = \frac{\delta_{{\rm r}0}\ta^{2}}{2} - \frac{\widehat{\beta}_{A}C_{1}^{2}}{4\ta^{2}\tcH^{2}} \, .
\end{equation}
This can be integrated by using the fact that
\begin{equation}
	\left(\frac{1}{\tcH}\right)' = 2 - \frac{\cE_{0}}{\ta^{2}\tcH^{2}} \, ,
\end{equation}
which follows from the background equation of motion~\eqref{eq:Ebkgd2}.  After some algebra, this yields the solution
\begin{equation}
	B = - \frac{\beta_{\tA}C_{1}^{2}\left(\eta - \eta_{0}\right)^{2}}{4\ta^{4}\tcH} + \frac{\beta_{B}}{2\ta^{2}} + \frac{\delta_{{\rm r}0}\cE_{0}\left(\eta - \eta_{0}\right)^{2}\left(C_{1}\left(\eta - \eta_{0}\right) + \ta^{2}\right)}{6\ta^{4}\tcH} \, , \label{eq:AscalaradtB}
\end{equation}
where $\beta_{\tB}$ is an integration constant.

Next we solve for $\delta\Phi$.  Integrating~\eqref{eq:scalaradscalarReEPhi} once after plugging in the background expression~\eqref{eq:scalaradbrPhi} gives
\begin{equation}
	\delta\Phi' = \frac{\alpha_{\delta\Phi}}{\ta^{2}} + \brPhi'\left(\tA - \alpha_{\tA}\right) \, , \label{eq:AscalaradddeltaPhi}
\end{equation}
where the integration constant $\alpha_{\delta\Phi}$ has been chosen such that the pure scalar field result~\eqref{eq:VlessscalardeltaPhi} is recovered in the limit $\cE_{0}\rightarrow 0$.  Using~\eqref{eq:AscalaradtA} as well as the fact that $\ta^{4}\cH^{2} = C_{1}^{2} + 4\cE_{0}\ta^{2}$, this can be integrated a second time to obtain 
\begin{equation}
	\delta\Phi = \frac{\alpha_{\delta\Phi}}{C_1}\ln\left(\frac{\eta - \eta_{0}}{1 + \frac{\cE_{0}}{C_{1}}\left(\eta - \eta_{0}\right)}\right) - \frac{2\alpha_{\brPhi}\widehat{\beta}_{\tA}\cE_{0}\left(\eta - \eta_{0}\right)}{C_{1}\ta^{2}\tcH} + \delta\Phi_{0} \, , \label{eq:AscalaraddeltaPhi}
\end{equation}
with the constant $\delta\Phi_{0}$ similarly chosen to match~\eqref{eq:VlessscalardeltaPhi}.

The dilaton perturbation $\delta\phi$ can be obtained from~\eqref{eq:scalaradscalarReE1}. 
Using~\eqref{eq:Ebkgd1},~\eqref{eq:Ebkgd3},~\eqref{eq:scalaradbrPhi},~\eqref{eq:hoC1scalarad},~\eqref{eq:AscalaradtA} and~\eqref{eq:AscalaradddeltaPhi}, equation~\eqref{eq:scalaradscalarReE1} can be expressed as
\begin{equation}
	-\left(\frac{\delta\phi}{\ta^{2}\brphi'}\right)' = \frac{\beta_{\tA}h_{\circ}^{2} + 16\pi G\alpha_{\brPhi}\alpha_{\delta\Phi}}{4\ta^{6}\brphi'{}^{2}} - \frac{\beta_{\tA}C_{1}^{2}}{4\ta^{6}\tcH^{2}} + \frac{\delta_{{\rm r}0}\cE_{0}}{2\ta^{4}\tcH^{2}} \, .
\end{equation}
The right-hand side can be simplified using relations such as
\begin{equation}
	\frac{h_{\circ}^{2}}{2\ta^{6}\brphi'{}^{2}} = \frac{2}{\ta^{2}} - \left(\frac{1}{\ta^{2}\brphi'}\right)' \, , \qquad \frac{C_{1}^{2}}{4\ta^{6}\tcH^{2}} = \frac{1}{\ta^{2}} + \left(\frac{1}{\ta^{2}\tcH}\right)' \, ,
\end{equation}
which follow from the background equations of motion.  Integrating then yields the dilaton perturbation, 
\begin{equation}
	\delta\phi = \frac{2\ta^{2}\brphi'}{h_{\circ}}\left[\widehat{\beta}_{\delta\phi} - \frac{8\pi G\alpha_{\brPhi}\alpha_{\delta\Phi}}{h_{\circ}C_{1}}\ln\left(\frac{\eta - \eta_{0}}{1 + \frac{\cE_{0}}{C_{1}}\left(\eta - \eta_{0}\right)}\right) + \frac{h_{\circ}\widehat{\beta}_{\tA}}{2\ta^{2}\tcH}\right] + \frac{\beta_{\tA}}{2} + \frac{8\pi G\alpha_{\brPhi}\alpha_{\delta\Phi}}{h_{\circ}^{2}} \, ,
\end{equation}
which matches the massless scalar field case~\eqref{eq:Vlessscalardeltaphi} when $\cE_{0} \rightarrow 0$, provided we identify $\beta_{\delta\phi} \equiv \widehat{\beta}_{\delta\phi} + \frac{h_{\circ}}{C_{1}}\widehat{\beta}_{\tA}$, and also reduces to the pure radiation solution~\eqref{eq:raddeltaphi} when $h_{\circ}\rightarrow\pm\sqrt{3}C_{1}$.

\section{Massless scalar field: $H$-flux perturbation solutions for $h\neq 0$} \label{sec:m}
In this Appendix we present the general solution for $\widehat{m}$ in the case of a massless scalar field, as obtained from~\eqref{eq:EDFEscalarE2k}.  For non-zero $h$, the general expression for $\widehat{m}'$ reads
\begin{align}
	\widehat{m}' = &-h^{2}\alpha_{\tA}\left(\eta - \eta_{0}\right) + \frac{h^{2}\beta_{\tB}}{2C_{1}\left(\eta - \eta_{0}\right)} - \frac{16\pi G\alpha_{\brPhi}\alpha_{\delta\Phi}h^{2}}{2h_{\circ}^{2}}\p_{0}\left[\left(\eta - \eta_{0}\right)^{2}\ln\left(\eta - \eta_{0}\right)\right] \nn \\ &+ \left(2\alpha_{\tA}C_{1} - 16\pi G\alpha_{\brPhi}\delta\Phi_{0} - 2\beta_{\delta\phi} -\alpha_{\tA}h_{\circ} - \frac{16\pi G\alpha_{\brPhi}\alpha_{\delta\Phi}}{h_{\circ}}\right)C_{1}\left(\eta - \eta_{0}\right)x^{2} \nn \\
	&+ \left(2\alpha_{\tA}C_{1} - 16\pi G\alpha_{\brPhi}\delta\Phi_{0} + 2\beta_{\delta\phi} + \frac{\alpha_{\tA}h_{\circ}^{2}}{C_{1}} + \frac{16\pi G\alpha_{\brPhi}\alpha_{\delta\Phi}}{C_{1}}\right)\frac{h^{2}C_{1}\left(\eta - \eta_{0}\right)}{2h_{\circ}^{2}} \nn \\
	&+ \left(2\alpha_{\tA}C_{1} - 16\pi G\alpha_{\brPhi}\delta\Phi_{0} - 2\beta_{\delta\phi} + \alpha_{\tA}h_{\circ} + \frac{16\pi G\alpha_{\brPhi}\alpha_{\delta\Phi}}{h_{\circ}}\right)\frac{h^{4}C_{1}\left(\eta - \eta_{0}\right)}{16h_{\circ}^4x^{2}} \, .
\end{align}
This leads to three different expressions for $\widehat{m}$, depending on whether $|h_{\circ}/C_{1}| \neq 1$ or $h_{\circ}/C_{1} = \pm 1$.
\begin{itemize}
	\item $|h_{\circ}/C_{1}| \neq 1$:
	\begin{align}
		\widehat{m} = &\; \widehat{m}_{0} + \frac{h^{2}}{2}\left[- \alpha_{\tA}\left(\eta - \eta_{0}\right)^{2} + \frac{\beta_{\tB}}{C_{1}}\ln\left(\eta - \eta_{0}\right) - \frac{16\pi G\alpha_{\brPhi}\alpha_{\delta\Phi}}{h_{\circ}^{2}}\left(\eta - \eta_{0}\right)^{2}\ln\left(\eta - \eta_{0}\right)\right] \nn \\
		&+ \left(2\alpha_{\tA}C_{1} - 16\pi G\alpha_{\brPhi}\delta\Phi_{0} - 2\beta_{\delta\phi} -\alpha_{\tA}h_{\circ} - \frac{16\pi G\alpha_{\brPhi}\alpha_{\delta\Phi}}{h_{\circ}}\right)\frac{C_{1}^{2}\left(\eta - \eta_{0}\right)^{2}x^{2}}{2\left(C_{1} + h_{\circ}\right)} \nn \\
		&+ \left(2\alpha_{\tA}C_{1} - 16\pi G\alpha_{\brPhi}\delta\Phi_{0} + 2\beta_{\delta\phi} + \frac{\alpha_{\tA}h_{\circ}^{2}}{C_{1}} + \frac{16\pi G\alpha_{\brPhi}\alpha_{\delta\Phi}}{C_{1}}\right)\frac{h^{2}C_{1}\left(\eta - \eta_{0}\right)^{2}}{4h_{\circ}^{2}} \nn \\
		&+ \left(2\alpha_{\tA}C_{1} - 16\pi G\alpha_{\brPhi}\delta\Phi_{0} - 2\beta_{\delta\phi} + \alpha_{\tA}h_{\circ} + \frac{16\pi G\alpha_{\brPhi}\alpha_{\delta\Phi}}{h_{\circ}}\right)\frac{h^{4}C_{1}^{2}\left(\eta - \eta_{0}\right)^{2}}{32h_{\circ}^4\left(C_{1} - h_{\circ}\right)x^{2}} \, ; \label{eq:Vlessscalarm1}
	\end{align}
	\item $h_{\circ}/C_{1} = 1$:
	\begin{align}
		\widehat{m} = &\; \widehat{m}_{0} + \frac{h^{2}}{2}\left[- \alpha_{\tA}\left(\eta - \eta_{0}\right)^{2} + \frac{\beta_{\tB}}{C_{1}}\ln\left(\eta - \eta_{0}\right) - \frac{16\pi G\alpha_{\brPhi}\alpha_{\delta\Phi}}{h_{\circ}^{2}}\left(\eta - \eta_{0}\right)^{2}\ln\left(\eta - \eta_{0}\right)\right] \nn \\
		&+ \left(2\alpha_{\tA}C_{1} - 16\pi G\alpha_{\brPhi}\delta\Phi_{0} - 2\beta_{\delta\phi} -\alpha_{\tA}h_{\circ} - \frac{16\pi G\alpha_{\brPhi}\alpha_{\delta\Phi}}{h_{\circ}}\right)\frac{C_{1}\left(\eta - \eta_{0}\right)^{2}x^{2}}{4} \nn \\
		&+ \left(2\alpha_{\tA}C_{1} - 16\pi G\alpha_{\brPhi}\delta\Phi_{0} + 2\beta_{\delta\phi} + \frac{\alpha_{\tA}h_{\circ}^{2}}{C_{1}} + \frac{16\pi G\alpha_{\brPhi}\alpha_{\delta\Phi}}{C_{1}}\right)\frac{h^{2}C_{1}\left(\eta - \eta_{0}\right)^{2}}{4h_{\circ}^{2}} \nn \\
		&+ \left(2\alpha_{\tA}C_{1} - 16\pi G\alpha_{\brPhi}\delta\Phi_{0} - 2\beta_{\delta\phi} + \alpha_{\tA}h_{\circ} + \frac{16\pi G\alpha_{\brPhi}\alpha_{\delta\Phi}}{h_{\circ}}\right)\frac{h^{4}C_{1}\eta_{\ast}^{2}}{16h_{\circ}^4}\ln\left(\eta - \eta_{0}\right) \, ; \label{eq:Vlessscalarm2}
	\end{align}
	\item $h_{\circ}/C_{1} = -1$:
	\begin{align}
		\widehat{m} = &\; \widehat{m}_{0} + \frac{h^{2}}{2}\left[- \alpha_{\tA}\left(\eta - \eta_{0}\right)^{2} + \frac{\beta_{\tB}}{C_{1}}\ln\left(\eta - \eta_{0}\right) - \frac{16\pi G\alpha_{\brPhi}\alpha_{\delta\Phi}}{h_{\circ}^{2}}\left(\eta - \eta_{0}\right)^{2}\ln\left(\eta - \eta_{0}\right)\right] \nn \\
		&+ \left(2\alpha_{\tA}C_{1} - 16\pi G\alpha_{\brPhi}\delta\Phi_{0} - 2\beta_{\delta\phi} -\alpha_{\tA}h_{\circ} - \frac{16\pi G\alpha_{\brPhi}\alpha_{\delta\Phi}}{h_{\circ}}\right)C_{1}\eta_{\ast}^{2}\ln\left(\eta - \eta_{0}\right) \nn \\
		&+ \left(2\alpha_{\tA}C_{1} - 16\pi G\alpha_{\brPhi}\delta\Phi_{0} + 2\beta_{\delta\phi} + \frac{\alpha_{\tA}h_{\circ}^{2}}{C_{1}} + \frac{16\pi G\alpha_{\brPhi}\alpha_{\delta\Phi}}{C_{1}}\right)\frac{h^{2}C_{1}\left(\eta - \eta_{0}\right)^{2}}{4h_{\circ}^{2}} \nn \\
		&+ \left(2\alpha_{\tA}C_{1} - 16\pi G\alpha_{\brPhi}\delta\Phi_{0} - 2\beta_{\delta\phi} + \alpha_{\tA}h_{\circ} + \frac{16\pi G\alpha_{\brPhi}\alpha_{\delta\Phi}}{h_{\circ}}\right)\frac{h^{4}C_{1}\left(\eta - \eta_{0}\right)^{2}}{64h_{\circ}^4x^{2}} \, . \label{eq:Vlessscalarm3}
	\end{align}
\end{itemize}
Note that in the limit $h\rightarrow 0$,~\eqref{eq:Vlessscalarm1} and~\eqref{eq:Vlessscalarm2} both converge to the upper equation of~\eqref{eq:Vlessscalarmh0}, while~\eqref{eq:Vlessscalarm3} reduces to the lower expression of~\eqref{eq:Vlessscalarmh0}.


\begin{thebibliography}{99}
	\bibitem{Planck:2018vyg}
	N.~Aghanim \textit{et al.} [Planck],
	``Planck 2018 results. VI. Cosmological parameters,''
	Astron. Astrophys. \textbf{641}, A6 (2020)
	[erratum: Astron. Astrophys. \textbf{652}, C4 (2021)]
	doi:10.1051/0004-6361/201833910
	[arXiv:1807.06209 [astro-ph.CO]].
	
	\bibitem{DESI:2024mwx}
	A.~G.~Adame \textit{et al.} [DESI],
	``DESI 2024 VI: Cosmological Constraints from the Measurements of Baryon Acoustic Oscillations,''
	[arXiv:2404.03002 [astro-ph.CO]].
	
	\bibitem{Riess:2021jrx}
	A.~G.~Riess, W.~Yuan, L.~M.~Macri, D.~Scolnic, D.~Brout, S.~Casertano, D.~O.~Jones, Y.~Murakami, L.~Breuval and T.~G.~Brink, \textit{et al.}
	``A Comprehensive Measurement of the Local Value of the Hubble Constant with 1 km s$^{-1}$ Mpc$^{-1}$ Uncertainty from the Hubble Space Telescope and the SH0ES Team,''
	Astrophys. J. Lett. \textbf{934}, no.1, L7 (2022)
	doi:10.3847/2041-8213/ac5c5b
	[arXiv:2112.04510 [astro-ph.CO]].
	
	\bibitem{Labbe:2022ahb}
	I.~Labbe, P.~van Dokkum, E.~Nelson, R.~Bezanson, K.~A.~Suess, J.~Leja, G.~Brammer, K.~Whitaker, E.~Mathews and M.~Stefanon, \textit{et al.}
	``A population of red candidate massive galaxies \textasciitilde{}600 Myr after the Big Bang,''
	Nature \textbf{616}, no.7956, 266-269 (2023)
	doi:10.1038/s41586-023-05786-2
	[arXiv:2207.12446 [astro-ph.GA]].
	
	\bibitem{Giddings:2001yu}
	S.~B.~Giddings, S.~Kachru and J.~Polchinski,
	``Hierarchies from fluxes in string compactifications,''
	Phys. Rev. D \textbf{66}, 106006 (2002)
	doi:10.1103/PhysRevD.66.106006
	[arXiv:hep-th/0105097 [hep-th]].
	
	\bibitem{Meissner:1991zj}
	K.~A.~Meissner and G.~Veneziano,
	``Symmetries of cosmological superstring vacua,''
	Phys. Lett. B \textbf{267}, 33-36 (1991)
	doi:10.1016/0370-2693(91)90520-Z
	
	\bibitem{Meissner:1991ge}
	K.~A.~Meissner and G.~Veneziano,
	``Manifestly O(d,d) invariant approach to space-time dependent string vacua,''
	Mod. Phys. Lett. A \textbf{6}, 3397-3404 (1991)
	doi:10.1142/S0217732391003924
	[arXiv:hep-th/9110004 [hep-th]].
	
	\bibitem{Gasperini:1991ak}
	M.~Gasperini and G.~Veneziano,
	``O(d,d) covariant string cosmology,''
	Phys. Lett. B \textbf{277}, 256-264 (1992)
	doi:10.1016/0370-2693(92)90744-O
	[arXiv:hep-th/9112044 [hep-th]].
	
	\bibitem{Buscher:1987sk}
	T.~H.~Buscher,
	``A Symmetry of the String Background Field Equations,''
	Phys. Lett. B \textbf{194}, 59-62 (1987)
	doi:10.1016/0370-2693(87)90769-6
	
	\bibitem{Buscher:1987qj}
	T.~H.~Buscher,
	``Path Integral Derivation of Quantum Duality in Nonlinear Sigma Models,''
	Phys. Lett. B \textbf{201}, 466-472 (1988)
	doi:10.1016/0370-2693(88)90602-8
	
	\bibitem{Siegel:1993xq}
	W.~Siegel,
	``Two vierbein formalism for string inspired axionic gravity,''
	Phys. Rev. D \textbf{47}, 5453-5459 (1993)
	doi:10.1103/PhysRevD.47.5453
	[arXiv:hep-th/9302036 [hep-th]].
	
	\bibitem{Siegel:1993th}
	W.~Siegel,
	``Superspace duality in low-energy superstrings,''
	Phys. Rev. D \textbf{48}, 2826-2837 (1993)
	doi:10.1103/PhysRevD.48.2826
	[arXiv:hep-th/9305073 [hep-th]].
	
	\bibitem{Hull:2009mi}
	C.~Hull and B.~Zwiebach,
	``Double Field Theory,''
	JHEP \textbf{09}, 099 (2009)
	doi:10.1088/1126-6708/2009/09/099
	[arXiv:0904.4664 [hep-th]].
	
	\bibitem{Hull:2009zb}
	C.~Hull and B.~Zwiebach,
	``The Gauge algebra of double field theory and Courant brackets,''
	JHEP \textbf{09}, 090 (2009)
	doi:10.1088/1126-6708/2009/09/090
	[arXiv:0908.1792 [hep-th]].
	
	\bibitem{Hohm:2010jy}
	O.~Hohm, C.~Hull and B.~Zwiebach,
	``Background independent action for double field theory,''
	JHEP \textbf{07}, 016 (2010)
	doi:10.1007/JHEP07(2010)016
	[arXiv:1003.5027 [hep-th]].
	
	\bibitem{Hohm:2010pp}
	O.~Hohm, C.~Hull and B.~Zwiebach,
	``Generalized metric formulation of double field theory,''
	JHEP \textbf{08}, 008 (2010)
	doi:10.1007/JHEP08(2010)008
	[arXiv:1006.4823 [hep-th]].
	
	\bibitem{Hitchin:2003cxu}
	N.~Hitchin,
	``Generalized Calabi-Yau manifolds,''
	Quart. J. Math. Oxford Ser. \textbf{54}, 281-308 (2003)
	doi:10.1093/qjmath/54.3.281
	[arXiv:math/0209099 [math.DG]].
	
	\bibitem{Gualtieri:2003dx}
	M.~Gualtieri,
	``Generalized complex geometry,''
	[arXiv:math/0401221 [math.DG]].
	
	\bibitem{Hitchin:2010qz}
	N.~Hitchin,
	``Lectures on generalized geometry,''
	[arXiv:1008.0973 [math.DG]].
	
	\bibitem{Coimbra:2011nw}
	A.~Coimbra, C.~Strickland-Constable and D.~Waldram,
	``Supergravity as Generalised Geometry I: Type II Theories,''
	JHEP \textbf{11}, 091 (2011)
	doi:10.1007/JHEP11(2011)091
	[arXiv:1107.1733 [hep-th]].
	
	\bibitem{Coimbra:2012yy}
	A.~Coimbra, C.~Strickland-Constable and D.~Waldram,
	``Generalised Geometry and type II Supergravity,''
	Fortsch. Phys. \textbf{60}, 982-986 (2012)
	doi:10.1002/prop.201100096
	[arXiv:1202.3170 [hep-th]].
	
	\bibitem{Vaisman:2012ke}
	I.~Vaisman,
	``On the geometry of double field theory,''
	J. Math. Phys. \textbf{53}, 033509 (2012)
	doi:10.1063/1.3694739
	[arXiv:1203.0836 [math.DG]].
	
	\bibitem{Berman:2013uda}
	D.~S.~Berman, C.~D.~A.~Blair, E.~Malek and M.~J.~Perry,
	``The $O_{D,D}$ geometry of string theory,''
	Int. J. Mod. Phys. A \textbf{29}, 1450080 (2014)
	doi:10.1142/S0217751X14500808
	[arXiv:1303.6727 [hep-th]].
	
	\bibitem{Garcia-Fernandez:2013gja}
	M.~Garcia-Fernandez,
	``Torsion-free generalized connections and Heterotic Supergravity,''
	Commun. Math. Phys. \textbf{332}, no.1, 89-115 (2014)
	doi:10.1007/s00220-014-2143-5
	[arXiv:1304.4294 [math.DG]].
	
	\bibitem{Cederwall:2014kxa}
	M.~Cederwall,
	``The geometry behind double geometry,''
	JHEP \textbf{09}, 070 (2014)
	doi:10.1007/JHEP09(2014)070
	[arXiv:1402.2513 [hep-th]].
	
	\bibitem{Cederwall:2016ukd}
	M.~Cederwall,
	``Double supergeometry,''
	JHEP \textbf{06}, 155 (2016)
	doi:10.1007/JHEP06(2016)155
	[arXiv:1603.04684 [hep-th]].
	
	\bibitem{Deser:2016qkw}
	A.~Deser and C.~S\"amann,
	``Extended Riemannian Geometry I: Local Double Field Theory,''
	doi:10.1007/s00023-018-0694-2
	[arXiv:1611.02772 [hep-th]].
	
	\bibitem{Sakamoto:2017cpu}
	J.~i.~Sakamoto, Y.~Sakatani and K.~Yoshida,
	``Homogeneous Yang-Baxter deformations as generalized diffeomorphisms,''
	J. Phys. A \textbf{50}, no.41, 415401 (2017)
	doi:10.1088/1751-8121/aa8896
	[arXiv:1705.07116 [hep-th]].
	
	\bibitem{Cederwall:2017fjm}
	M.~Cederwall and J.~Palmkvist,
	``Extended geometries,''
	JHEP \textbf{02}, 071 (2018)
	doi:10.1007/JHEP02(2018)071
	[arXiv:1711.07694 [hep-th]].
	
	\bibitem{Freidel:2018tkj}
	L.~Freidel, F.~J.~Rudolph and D.~Svoboda,
	``A Unique Connection for Born Geometry,''
	Commun. Math. Phys. \textbf{372}, no.1, 119-150 (2019)
	doi:10.1007/s00220-019-03379-7
	[arXiv:1806.05992 [hep-th]].
	
	\bibitem{Chatzistavrakidis:2019huz}
	A.~Chatzistavrakidis, L.~Jonke, F.~S.~Khoo and R.~J.~Szabo,
	``The Algebroid Structure of Double Field Theory,''
	PoS \textbf{CORFU2018}, 132 (2019)
	doi:10.22323/1.347.0132
	[arXiv:1903.01765 [hep-th]].
	
	\bibitem{Jeon:2011cn}
	I.~Jeon, K.~Lee and J.~H.~Park,
	``Stringy differential geometry, beyond Riemann,''
	Phys. Rev. D \textbf{84}, 044022 (2011)
	doi:10.1103/PhysRevD.84.044022
	[arXiv:1105.6294 [hep-th]].
	
	\bibitem{Jeon:2010rw}
	I.~Jeon, K.~Lee and J.~H.~Park,
	``Differential geometry with a projection: Application to double field theory,''
	JHEP \textbf{04}, 014 (2011)
	doi:10.1007/JHEP04(2011)014
	[arXiv:1011.1324 [hep-th]].
	
	\bibitem{Hohm:2011si}
	O.~Hohm and B.~Zwiebach,
	``On the Riemann Tensor in Double Field Theory,''
	JHEP \textbf{05}, 126 (2012)
	doi:10.1007/JHEP05(2012)126
	[arXiv:1112.5296 [hep-th]].
	
	\bibitem{Choi:2015bga}
	K.~S.~Choi and J.~H.~Park,
	``Standard Model as a Double Field Theory,''
	Phys. Rev. Lett. \textbf{115}, no.17, 171603 (2015)
	doi:10.1103/PhysRevLett.115.171603
	[arXiv:1506.05277 [hep-th]].
	
	\bibitem{Jeon:2011vx}
	I.~Jeon, K.~Lee and J.~H.~Park,
	``Incorporation of fermions into double field theory,''
	JHEP \textbf{11}, 025 (2011)
	doi:10.1007/JHEP11(2011)025
	[arXiv:1109.2035 [hep-th]].
	
	\bibitem{Jeon:2011kp}
	I.~Jeon, K.~Lee and J.~H.~Park,
	``Double field formulation of Yang-Mills theory,''
	Phys. Lett. B \textbf{701}, 260-264 (2011)
	doi:10.1016/j.physletb.2011.05.051
	[arXiv:1102.0419 [hep-th]].
	
	\bibitem{Hohm:2011ex}
	O.~Hohm and S.~K.~Kwak,
	``Double Field Theory Formulation of Heterotic Strings,''
	JHEP \textbf{06}, 096 (2011)
	doi:10.1007/JHEP06(2011)096
	[arXiv:1103.2136 [hep-th]].
	
	\bibitem{Hohm:2014sxa}
	O.~Hohm, A.~Sen and B.~Zwiebach,
	``Heterotic Effective Action and Duality Symmetries Revisited,''
	JHEP \textbf{02}, 079 (2015)
	doi:10.1007/JHEP02(2015)079
	[arXiv:1411.5696 [hep-th]].
	
	\bibitem{Cho:2018alk}
	K.~Cho, K.~Morand and J.~H.~Park,
	``Kaluza\textendash{}Klein reduction on a maximally non-Riemannian space is moduli-free,''
	Phys. Lett. B \textbf{793}, 65-69 (2019)
	doi:10.1016/j.physletb.2019.04.042
	[arXiv:1808.10605 [hep-th]].
	
	\bibitem{Jeon:2012kd}
	I.~Jeon, K.~Lee and J.~H.~Park,
	``Ramond-Ramond Cohomology and O(D,D) T-duality,''
	JHEP \textbf{09}, 079 (2012)
	doi:10.1007/JHEP09(2012)079
	[arXiv:1206.3478 [hep-th]].
	
	\bibitem{Ko:2016dxa}
	S.~M.~Ko, J.~H.~Park and M.~Suh,
	``The rotation curve of a point particle in stringy gravity,''
	JCAP \textbf{06}, 002 (2017)
	doi:10.1088/1475-7516/2017/06/002
	[arXiv:1606.09307 [hep-th]].
	
	\bibitem{Blair:2017gwn}
	C.~D.~A.~Blair,
	``Particle actions and brane tensions from double and exceptional geometry,''
	JHEP \textbf{10}, 004 (2017)
	doi:10.1007/JHEP10(2017)004
	[arXiv:1707.07572 [hep-th]].
	
	\bibitem{Hull:2006va}
	C.~M.~Hull,
	``Doubled Geometry and T-Folds,''
	JHEP \textbf{07}, 080 (2007)
	doi:10.1088/1126-6708/2007/07/080
	[arXiv:hep-th/0605149 [hep-th]].
	
	\bibitem{Lee:2013hma}
	K.~Lee and J.~H.~Park,
	``Covariant action for a string in ''doubled yet gauged'' spacetime,''
	Nucl. Phys. B \textbf{880}, 134-154 (2014)
	doi:10.1016/j.nuclphysb.2014.01.003
	[arXiv:1307.8377 [hep-th]].
	
	\bibitem{Park:2016sbw}
	J.~H.~Park,
	``Green-Schwarz superstring on doubled-yet-gauged spacetime,''
	JHEP \textbf{11}, 005 (2016)
	doi:10.1007/JHEP11(2016)005
	[arXiv:1609.04265 [hep-th]].
	
	\bibitem{Sakamoto:2018krs}
	J.~I.~Sakamoto and Y.~Sakatani,
	``Local $\beta$-deformations and Yang-Baxter sigma model,''
	JHEP \textbf{06}, 147 (2018)
	doi:10.1007/JHEP06(2018)147
	[arXiv:1803.05903 [hep-th]].
	
	\bibitem{Angus:2018mep}
	S.~Angus, K.~Cho and J.~H.~Park,
	``Einstein Double Field Equations,''
	Eur. Phys. J. C \textbf{78}, no.6, 500 (2018)
	doi:10.1140/epjc/s10052-018-5982-y
	[arXiv:1804.00964 [hep-th]].
	
	\bibitem{Gasperini:2004ss}
	M.~Gasperini, M.~Giovannini and G.~Veneziano,
	``Cosmological perturbations across a curvature bounce,''
	Nucl. Phys. B \textbf{694}, 206-238 (2004)
	doi:10.1016/j.nuclphysb.2004.06.020
	[arXiv:hep-th/0401112 [hep-th]].
	
	\bibitem{Gasperini:2007ar}
	M.~Gasperini,
	``Dilaton cosmology and phenomenology,''
	Lect. Notes Phys. \textbf{737}, 787-844 (2008)
	[arXiv:hep-th/0702166 [hep-th]].
	
	\bibitem{Gasperini:2002bn}
	M.~Gasperini and G.~Veneziano,
	``The Pre - big bang scenario in string cosmology,''
	Phys. Rept. \textbf{373}, 1-212 (2003)
	doi:10.1016/S0370-1573(02)00389-7
	[arXiv:hep-th/0207130 [hep-th]].
	
	\bibitem{Tseytlin:1991xk}
	A.~A.~Tseytlin and C.~Vafa,
	``Elements of string cosmology,''
	Nucl. Phys. B \textbf{372}, 443-466 (1992)
	doi:10.1016/0550-3213(92)90327-8
	[arXiv:hep-th/9109048 [hep-th]].
	
	\bibitem{Brustein:1997ny}
	R.~Brustein and R.~Madden,
	``Graceful exit and energy conditions in string cosmology,''
	Phys. Lett. B \textbf{410}, 110-118 (1997)
	doi:10.1016/S0370-2693(97)00945-3
	[arXiv:hep-th/9702043 [hep-th]].
	
	\bibitem{Gasperini:2007zz}
	M.~Gasperini,
	``Elements of string cosmology,''
	Cambridge University Press, 2007,
	ISBN 978-0-511-33229-6, 978-0-521-18798-5, 978-0-521-86875-4
	
	\bibitem{Copeland:1994vi}
	E.~J.~Copeland, A.~Lahiri and D.~Wands,
	``Low-energy effective string cosmology,''
	Phys. Rev. D \textbf{50}, 4868-4880 (1994)
	doi:10.1103/PhysRevD.50.4868
	[arXiv:hep-th/9406216 [hep-th]].
	
	\bibitem{Lidsey:1999mc}
	J.~E.~Lidsey, D.~Wands and E.~J.~Copeland,
	``Superstring cosmology,''
	Phys. Rept. \textbf{337}, 343-492 (2000)
	doi:10.1016/S0370-1573(00)00064-8
	[arXiv:hep-th/9909061 [hep-th]].
	
	\bibitem{Cicoli:2023opf}
	M.~Cicoli, J.~P.~Conlon, A.~Maharana, S.~Parameswaran, F.~Quevedo and I.~Zavala,
	``String cosmology: From the early universe to today,''
	Phys. Rept. \textbf{1059}, 1-155 (2024)
	doi:10.1016/j.physrep.2024.01.002
	[arXiv:2303.04819 [hep-th]].
	
	\bibitem{Angus:2019bqs}
	S.~Angus, K.~Cho, G.~Franzmann, S.~Mukohyama and J.~H.~Park,
	``$\mathbf {O}(D,D)$ completion of the Friedmann equations,''
	Eur. Phys. J. C \textbf{80}, no.9, 830 (2020)
	doi:10.1140/epjc/s10052-020-8379-7
	[arXiv:1905.03620 [hep-th]].
	
	\bibitem{Wu:2013sha}
	H.~Wu and H.~Yang,
	``Double Field Theory Inspired Cosmology,''
	JCAP \textbf{07}, 024 (2014)
	doi:10.1088/1475-7516/2014/07/024
	[arXiv:1307.0159 [hep-th]].
	
	\bibitem{Wu:2013ixa}
	H.~Wu and H.~Yang,
	``New Cosmological Signatures from Double Field Theory,''
	[arXiv:1312.5580 [hep-th]].
	
	\bibitem{Ma:2014ala}
	C.~T.~Ma and C.~M.~Shen,
	``Cosmological Implications from O(D,D),''
	Fortsch. Phys. \textbf{62}, 921-941 (2014)
	doi:10.1002/prop.201400049
	[arXiv:1405.4073 [hep-th]].
	
	\bibitem{Brandenberger:2017umf}
	R.~Brandenberger, R.~Costa, G.~Franzmann and A.~Weltman,
	``Point particle motion in double field theory and a singularity-free cosmological solution,''
	Phys. Rev. D \textbf{97}, no.6, 063530 (2018)
	doi:10.1103/PhysRevD.97.063530
	[arXiv:1710.02412 [hep-th]].
	
	\bibitem{Brandenberger:2018xwl}
	R.~Brandenberger, R.~Costa, G.~Franzmann and A.~Weltman,
	``Dual spacetime and nonsingular string cosmology,''
	Phys. Rev. D \textbf{98}, no.6, 063521 (2018)
	doi:10.1103/PhysRevD.98.063521
	[arXiv:1805.06321 [hep-th]].
	
	\bibitem{Brandenberger:2018bdc}
	R.~Brandenberger, R.~Costa, G.~Franzmann and A.~Weltman,
	``T-dual cosmological solutions in double field theory,''
	Phys. Rev. D \textbf{99}, no.2, 023531 (2019)
	doi:10.1103/PhysRevD.99.023531
	[arXiv:1809.03482 [hep-th]].
	
	\bibitem{Bernardo:2019pnq}
	H.~Bernardo, R.~Brandenberger and G.~Franzmann,
	``$T$-dual cosmological solutions in double field theory. II.,''
	Phys. Rev. D \textbf{99}, no.6, 063521 (2019)
	doi:10.1103/PhysRevD.99.063521
	[arXiv:1901.01209 [hep-th]].
	
	\bibitem{Hohm:2019jgu}
	O.~Hohm and B.~Zwiebach,
	``Duality invariant cosmology to all orders in $\alpha$',''
	Phys. Rev. D \textbf{100}, no.12, 126011 (2019)
	doi:10.1103/PhysRevD.100.126011
	[arXiv:1905.06963 [hep-th]].
	
	\bibitem{Hohm:2019ccp}
	O.~Hohm and B.~Zwiebach,
	``Non-perturbative de Sitter vacua via $\alpha'$ corrections,''
	Int. J. Mod. Phys. D \textbf{28}, no.14, 1943002 (2019)
	doi:10.1142/S0218271819430028
	[arXiv:1905.06583 [hep-th]].
	
	\bibitem{Bernardo:2019bkz}
	H.~Bernardo, R.~Brandenberger and G.~Franzmann,
	``O$(d,d)$ covariant string cosmology to all orders in $\alpha^{\prime}$,''
	JHEP \textbf{02}, 178 (2020)
	doi:10.1007/JHEP02(2020)178
	[arXiv:1911.00088 [hep-th]].
	
	\bibitem{Codina:2021cxh}
	T.~Codina, O.~Hohm and D.~Marques,
	``General string cosmologies at order \ensuremath{\alpha}'3,''
	Phys. Rev. D \textbf{104}, no.10, 106007 (2021)
	doi:10.1103/PhysRevD.104.106007
	[arXiv:2107.00053 [hep-th]].
	
	\bibitem{Bernardo:2020zlc}
	H.~Bernardo and G.~Franzmann,
	``$\alpha'$-Cosmology: solutions and stability analysis,''
	JHEP \textbf{05}, 073 (2020)
	doi:10.1007/JHEP05(2020)073
	[arXiv:2002.09856 [hep-th]].
	
	\bibitem{Quintin:2021eup}
	J.~Quintin, H.~Bernardo and G.~Franzmann,
	``Cosmology at the top of the \ensuremath{\alpha}' tower,''
	JHEP \textbf{07}, 149 (2021)
	doi:10.1007/JHEP07(2021)149
	[arXiv:2105.01083 [hep-th]].
	
	\bibitem{Bernardo:2022nex}
	H.~Bernardo, J.~Chojnacki and V.~Comeau,
	``Non-linear stability of \ensuremath{\alpha}'-corrected Friedmann equations,''
	JHEP \textbf{03}, 119 (2023)
	doi:10.1007/JHEP03(2023)119
	[arXiv:2212.11392 [hep-th]].
	
	\bibitem{Nunez:2020hxx}
	C.~A.~N\'u\~nez and F.~E.~Rost,
	``New non-perturbative de Sitter vacua in $\alpha'$-complete cosmology,''
	JHEP \textbf{03}, 007 (2021)
	doi:10.1007/JHEP03(2021)007
	[arXiv:2011.10091 [hep-th]].
	
	\bibitem{Bernardo:2021xtr}
	H.~Bernardo, P.~R.~Chouha and G.~Franzmann,
	``Kalb-Ramond backgrounds in $\alpha$'-complete cosmology,''
	JHEP \textbf{09}, 109 (2021)
	doi:10.1007/JHEP09(2021)109
	[arXiv:2104.15131 [hep-th]].
	
	\bibitem{Gasperini:2023tus}
	M.~Gasperini and G.~Veneziano,
	``Non-singular pre-big bang scenarios from all-order \ensuremath{\alpha}' corrections,''
	JHEP \textbf{07}, 144 (2023)
	doi:10.1007/JHEP07(2023)144
	[arXiv:2305.00222 [hep-th]].
	
	\bibitem{Lescano:2021nju}
	E.~Lescano and N.~Mir\'on-Granese,
	``Double field theory with matter and its cosmological application,''
	Phys. Rev. D \textbf{107}, no.4, 046016 (2023)
	doi:10.1103/PhysRevD.107.046016
	[arXiv:2111.03682 [hep-th]].
	
	\bibitem{Lescano:2023gge}
	E.~Lescano, N.~Mir\'on-Granese and Y.~Sakatani,
	``O(D,D)-covariant formulation of perfect and imperfect fluids in the double geometry,''
	Phys. Rev. D \textbf{109}, no.8, 086006 (2024)
	doi:10.1103/PhysRevD.109.086006
	[arXiv:2312.03610 [hep-th]].
	
	\bibitem{Arapoglu:2024umz}
	A.~S.~Arapo\u{g}lu, S.~\c{C}a\u{g}an and A.~\c{C}atal-\"Ozer,
	``Stability analysis of the cosmological dynamics of O(D,~D)-complete stringy gravity,''
	Eur. Phys. J. C \textbf{84}, no.8, 848 (2024)
	doi:10.1140/epjc/s10052-024-13213-7
	[arXiv:2405.07825 [gr-qc]].
	
	\bibitem{Brandenberger:2023ver}
	R.~Brandenberger,
	``Superstring cosmology \textemdash{} a complementary review,''
	JCAP \textbf{11}, 019 (2023)
	doi:10.1088/1475-7516/2023/11/019
	[arXiv:2306.12458 [hep-th]].
	
	\bibitem{Hohm:2022pfi}
	O.~Hohm and A.~F.~Pinto,
	``Cosmological Perturbations in Double Field Theory,''
	JHEP \textbf{04}, 073 (2023)
	doi:10.1007/JHEP04(2023)073
	[arXiv:2207.14788 [hep-th]].
	
	\bibitem{Morand:2017fnv}
	K.~Morand and J.~H.~Park,
	``Classification of non-Riemannian doubled-yet-gauged spacetime,''
	Eur. Phys. J. C \textbf{77}, no.10, 685 (2017)
	[erratum: Eur. Phys. J. C \textbf{78}, no.11, 901 (2018)]
	doi:10.1140/epjc/s10052-017-5257-z, 10.1140/epjc/s10052-018-6394-8
	[arXiv:1707.03713 [hep-th]].
	
	\bibitem{Angus:2021jvm}
	S.~Angus, M.~Kim and J.~H.~Park,
	``Fractons, non-Riemannian geometry, and double field theory,''
	Phys. Rev. Res. \textbf{4}, no.3, 033186 (2022)
	doi:10.1103/PhysRevResearch.4.033186
	[arXiv:2111.07947 [hep-th]].
	
	\bibitem{Park:2015bza}
	J.~H.~Park, S.~J.~Rey, W.~Rim and Y.~Sakatani,
	``O(D, D) covariant Noether currents and global charges in double field theory,''
	JHEP \textbf{11}, 131 (2015)
	doi:10.1007/JHEP11(2015)131
	[arXiv:1507.07545 [hep-th]].
	
	\bibitem{Lescano:2024vrq}
	E.~Lescano,
	``On the inclusion of statistical matter in the non-relativistic limit of NS-NS supergravity,''
	[arXiv:2406.09497 [hep-th]].
	
	\bibitem{Bardeen:1980kt}
	J.~M.~Bardeen,
	``Gauge Invariant Cosmological Perturbations,''
	Phys. Rev. D \textbf{22}, 1882-1905 (1980)
	doi:10.1103/PhysRevD.22.1882
	
	\bibitem{Caprini:2018mtu}
	C.~Caprini and D.~G.~Figueroa,
	``Cosmological Backgrounds of Gravitational Waves,''
	Class. Quant. Grav. \textbf{35}, no.16, 163001 (2018)
	doi:10.1088/1361-6382/aac608
	[arXiv:1801.04268 [astro-ph.CO]].
	
	\bibitem{Mueller:1989in}
	M.~T.~Mueller,
	``Rolling Radii and a Time Dependent Dilaton,''
	Nucl. Phys. B \textbf{337}, 37-48 (1990)
	doi:10.1016/0550-3213(90)90249-D
	
	\bibitem{Apers:2024ffe}
	F.~Apers, J.~P.~Conlon, E.~J.~Copeland, M.~Mosny and F.~Revello,
	``String theory and the first half of the universe,''
	JCAP \textbf{08}, 018 (2024)
	doi:10.1088/1475-7516/2024/08/018
	[arXiv:2401.04064 [hep-th]].
	
	\bibitem{Choi:2022srv}
	K.~S.~Choi and J.~H.~Park,
	``Post-Newtonian Feasibility of the Closed String Massless Sector,''
	Phys. Rev. Lett. \textbf{129}, no.6, 061603 (2022)
	doi:10.1103/PhysRevLett.129.061603
	[arXiv:2202.07413 [hep-th]].
	
	\bibitem{Lee:2023boi}
	H.~Lee, J.~H.~Park, L.~Velasco-Sevilla and L.~Yin,
	``Late-time Cosmology without Dark Sector but with Closed String Massless Sector,''
	[arXiv:2308.07149 [hep-th]].
	
	\bibitem{Bartolo:2004if}
	N.~Bartolo, E.~Komatsu, S.~Matarrese and A.~Riotto,
	``Non-Gaussianity from inflation: Theory and observations,''
	Phys. Rept. \textbf{402}, 103-266 (2004)
	doi:10.1016/j.physrep.2004.08.022
	[arXiv:astro-ph/0406398 [astro-ph]].
	
	\bibitem{Minami:2020odp}
	Y.~Minami and E.~Komatsu,
	``New Extraction of the Cosmic Birefringence from the Planck 2018 Polarization Data,''
	Phys. Rev. Lett. \textbf{125}, no.22, 221301 (2020)
	doi:10.1103/PhysRevLett.125.221301
	[arXiv:2011.11254 [astro-ph.CO]].
	
	\bibitem{Morand:2021xeq}
	K.~Morand, J.~H.~Park and M.~Park,
	``Identifying Riemannian Singularities with Regular Non-Riemannian Geometry,''
	Phys. Rev. Lett. \textbf{128}, no.4, 041602 (2022)
	doi:10.1103/PhysRevLett.128.041602
	[arXiv:2106.01758 [hep-th]].
	
	\bibitem{deBoer:2021jej}
	J.~de Boer, J.~Hartong, N.~A.~Obers, W.~Sybesma and S.~Vandoren,
	``Carroll Symmetry, Dark Energy and Inflation,''
	Front. in Phys. \textbf{10}, 810405 (2022)
	doi:10.3389/fphy.2022.810405
	[arXiv:2110.02319 [hep-th]].
	
\end{thebibliography}
\end{document}